\documentclass[fleqn,usenatbib]{mnras}

\usepackage{newtxtext,newtxmath}

\usepackage[T1]{fontenc}

\DeclareRobustCommand{\VAN}[3]{#2}
\let\VANthebibliography\thebibliography
\def\thebibliography{\DeclareRobustCommand{\VAN}[3]{##3}\VANthebibliography}

\usepackage{graphicx}	
\usepackage{amsmath}	
\usepackage{xspace}
\xspaceaddexceptions{]\}}
\usepackage{multirow}
\usepackage{booktabs}
\usepackage{pdflscape}

\newcommand{\logg}{$\log g$\xspace}
\newcommand{\Teff}{\ensuremath{T_\mathrm{eff}}\xspace}
\newcommand{\rprstar}{\ensuremath{R_\mathrm{p}/R_\mathrm{\star}}\xspace}
\newcommand{\Rp}{\ensuremath{R_\mathrm{p}}\xspace}
\newcommand{\Rstar}{\ensuremath{R_\mathrm{\star}}\xspace}
\newcommand{\Porb}{\ensuremath{P_\mathrm{orb}}\xspace}

\newcommand{\Rsun}{\ensuremath{\mathrm{R_{\odot}}}\xspace}

\newcommand{\Rearth}{\ensuremath{\mathrm{R_{\oplus}}}\xspace}

\newcommand{\degrees}{\ensuremath{\mathrm{^{\circ}}}\xspace}

\newcommand{\tess}{\textit{TESS}\xspace}
\newcommand{\kepler}{\textit{Kepler}\xspace}

\newcommand{\gaia}{\textit{Gaia}\xspace}
\newcommand{\raven}{\texttt{RAVEN}\xspace}
\newcommand{\juliet}{\texttt{juliet}\xspace}

\defcitealias{hadjigeorghiou2025raven}{H25}

\title[\raven candidates]{Automatic search for transiting planets in \tess-SPOC FFIs with \raven: over 100 newly validated planets and over 2000 vetted candidates}

\author[M. Lafarga et al.]{
M.~Lafarga,$^{1,2}$\thanks{E-mail: marina.lafarga-magro@warwick.ac.uk}
D.~J.~Armstrong,$^{1,2}$
K.~Cui,$^{1,2}$
A.~Hadjigeorghiou,$^{1,2}$
V.~Kunovac,$^{1,2}$
L.~Doyle,$^{1,2}$
\newauthor
E.~M.~Bryant,$^{1,2}$
R.~F.~Díaz,$^{3, 4}$
L.~A.~Nieto$^{3, 5}$ and 
A.~Osborn$^{1,2}$
\\
$^{1}$Department of Physics, University of Warwick, Gibbet Hill Road, Coventry CV4 7AL, United Kingdom\\
$^{2}$Centre for Exoplanets and Habitability, University of Warwick, Coventry, CV4 7AL, UK\\
$^{3}$Instituto de Ciencias Físicas (CONICET / ECyT-UNSAM), Campus Miguelete, 25 de Mayo y Francia, (1650) Buenos Aires, Argentina\\
$^{4}$Instituto Tecnológico de Buenos Aires (ITBA), Iguazú 341, Buenos Aires, CABA C1437, Argentina\\
$^{5}$Gerencia de Tecnología de la información y de las Comunicaciones (GTIC), Subgerencia Vinculación y Desarrollo de Nuevas Tecnologías de la Información, \\DTE-CNEA. Centro Atómico Constituyentes, Av. Gral. Paz 1499, (1650) Buenos Aires, Argentina\\
}

\date{Accepted XXX. Received YYY; in original form ZZZ}

\pubyear{2025}

\begin{document}
\label{firstpage}
\pagerange{\pageref{firstpage}--\pageref{lastpage}}
\maketitle

\begin{abstract}
%
Space-based missions such as \tess are identifying a wealth of short-period ($\lesssim30$~d) transiting planets. Despite the growing number of confirmed and candidate planets, the sample is still incomplete and highly biased, challenging demographic studies. Moreover, there are still a large number of unconfirmed candidates that can end up being false positives.
We use the new pipeline \raven to perform a uniform search and validation of transiting planet candidates in \tess data. We focus on a magnitude-limited sample of over 2.2 million main sequence stars well characterised by \gaia and observed by \tess in its Full Frame Images during its first 4 years of operations (sectors 1 to 55).
We aim to detect candidates with periods within $0.5-16$~days.
\raven detects candidates with a box least squares algorithm, classifies them into transiting planets and false positives using machine learning models trained with realistic simulations, and performs statistical validation.
We present several samples of candidates with different levels of vetting and validation. 
We newly validate 118 planets, including 31 newly detected here.
We also present a sample of over 2000 candidates not validated but with high probability of being planets, including $\sim1000$ new candidates, a small sample of newly identified mono- and duo-transiting candidates, and a sample of large radii ($>8~\Rearth$) candidates with high planet probability suited for further follow-up.
Our samples of vetted and validated transiting planet candidates represent a major effort towards improving the candidate sample from \tess.
\end{abstract}




\section{Introduction}

Space-based missions such as \kepler/\textit{K2} \citep{borucki2010kepler,howell2014k2} and \tess \citep{ricker2015tess} have significantly increased the number of confirmed and candidate transiting planets in the last two decades.
\kepler found over 4700 transiting planet candidates by continuously monitoring over 110\,000 stars for four years, of which over 2700 are confirmed or validated\footnote{NASA Exoplanet Archive (NEA) \url{https://exoplanetarchive.ipac.caltech.edu/docs/counts_detail.html}, accessed on 13/08/2025}.
After being re-purposed into the \textit{K2} mission, the telescope performed 20 80-day long campaigns targeting different pointings along the ecliptic, and detected over 1500 candidates, about 500 of them confirmed or validated.
\tess has been performing a nearly all-sky survey in the form of 27-day sectors for the past seven years. It has found over 7000 candidates and confirmed or validated as true exoplanets almost 700.
Transiting exoplanets from these three missions make up over two thirds of all known exoplanets, which highlights the importance of large space-based surveys.

This large number of confirmed exoplanets has enabled the first exoplanet demographic studies, which have uncovered interesting features such as the hot Jupiter pile-up \citep[a clustering of giant planets at short periods, see e.g.][]{cumming2008distribution,wright2012hj}, the Neptunian desert \citep[a lack of Neptune-size planets at short periods, $\lesssim3$~d, e.g.][]{mazeh2016desert,lundkvist2016desert}, and the radius valley \citep[a lack of planets between 1.5 and 2 Earth radii at periods $<100$~d, e.g.][]{fulton2017radiusvalley,vaneylen2018radiusvalley}. 
Transiting planets allow for direct measurement of radii, which can be combined with masses derived from spectroscopic observations to obtain bulk density estimates.
These planets are also high-priority targets for follow-up atmospheric surveys. 
Demographic studies, densities, and atmospheric constraints are all key to improving our understanding of planet formation and evolution theories, and to uncover underlying exoplanet occurrence rates.

Despite the large number of confirmed exoplanets, there is an even higher number of candidates yet to be confirmed. 
One of the main challenges in the confirmation of candidate transiting planets is the numerous false positives (FPs) common in these kinds of searches. These FPs include eclipsing binaries, hierarchical systems producing transits in background or nearby stars to the target star, and signals due to stellar variability or instrument systematics, which can appear as transiting planets and are detected by processing pipelines. Hence, searches for transiting planets tend to include some sort of FP vetting.
Generally, most pipelines perform (a subset of) the following steps: extracting light curves from photometric observations, detrending the extracted light curves to remove instrument systematics and stellar variability while preserving the transit signals, searching for periodic transit-like signals (threshold-crossing events, TCEs), and applying a diverse range of vetting (and validation) procedures to distinguish between true transiting planets and FPs.

There are many different pipelines that provide different levels of candidate vetting from \tess light curves.
\tess TCEs are identified by the Science Processing Operations Center (SPOC) pipeline \citep{jenkins2016tessspoc} and the Quick Look Pipeline \citep[QLP,][]{huang2020qlp1,huang2020qlp2}.
The SPOC search is based on a wavelet, adaptive matched filter, and is applied to both short cadence data of the primary mission (target pixel files) and longer cadence full frame images (FFIs), while QLP uses a box least squares algorithm \citep[BLS,][]{kovacs2002bls} and is applied to FFIs.
These pipelines use several diagnostics to provide vetting of the TCEs and flag FPs, some of which were originally developed for \kepler data \citep[e.g.][]{mullally2015robovetter,mccauliff2015autovetter,coughlin2016robovetter,thompson2018kepler25,kostov2019dave,guerrero2021toi}.
Due to the large amount of data provided by these missions, pipelines based on machine learning methods such as random forest classifiers, Gaussian process (GP) classifiers, self-organising maps (SOMs), or neural networks have also gained popularity to vet and classify candidates \citep[e.g.][]{mccauliff2015autovetter,armstrong2017som,shallue2018astronet,ansdell2018exonet,dattilo2019astronetK2,yu2019astronettriage,caceres2019arps,osborn2020nn,montalto2020diamante,valizadegan2022exominer,valizadegan2023exominer,salinas2023nn,tey2023astronettriage,melton2024dtarps1,melton2024dtarps2,tardugnopoleo2024,wang2024gpfc,valizadegan2025exominer,fiscale2025dartvetter,hernandezcarnerero2025attention}.
Vetting procedures are key to rank and prioritise candidates for further characterisation.
Promising \tess candidates are released to the community as \tess Objects of Interest \citep[TOI,][]{guerrero2021toi} and Community TOIs (CTOIs) for further follow-up and confirmation.

Once a likely transiting planet candidate is identified, confirmation of its planetary origin is ideally provided by time-series radial velocity (RV) measurements from high-resolution spectroscopy, from which the mass of the candidate can be derived.
If these are not available, some FPs can be ruled out with different kinds of follow-up observations, mainly: reconnaissance spectroscopy can identify multiple sets of stellar spectral lines and/or large RV shifts due to a stellar-mass companion, multi-band transit photometry can identify wavelength-dependent transit depths due to unresolved stars, and high-resolution imaging can identify close (within fractions of an arcsecond) unresolved stars.

A different approach is to use statistical validation, which relies on probabilistic models to estimate the likelihood of a candidate being a true transiting planet compared to FP scenarios. Candidates that achieve a high probability (with a typical threshold of $\geq0.99$ probability) are considered validated, that is, they have a high degree of certainty, close to that of planets with a mass measurement. 
Validation-based methods are useful and efficient because they do not rely on follow-up observations, which are generally expensive (requiring relatively large amounts of time in over-subscribed observatories) and might not be possible for all candidates, and can be efficiently applied to the large number of candidates provided by large space-based missions.
There are several pipelines that can provide validation of \kepler/\textit{K2} candidates such as \texttt{BLENDER} \citep{torres2011blender}, \texttt{vespa} \citep{morton2012validation,morton2016vespa}, \texttt{PASTIS} \citep{diaz2014pastis,santerne2015pastis}, a combination of several machine learning models \citep{armstrong2021kepler}, or \texttt{ExoMiner} \citep{valizadegan2022exominer}.
A common pipeline to vet and validate \tess TOIs is \texttt{TRICERATOPS} \citep{giacalone2021triceratops}. \texttt{TRICERATOPS} computes the probability of a given candidate being due to a true transiting planet and several FPs, incorporating prior knowledge of the host and nearby stars, and current understanding of planet and stellar rates. \texttt{TRICERATOPS} was originally used to validate 12 TOIs and has since then been used as a validation tool by other searches \citep[e.g.][]{devorapajares2024sherlock,giacalone2022terrestrial}. An extended version of the pipeline also incorporates the usage of ground-based multi-colour photometry commonly used for vetting \citep{gomezbarrientos2025triceratops}.

Other than the official \tess TCE and TOI/CTOI releases, there are alternative catalogues of candidates at different stages of vetting and validation detected using a range of different pipelines. These exploit the vast amounts of \tess data using different light curve extractions, TCE finding methods, vetting checks with a range of thresholds, and/or validation. By using alternative approaches, these works can find new candidates missed by the main pipelines. Another motivation for these searches is to use end-to-end pipelines to be able to quantify each step of the process, understand biases, and learn about the underlying distribution of exoplanets from well characterised samples of stars and candidates.

For instance, the \tess Faint Star Search \citep{kunimoto2022qlpfaint} uses the \texttt{LEO-Vetter} pipeline \citep{Kunimoto2025} to regularly vet QLP candidates in faint stars. It currently accounts for almost half of the reported TOI, over 3700 candidates.

\citet{cloutier2019orion} presented an early search for candidates on low-mass stars with the \texttt{ORION} pipeline on \tess SPOC 2-min cadence data of sectors 1 and 2. The pipeline performs its own transit search, several vetting checks, and validates candidates with \texttt{vespa}. It detected seven new candidates.

\citet{feliz2021nemesis} focused on close-in planets orbiting a sample of 33\,000 nearby ($<100$~pc) M dwarfs and applied the \texttt{NEMESIS} pipeline to \tess FFIs (sectors 1 to 5) to extract detrended photometry, search for transits in single sectors, and perform vetting checks. This work detected 29 candidate planets amongst 183 TCEs, 24 of which are new.

\citet{rao2021nigraha} identified 38 new candidates in seven sectors of \tess data with the \texttt{Nigraha} pipeline, which performs its own search for TCEs with the transit least squares algorithm \citep[TLS,][a BLS alternative]{hippke2019tls} and uses the pipeline \texttt{DAVE} \citep{kostov2019dave} and several vetting steps, including its own neural network, to rank candidates.

In the citizen science project Planet Hunters \tess \citep{eisner2021pht}, citizen scientists identified TCEs in Year 1 of \tess data (sectors 1 to 26) and detected 90 new candidates. TCEs were ranked by a clustering algorithm and visual vetting was performed to the highest ranked candidates.

\citet{olmschenk2021nn} used a neural network for transit identification in \texttt{eleanor} FFI light curves \citep[an alternative extraction to the SPOC and QLP pipelines,][]{feinstein2019eleanor} of the first two years of \tess data, together with the pipeline \texttt{DAVE} and manual inspection for vetting. This process identified 181 new vetted transiting planet candidates.

\citet{nardiello2020pathos2} looked for candidates in stellar clusters and young associations in Year 1 of \tess data and found 33 new candidates by performing its custom light curve extraction and TLS search.
\citet{fernandes2022pterodactyls} presents the \texttt{PTERODACTYLS} pipeline, which is also focused on the detection of planets around young stars and uses \texttt{eleanor} FFI light curves, applies several vetting checks, and relies on \texttt{TRICERATOPS} for validation.

\citet{devorapajares2024sherlock} applied another end-to-end pipeline, \texttt{SHERLOCK}, to find new candidates in TOIs. \texttt{SHERLOCK} can use different light curve extractions, searches for planets with a TLS, applies several vetting checks, and performs validation with \texttt{TRICERATOPS}. The authors applied the pipeline on stars hosting TOIs and detected four new candidates, and presented future plans for an end-to-end search on nearby ($<50$~pc) K and M dwarfs to detect small exoplanets.

\citet{montalto2020diamante} and \citet{montalto2023diamante} used the \texttt{DIAmante} pipeline to search for candidates in a sample of 1.4 million FGKM stars in Years 1 and 2 of \tess FFIs (sectors 1-26). \texttt{DIAmante} extracts custom photometry from the FFIs, identifies signals with a BLS, and vets transiting planets from FPs with a random forest classifier trained on simulations injected on \tess light curves. This search resulted in 1160 candidates, 842 of which are newly discovered. These include 612 (516 new) candidates in the Neptunian desert \citep[as defined by][]{mazeh2016desert}, and several ultra-short period (USP, $<1$~d) super-Earths.

The same \texttt{DIAmante}-extracted light curves \citep[\tess Year 1 FFIs of $\sim$0.9 million FGKM stars,][]{montalto2020diamante} were used together with an update of the \texttt{ARPS} pipeline \citep[][originally developed for \kepler]{caceres2019arps,caceres2019kepler} in the DTARPS project presented in \citet{melton2024dtarps1,melton2024dtarps2,melton2024dtarps3}. \texttt{ARPS} detrends light curves with a differencing algorithm, further removes autocorrelated noise with ARIMA models, identifies candidates with a Transit Comb Filter (TCF), and uses a random forest classifier trained with simulations of transiting planets and FPs injected on light curves. Further vetting of the over 7000 candidates found is applied to reduce the number of FPs. This search resulted in a catalogue of 772 candidates (about 60\% new) with several vetting flags. A range of validation tests, including \texttt{TRICERATOPS} and limited reconnaissance spectroscopy, estimated the FP rate of the sample to be $\sim50\%$. Similarly to \citet{montalto2023diamante}, the DTARPS search found hundreds of candidates (about half of the sample, 387 candidates) in the Neptunian desert, as well as several USP candidates.

Recently, \citet{salinas2025transformer} employed a neural network to identify transits in light curves without the need of assuming a periodicity or phase-folding them. This model was applied to Year 1 of \tess FFIs and identified 214 new candidate systems.

In this work we apply the new vetting and validation pipeline \raven \citep[][hereafter \citetalias{hadjigeorghiou2025raven}]{hadjigeorghiou2025raven} to a sample of $\sim2.26$ million \tess FFI stars (observed during the first four years of \tess, sectors 1 to 55) well characterised by \gaia \citep[see][]{doyle2024tessgaia}. \raven identifies transiting planet candidates, classifies them into likely planets and a range of false positive scenarios, and can perform statistical validation. To perform the classification, \raven uses machine learning models trained with realistic simulations.
We limit our search to candidates with periods from 0.5 up to 16~d (as longer periods would require multiple sectors to be detected and the instrumental false positives are likely to be different than those of short period) and radii up to 16~\Rearth. These limits allow us to uniformly detect candidate transiting planets and study the demographics of short period planets, covering the Neptunian desert and surrounding regions.
We present several samples of vetted and validated transiting planet candidates, including hundreds of new vetted candidates and $\sim100$ newly validated planets.
This article is structured as follows. In Section \ref{sec:data} we describe the stellar sample and \tess data used. Section \ref{sec:raven} summarises the methodology of the \raven pipeline and its application to our sample of stars. In section \ref{sec:res_cand} we present the different samples of transiting planet candidates that we identified, and in Section \ref{sec:discussion} we discuss our main results. We conclude in Section \ref{sec:conclusion}.


\section{Data}\label{sec:data}

Our stellar sample consists of all main sequence stars present in the \tess FFI light curves of sectors 1 to 55 well-characterised by \gaia, presented in \citet{doyle2024tessgaia}.
\citet{doyle2024tessgaia} performed a cross-match between all stars in the FFI light curves produced by \tess SPOC \citep[][]{jenkins2016tessspoc,caldwell2020spocffi} for sectors 1 to 55 (corresponding to \tess Cycles 1 to 4) and all stars in \gaia Data Release 3 \citep[DR3,][]{gaia2016mission,gaia2023dr3} brighter than \textit{G} magnitude 14 and with relative parallax uncertainty better than 20\% (i.e. parameter \texttt{parallax\_over\_error} > 5). 
These constraints allow for reliable stellar parameters.
This cross-match resulted in $\sim$2.75 million targets. 
Additional FFI targets that failed the original cross-match were added to the sample by performing a radius search within the \gaia DR3 catalogue, resulting in a final sample of $\sim$2.9~million targets.
The main sequence targets were subsequently isolated by performing a cut at $\log g > 3.5$, resulting in a sample with $\sim$2.3~million main sequence targets. 
Some of the targets added by the radius search after the original cross-match do not meet the criteria of brightness $G < 14$ and relative parallax precision $< 20\%$. To have a uniform sample, we removed targets not meeting the criteria of the original cross-match from the main sequence sample presented in \citet{doyle2024tessgaia}. After removing those targets, we are left with a sample of 2\,259\,830 main sequence targets brighter than $G$ magnitude 14 and relative parallax uncertainty better than $20\%$. 

We obtained the FFI light curves of our $\sim$2.26 million targets for sectors 1 to 55 reduced with the SPOC pipeline\footnote{DOI \href{https://dx.doi.org/10.17909/t9-wpz1-8s54}{10.17909/t9-wpz1-8s54}} \citep{jenkins2016tessspoc,caldwell2020spocffi} from the Mikulski Archive for Space Telescopes (MAST\footnote{\url{https://archive.stsci.edu/}}).
FFIs from sectors $1-26$ (cycles 1 and 2) have an observational cadence of 30 minutes, and those from sectors $27-55$ (cycles 3 and 4), of 10 minutes.
We used the pre-search data conditioning simple aperture photometry (PDCSAP) flux time series, from which long term trends have been removed with co-trending basis vectors \citep{stumpe2012pdc,smith2012pdc,stumpe2014msmap}. The flux of the PDCSAP light curves has also been corrected for crowding effects (which, if not corrected, can impact results due to \tess' large pixels, since flux from nearby stars enters in the photometric aperture).

\begin{figure}
\centering
\includegraphics[width=0.5\textwidth]{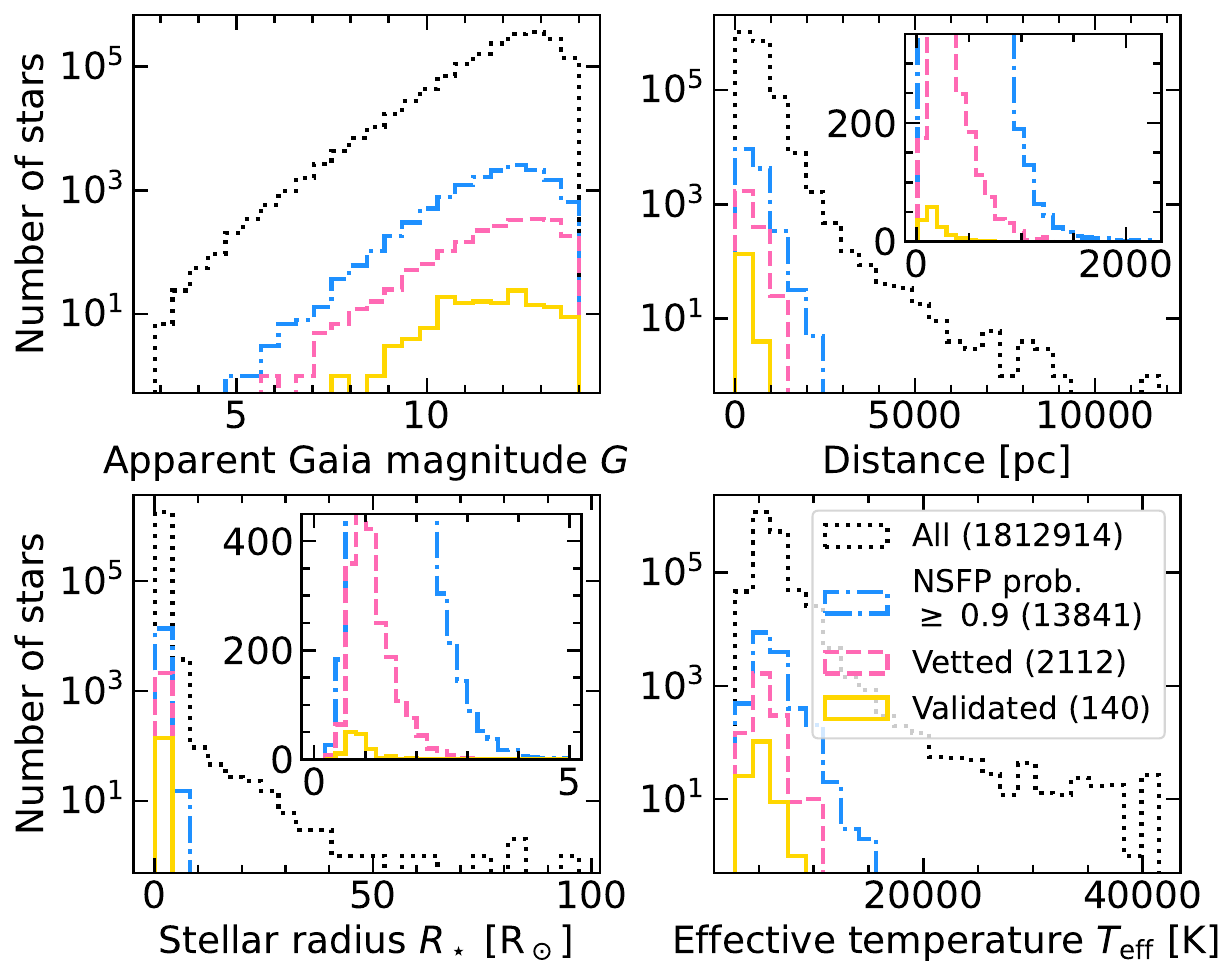}
\caption[]{\gaia properties of our stellar samples: apparent \gaia magnitude $G$ (from DR3, top left panel), distance (computed by inverting the DR3 parallax, top right, note that the parallax values of the stars in our sample have uncertainties better than 20\%), stellar radius (from DR2, bottom left), and effective temperature (from DR2, bottom right).
Different colours/line styles represent different samples used in this work: black dotted lines correspond to all stars (after SDE and MES cuts, see Sect. \ref{sec:bls}), blue dashed dotted lines show the subsample of stars with candidates with NSFP mean classifier probability $\geq0.9$ (see Sect. \ref{sec:nsfp_vetting}), pink dashed lines show the subsample stars with vetted candidates (see Sect. \ref{sec:res_cand_vet}), and yellow solid lines show the subsample stars with validated candidates (see Sect. \ref{sec:res_cand_val}).
The numbers in parenthesis in the legend show the number of stars in each sample.
The inset panels show a zoom in of the stars in the vetted and validated samples. Note that the y-axes of the main panels are in logarithmic scale while, in the insets, they are in linear scale.
}
\label{fig:stellar_sample}
\end{figure}


\section{Candidate identification, classification and ranking}\label{sec:raven}

We followed the methodology of the RAnking and Validation of ExoplaNets (\raven) pipeline\footnote{\url{https://github.com/ahadjigeorghiou/RAVEN/}}, presented in \citetalias{hadjigeorghiou2025raven} and based on the framework introduced by \citet{armstrong2021kepler} for \kepler candidates.
\raven first identifies candidate transiting planets in \tess-SPOC FFI light curves with a BLS search \citep[][]{kovacs2002bls}. For each candidate, \raven then computes relevant features that are subsequently used by several machine learning models that classify candidates into transiting planets (\emph{Planet} category) or FPs. The machine learning models have been trained with realistic simulations of planets and a range of common FPs. The probability obtained from each classifier is then combined with the appropriate prior to obtain the posterior probability of the candidate being a Planet or a FP. A final \raven probability is obtained by taking the minimum posterior probability of all FP scenarios, which makes sure that the highest ranked candidates have a high probability of being true planets for all FP scenarios considered.
In the following sections, we present our application of \raven to our sample of $\sim2.26$ million main sequence stars, and refer the reader to \citet{hadjigeorghiou2024positional} and \citetalias{hadjigeorghiou2025raven} for more details about the pipeline.


\subsection{\raven candidate identification}\label{sec:bls}

We started by performing a BLS search on all light curves of our stellar sample.
\raven uses the \texttt{cuvarbase} BLS implementation for GPUs \citep{hoffman2022cuvarbase}\footnote{\url{https://github.com/johnh2o2/cuvarbase}} to shorten computing times compared to other BLS implementations that work only on CPUs.
Before computing the BLS, each light curve is detrended with a Savitzky-Golay filter with a third-degree polynomial and a 4-day window, which removes low-frequency variations while preserving transit-like signals \citep{savitzky1964smooth}.
The BLS periods considered ranged from 0.5 to 16~days.
The BLS power spectrum obtained was then detrended with a minimum filter, and the signal detection efficiency \citep[SDE, originally defined in][]{kovacs2002bls} was computed by subtracting from the detrended power its median value and then dividing the result by the median absolute deviation of the detrended power. 
After transforming the BLS from power to SDE, the five highest peaks were selected, which are our candidate signals. 

It is common for the BLS to show significant peaks at the true period of a signal as well as multiples or integer fractions of that period (harmonics). To avoid selecting as candidates periods related to the same event, which could cause other events such as less significant transits in multi-planet systems to be missed, \raven attempts to identify and remove the peaks related to harmonics of already selected events. Once a significant peak, from strongest to weakest, was selected at a specific period, \raven checked subsequent strong peaks to test if they are the double or half of that period. When that is indeed the case, the peak was skipped and the selection continued to the next peak. 

Our full BLS run includes spurious signals and signals with very low signal-to-noise ratio (S/N) that are unlikely to be identified as true planetary companions. These candidates have low SDE and low multiple event statistic \citep[MES,][another significance statistic that estimates the event S/N]{twicken2018keplervalidation}.
Recovery tests performed on \tess-SPOC FFI light curves with injected planetary transits showed that the recovery rate drops for signals with SDE below 7 and MES below 0.8 \citepalias{hadjigeorghiou2025raven}. Therefore, following \citetalias{hadjigeorghiou2025raven}, we discarded all signals with SDE~$\leq$~7 and MES~$\leq$~0.8 to remove low S/N candidates and decrease the total number of candidates to be processed further, saving computational resources.
After performing these cuts we were left with 5\,664\,552 candidates (i.e. from the peaks identified in the BLS) from 1\,812\,914 different stars (which amounts to about 3 candidates or peaks per star on average). The main stellar properties of this sample are shown in Fig. \ref{fig:stellar_sample}.


\subsection{\raven machine learning classification}\label{sec:ml}

\subsubsection{Features}

After the BLS process, \raven prepares a range of different features that are used by the machine learning models. These features are taken from the \tess Input Catalogue \citep[TIC,][]{stassun2019tic} or \gaia archives, or directly computed from the \tess data.
The features include properties of the target star (e.g. magnitude, colour, radius, distance), transit metrics derived from the BLS, and from a trapezoid fit and \texttt{batman} \citep{kreidberg2015batman} model fit to the phase-folded light curve (e.g. period, transit durations, transit depths), transit significance metrics (e.g. S/N, quantities derived from the single event and multiple event statistics), metrics related to nearby sources, and metrics based on self-organising maps (SOMs), an unsupervised clustering algorithm trained to distinguish between Planets and FPs based on transit shape \citep{kohonen1982som,armstrong2017som}.

\subsubsection{Models and performance}

\raven includes two different types of machine learning classifiers: gradient boosted decision trees (GBDT), implemented using the extreme gradient boosting model \citep[\texttt{XGBoost},][]{chen2016xgboost}, and a Gaussian Process (GP) classifier \citep{rasmussen2006gp}.
These models have been trained with several sets of light curves containing simulated planetary transits (Planet, our target) and eight types of common FPs. These FPs include eclipsing binaries (EB), background eclipsing binaries (BEB), hierarchical eclipsing binaries (HEB), hierarchical transiting planets (HTP), nearby transiting planets (NTP), nearby eclipsing binaries (NEB), nearby hierarchical eclipsing binaries (NHEB), and an extra set of non-simulated false positives (NSFP).

The synthetic scenarios Planet, EB, BEB, HEB, and HTP were simulated with the \texttt{PASTIS} code \citep{diaz2014pastis, santerne2015pastis}. The target star for each event was randomly selected from the main sequence sample used here \citep{doyle2024tessgaia} with parameters from \gaia and the TIC.
The simulations were then injected in the \tess-SPOC FFI light curves of the corresponding target star. 
The nearby star synthetic sets (NTP, NEB, NHEB) were not directly simulated. 
Instead, simulated events of the Planet, EB, and HEB scenarios were first diluted in accordance with the selected target star and then injected \citepalias[for details and dilution factor distributions, see][]{hadjigeorghiou2025raven}.
The use of real \tess targets and light curves for \raven's synthetic sets ensure a realistic representation of the \tess data that we then used to identify our real candidates.
The simulated Planets represented our parameter space of interest: planets with radius up to $16~\Rearth$ and orbital periods from 0.5 to 16~d.

To account for false positives not included in our simulations, such as stellar variability and instrumental noise, \raven includes as a training set a sample of detected events from a random selection of \tess-SPOC FFI candidates. Given the fact that the majority of the candidates from the FFIs will not be a true planetary signal, we expect that these randomly selected NSFP candidates are representative of true false positives in our sample. To ensure this, known TOIs and high significance events (MES~$>100$) were excluded from the sample. We note here that the NSFP sample might also include other common FP scenarios, mainly EBs.

The synthetic set light curves were processed by \raven in a similar way as the light curves of our candidates. First, a BLS search and recovery was performed, selecting only  the signals that were correctly identified in the BLS.
We considered a signal as recovered if the BLS period matched the injected period within 2\% and if the BLS mid-transit time was within 0.5~d of the injected one (note that we allow for the BLS mid-transit time to be an integer number of periods away from the injected one). For scenarios including eclipsing binaries (i.e. EB, BEB, HEB, NEB, NHEB), signals whose BLS period matched half and double the injected periods were also considered successful recoveries, since these harmonics of the period are commonly detected FPs.
Features were computed for the recovered signals, which were subsequently used to train and test the GBDT and GP classifiers. 
raven uses a binary classification, where the simulated Planets dataset is paired with each FP dataset separately, resulting in eight different training (and test and validation) sets and hence eight different classifiers, one per FP scenario considered. For a given candidate, each of these classifiers returns the probability of the candidate being a Planet compared to a specific FP scenario. \raven defines the probability of a candidate being a Planet as 1, and the probability of the candidate being the considered FP as 0. Hence, candidates with probability close to 1 are classified as likely Planets, and those with probability close to 0 are classified as likely FPs.
\raven uses one GBDT and one GP classifier for each Planet-FP pair (16 classifiers in total).
For each Planet-FP pair, \raven computes the mean of the GBDT and GP probabilities, resulting in one mean probability per FP scenario.

\citetalias{hadjigeorghiou2025raven} showed that these models are effective at distinguishing Planets from FPs.
Models applied to test sets (simulations not used during the training) achieve a receiver operating characteristic area-under-curve (ROC-AUC\footnote{ROC-AUC: area under the ROC curve, which plots the rate of true positives (also called recall or sensitivity, fraction of correctly classified planets over all correct classifications, including both planets and false positives) against the rate of false positives (fraction of incorrectly classified FPs over all FPs in the set) at different probability thresholds.}) scores of $99\%$ for all but one classifier (the NTP, which has an AUC score of $97\%$), 
accuracy\footnote{Accuracy: fraction of correct classifications overall, including both planets and FPs.} above $96\%$ for all but one classifier (again the NTP, which has an accuracy of $92\%$) at a probability threshold of 0.5, and 
precision\footnote{Precision: fraction of correctly classified planets over all candidates (correct and incorrect) classified as planets.} scores greater than 99\% across all scenarios for a probability threshold of 0.9.
The full pipeline was also tested on an independent sample of 1367 pre-classified TOIs including known and confirmed planets and FPs within our parameter space of interest. 
The pipeline again demonstrated a strong performance, achieving a ROC-AUC score above 97\% and a precision above 97\% at a probability threshold of 0.9, and an accuracy of $91\%$ at a probability threshold of 0.5, despite the fact that these TOIs were primarily classified based on follow-up data which \raven did not have access to.

The ROC-AUC and the precision metrics are crucial in describing the pipeline's performance when it comes to ranking candidates and validating planets, our goals here. In particular, the precision defines how many instances of true FPs have high probabilities (i.e. incorrectly classified as planets) compared to the true planets. For a good validation pipeline, the validated sample should be as pure as possible, which is why a high precision is crucial. As for the AUC score, it defines the pipeline's ability to rank true planets above the true FPs. As ranking of candidates is paramount to the pipeline's goal, this score again should be as high as possible. For the test sets, both scores are above 99\% for almost all scenarios, while for the TOIs, both scores are above 97\%.
Regarding the accuracy, a score of $91\%$ on the TOI sample is a high number when considering the sample. The TOIs are a specially curated collection of \tess candidates, which have been detected through a variety of methods and data reductions, and have also passed different vetting stages, including manual vetting. This means that FPs which are clearly identifiable from the light curve alone have often been already excluded, leaving cases that require dedicated investigation and perhaps further follow-up to determine their FP status. The list also includes TOIs missed by regular period-based searches, such as promoted CTOIs and also candidates on young stars that require specialised de-trending for detection and analysis. As a result, the sample is a rather challenging test case for the \raven pipeline, and one that is rarely if ever used for testing similar tools. Achieving above 90\% accuracy but also high precision and ROC-AUC scores on this sample strongly demonstrates the pipeline's ability to provide meaningful and accurate probability scores for the various \tess candidates.


\subsubsection{Planet-NSFP classifier vetting}\label{sec:nsfp_vetting}

The \raven BLS search of candidates in our stellar sample of $\sim2.26$M targets resulted in 5\,664\,552 candidates. We expect that most of these candidates will be false positives, mainly signals caused by instrumental and stellar noise.
Therefore, we used the Planet-NSFP classifiers, which are primarily sensitive to these false positives, to perform an initial vetting of the candidates.
We note again that other FP scenarios such as EBs can also be classified as NSFP at this stage depending on how alike they are to the NSFP sample.

\begin{figure}
\centering
\includegraphics[width=0.5\textwidth]{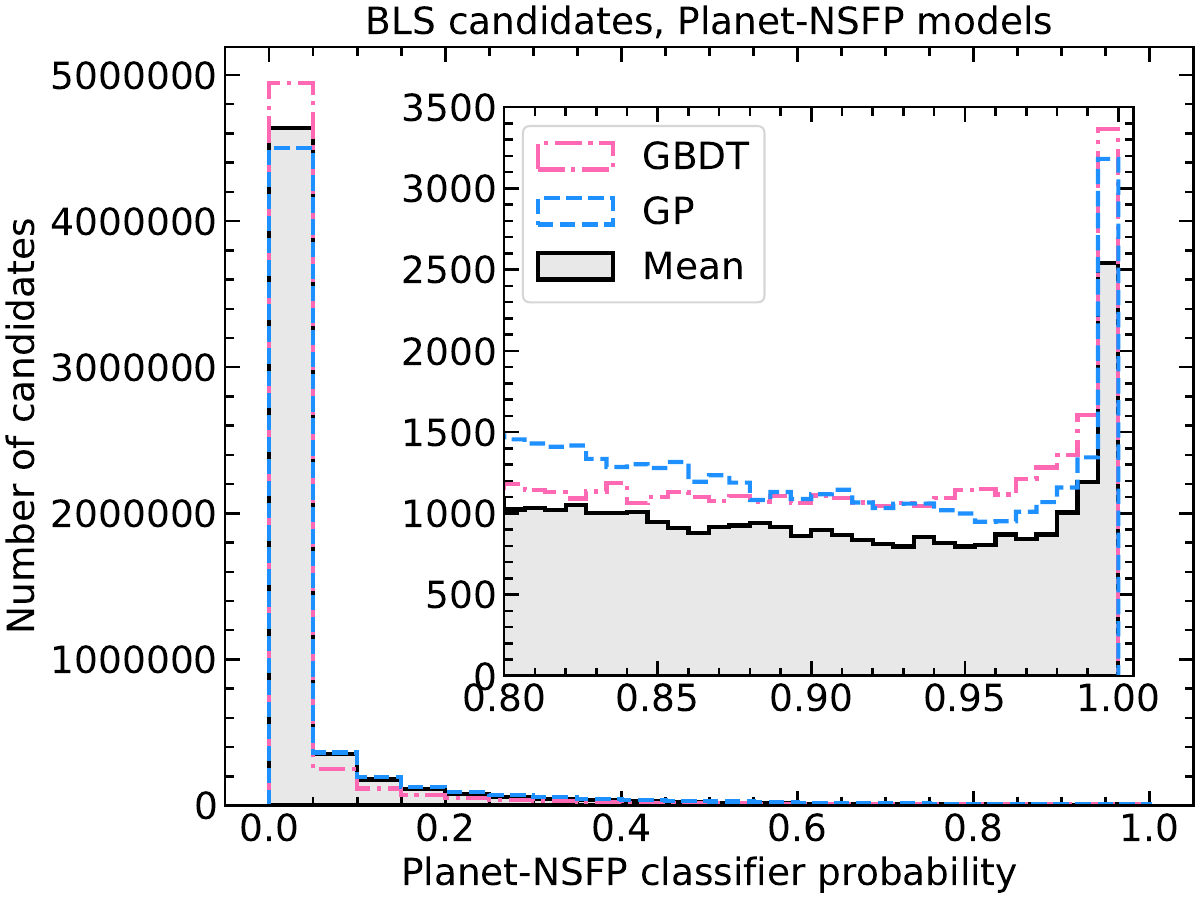}
\caption{Classification results of all candidates for the Planet-NSFP classifiers. Each histogram shows the posterior probability of a different classifier: pink dash-dotted line for the GBDT, blue dashed line for the GP, and solid black line and grey-filled for the mean of both GBDT and GP. Candidates with probability close to 1 are classified as Planet, and candidates with probability close to 0 are classified as NSFP. The inset panel shows a zoom in on the candidates with probabilities above 0.8.}
\label{fig:cand_pla_nsfp_prob}
\end{figure}

The Planet-NSFP classifier probabilities for our BLS candidates are shown in Fig. \ref{fig:cand_pla_nsfp_prob}. In this case, a classifier probability close to 1 means that the candidate is classified as a Planet, while a probability close to 0 means that the candidate is classified as an NSFP. 
As expected, most of the BLS candidates are consistent with being a false positive (probability $\sim0$) rather than a planetary candidate (probability $\sim1$).
We used the mean Planet-NSFP probability to apply an initial vetting step by discarding candidates with probability below 0.9. This threshold value is based on the application of this classifier to the test set used to assess the performance of the models, where all the candidates with probability $\geq0.87$ showed a clear transit-like signal, rather than something consistent with instrumental or stellar variability \citepalias{hadjigeorghiou2025raven}.
This cut leaves us with 14\,815 candidates from 13\,841 different stars.

\begin{figure*}
\centering
\includegraphics[width=\textwidth]{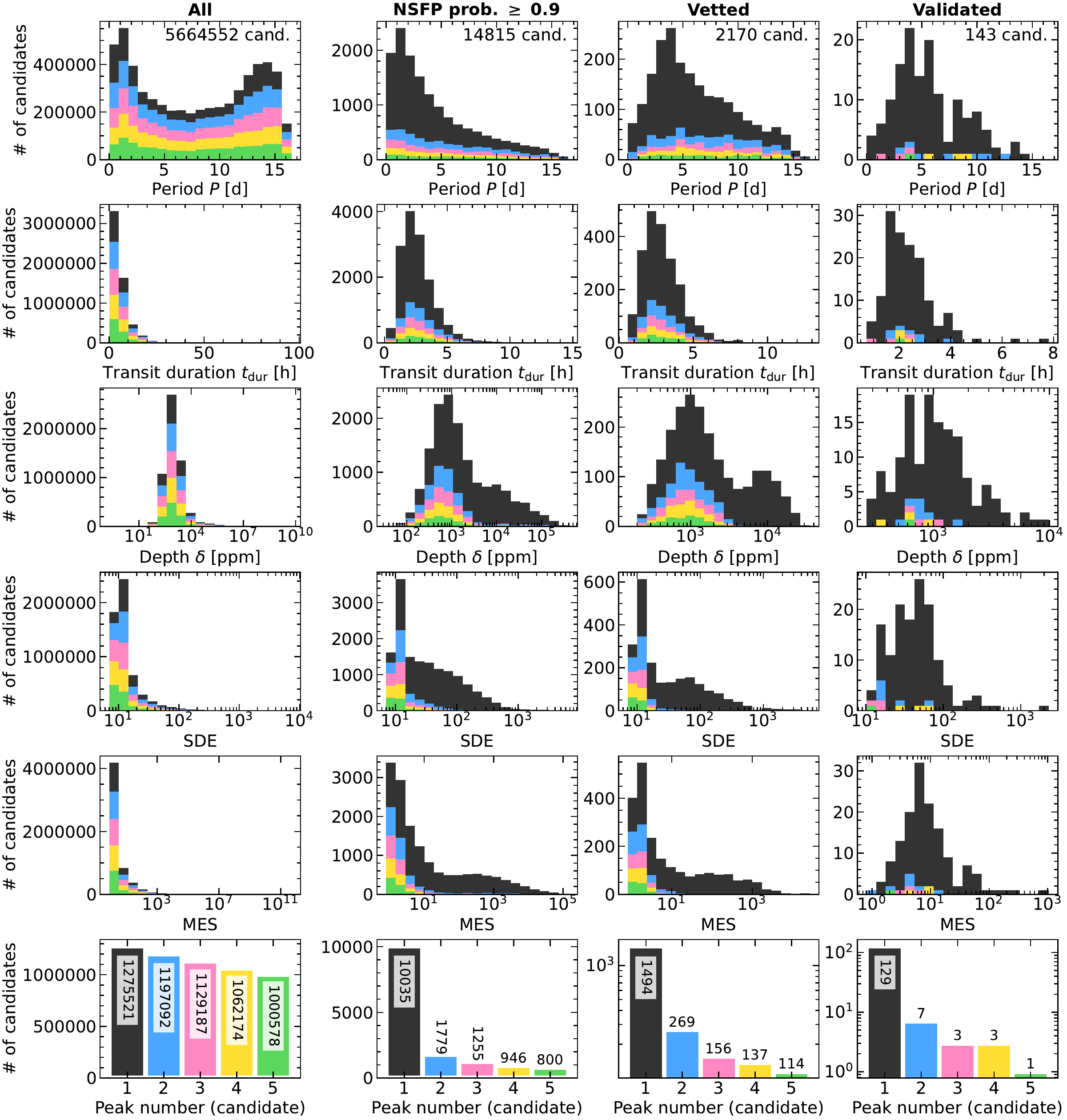}
\caption{Distribution of the main BLS properties, from top to bottom: period $P$, transit duration $t_\mathrm{dur}$, depth $\delta$, signal detection efficiency SDE, multiple event statistics MES, and candidate peak number, colour-coded with the candidate peak number. 
The different columns show different samples of candidates, from left to right: all candidates (after SDE and MES cuts), subsample of candidates with non-simulated false positive (NSFP) mean classifier probability $\geq0.9$ (see Sect. \ref{sec:nsfp_vetting}), vetted candidates (see Sect. \ref{sec:res_cand_vet}), and validated candidates (see Sect. \ref{sec:res_cand_val}).
In the first row of panels, the top right numbers show the total number of candidates in each sample.
In the last row of panels, the numbers on each bin show the number of candidates per BLS peak number.}
\label{fig:cand_bls_hist}
\end{figure*}

In Fig. \ref{fig:cand_bls_hist} we show the distribution of the main BLS properties of all candidates {(first column)} and of the sample of candidates with Planet-NSFP mean probability $\geq0.9$ (second column).
The period distribution of the full BLS sample has a significant number of candidates with periodicities around 14~days. We expect most of these periodicities to be of instrumental origin, since this period is close to half the length of a \tess sector (27.4/2 days). The Planet-NSFP classifier cut removes most of these candidates.
The Planet-NSFP classifier also removes most of the candidates found in the second to the fifth peak identified in the periodogram, which tend to be candidates with low S/N.
We also see that extreme outliers, such as candidates with long transit duration, very shallow or deep depths, and very high SDE and MES, are also removed by the Planet-NSFP cut.

We make available all BLS candidates with Planet-NSFP mean probability $\geq0.9$ (14\,815 TCEs identified by our pipeline) in Table~\ref{tab:cand_nsfp0.9} (only column descriptions shown here, full table available online). The table includes the main BLS parameters and the mean Planet-NSFP machine learning classification probability.

\begin{table}
\caption{BLS sample candidates with Planet-NSFP mean probability~$\geq0.9$. The table shows stellar properties from \gaia (see Sect \ref{sec:data}), the BLS results (see Sect. \ref{sec:bls}) and mean Planet-NSFP classifier probability (see Sect \ref{sec:nsfp_vetting}). The candidate number indicates the peak number in order of significance in the BLS. Candidates are sorted by TIC id and candidate number.}
\label{tab:cand_nsfp0.9}
\begin{tabular}{p{0.27\linewidth}p{0.63\linewidth}}
\hline
Column & Description \\
\hline
TIC & TIC identifier. \\
Candidate & Candidate number, i.e., peak number in the BLS (from 1 to 5). \\
TOI & TOI identifier, if the star has been flagged as having any TOI. The specific candidate found in our search does not necessarily match any of the TOI candidates. \\
CTOI & CTOI identifier, if the star has been flagged as having any CTOI, not promoted to TOI. The specific candidate found in our search does not necessarily match any of the CTOI candidates. \\
\gaia DR3 & \gaia DR3 identifier. \\
\gaia DR2 & \gaia DR2 identifier. \\
RA & \gaia right ascension. \\
Dec & \gaia declination. \\
\hline
$G$ & Apparent \gaia $G$ magnitude. \\
$d$ & Distance derived from \gaia DR3 parallax [pc]. \\
$R_\star$ & Stellar radius from \gaia DR2 [\Rsun]. \\
$T_\mathrm{eff}$ & Stellar effective temperature from \gaia DR2 [K]. \\
\logg & Stellar surface gravity from \gaia DR3 [dex]. \\
\hline
$P$ & BLS period [d]. \\
$t_0$ & BLS mid-transit time [BJD]. \\
$t_\mathrm{dur}$ & BLS transit duration [h]. \\
$\delta$ & Transit depth [ppm]. \\
SDE & Signal detection efficiency. \\
MES & Multiple event statistic. \\
\hline
Planet-NSFP mean classifier probability & Mean probability from the Planet-NSFP GBDT and GP classifiers. \\
\hline
\end{tabular}
\begin{flushleft}
{\it Notes:} Full table available online, this table describes the available columns.
\end{flushleft}
\end{table}

\subsubsection{Planet-simulated-FP classification, priors and RAVEN probability}

After cleaning our sample of NSFPs, we then applied the rest of the classifiers (i.e. those based on simulated astrophysical FPs) and obtained a mean probability for each Planet-simulated-FP pair.

The machine learning models assume no prior knowledge on the overall occurrence of the different scenarios considered (Planet and FPs) because the training sets are balanced (i.e. each category of a given pair has the same number of samples). Therefore, for each candidate, \raven needs to combine the posterior probability obtained from the machine learning models with a prior probability of the candidate belonging to any of the considered scenarios. These prior probabilities include the probability that the scenario occurs (based on empirical occurrence rates), the probability that the candidate is detected in the \tess data (computed during the creation of the simulation sets), the probability that the candidate can be recovered (based on \raven's BLS recovery of the simulation sets), and the probability that the candidate is located on the target star as opposed to on a nearby source \cite[i.e. the positional probability, computed as introduced in ][with a method based on comparing the observed photometric centroid offset of the target star with offsets modelled assuming that the event occurs on the target or on any of the nearby sources]{hadjigeorghiou2024positional}.
Note that for the NSFP case, a prior probability cannot be computed, since this model includes events (instrumental noise, stellar variability, and some astrophysical FPs such as EBs) whose occurrence is hard to constrain. Therefore, for the Planet-NSFP classifier, we only consider the classifier posterior probability.

This step resulted in a final posterior probability for each FP scenario. Finally, \raven combined all posterior probabilities into a single final posterior probability by taking the minimum amongst all Planet-simulated-FP probabilities. 
We call this final probability the \raven probability, and used it to rank our candidates.
For example, if a given candidate has a probability of 0.8 for the BEB scenario and a probability above 0.99 for the rest of scenarios, the \raven probability will be 0.8, reflecting the fact that the candidate has a relatively high probability of being a FP.
By taking the minimum of all scenarios, we ensure that a highly ranked candidate will have a high probability for all scenarios.
In the following, we consider only candidates with Planet-NSFP probability $\geq0.9$ and with \raven probability $\geq0.9$, which represents a sample of 3899 highly ranked candidates.


\subsection{Candidate properties refinement}\label{sec:juliet}

To refine the transit parameters derived for our candidates, we used \texttt{juliet} \citep{espinoza2019juliet}, a widely-used package that models data using nested samplers to explore the prior volume and efficiently compute Bayesian model evidence.
We used the nested sampling algorithm implemented in the \texttt{dynesty} sampler \citep{speagle2020dynesty,koposov2024dynesty}.
Before modeling the data, we cleaned the light curve of significant outliers by removing data points with flux higher than 3 times the standard deviation from the median of all points (i.e. we performed one iteration of a 3~sigma-clipping procedure on the upper part of the light curve).

\texttt{juliet} models planetary transits with \texttt{batman} \citep{kreidberg2015batman}, a package for fast calculation of exoplanet transit light curves. We assumed a single-planet system with a circular orbit (i.e., eccentricity $e$ and argument of periastron passage $\omega$ fixed to 0 and 90\,\degrees, respectively) and fit for the planet's orbital period $P$, mid-transit time $t_0$, planet-to-star radius ratio \rprstar, impact parameter $b$, stellar density $\rho_\star$, and stellar limb darkening. We assumed wide priors for all parameters.
In particular, we used normal priors on $P$, $t_0$, and \rprstar, with central values set as those found by the BLS search for $P$ and $t_0$, and from the trapezoid fit (performed while computing the features) for \rprstar, and sigma set to 0.1 for the three parameters.
We used a uniform prior for $b$ within $[0,\,2]$, and a uniform prior in logarithmic space for $\rho$ within $[100,\,10000]$ (in kg~m$^{-3}$).
For the stellar limb darkening, we assumed a quadratic law\footnote{The intensity profile of the stellar disc is given by
$I(\mu) = I_0 \left[ 1 - u_1 \left( 1 - \mu \right) - u_2 \left( 1 - \mu \right)^2 \right]$
where $\mu=\sqrt{1-x^2}$, $x$ is the normalised radial coordinate ($0\leq x \leq1$), $I_0$ is a normalisation constant such that integrated stellar intensity equals unity, and $u_1$ and $u_2$ are limb darkening coefficients we want to estimate.} (with coefficients $u_1$ and $u_2$) parametrised by $q_1$ and $q_2$ following \citet{kipping2013ld}\footnote{
where $q_1=\left( u_1 + u_2 \right)^2$ and $q_2=0.5 u_1\left(u_1 + u_2 \right)^{-1}$}, which samples these parameters efficiently with uniform priors within $[0,\,1]$.

To detrend time-correlated noise (due to stellar variability and/or instrumental noise), we used a GP with the approximate Matérn-3/2 kernel implemented in \texttt{celerite} \citep{foremanmackey2017celerite}, which allows for both smooth and rough signals, since we have a large variety of signals among our relatively large number of candidates.
This kernel depends on parameters $\sigma_\mathrm{GP}$, which represents the amplitude of the GP (in ppm), and $\rho_\mathrm{GP}$, the length-scale (in our case, time-scale, in days) of the GP modulations, which controls the smoothness of the returned functions. For both parameters we used a uniform prior in logarithmic space within $[10^{-6},\,10^{6}]$ for $\sigma_\mathrm{GP}$ and $[10^{-3},\,10^{3}]$ for $\rho_\mathrm{GP}$.
We note that now, the light curves have not been detrended with the Savitzky-Golay filter, as done before computing the BLS and the features, because we expect the GP to account for any variability.
We also considered an additional term for unknown extra jitter, $\sigma_{w}$, added in quadrature to the nominal flux uncertainties, with a uniform prior in logarithmic space within $[0.1, \,1000]$ (in ppm).
We also fit for a mean instrumental offset of the out-of-transit flux (mflux) with a normal prior centred at 0 and standard deviation 0.1 (in ppm).
\texttt{juliet} can also fit for a dilution factor (mdilution) in the light curve that accounts for contaminating flux of nearby sources, which can affect the derived transit parameters. This dilution has already been taken into account in the PDCSAP light curves that we use, hence, we fixed this parameter to 1.
Note that the transit and GP parameters are shared for both sectors with 30~min and 10~min exposure times, while we fit the jitter and instrumental offset for the 30~min and 10~min sectors separately.
We set the nested sampling samplers to use 3000 live points.

The \tess FFIs have relatively long exposure times (30~min for sectors 1-26 and 10~min for sectors 27-55) that smear the transit shape \citep[e.g.][]{kipping2010binning}. We account for these exposure times in the fitted \texttt{batman} models by computing them at 1 min intervals (i.e. oversampling factor of 30 for the 30~min exposure times, and factor of 10 for the 10~min ones) and integrating them to the corresponding observation exposure time.
The transit shape smearing is especially notable for short ingress times and long exposures. To make sure that the computed models correctly represent the observed transit shape while saving computing time, we compute models at shorter intervals of 30~s (i.e. oversampling factor of 60 and 20 for the 30 and 10 min exposure times, respectively) for candidates with transit durations shorter or equal than 1~h (as computed from the initial BLS parameters). We use the transit duration rather than the ingress (or egress) time to select these candidates because the total duration can be computed more reliably.

Aside from the fitted parameters, we also derive the orbital semi-major axis $a$, orbital inclination $i_\mathrm{p}$, transit duration $t_\mathrm{dur}$, original limb-darkening coefficients $u_1$ and $u_2$, and planetary radius from the posterior distributions of the corresponding fitted parameters.
To derive the planetary radius from the fitted transit depth, we assumed a normal distribution of stellar radii with mean equal to the \gaia radius value and uncertainty of $6\%$ of the radius value \citep{andrae2018gaia}.
%
Our best estimates from the fitted and derived parameters are the medians of the posterior distribution of all samples. We assumed $1\sigma$ uncertainties from the $16\%$ and $84\%$ percentiles of the posterior distributions.


\subsection{Further vetting checks}\label{sec:further_checks}

We performed extra vetting checks to ensure that our highest ranked candidates contain mostly true transiting planets. We detail the checks below, and mention how many candidates are discarded by each successive check.

\begin{itemize}
\item We expect most our highest ranked candidates to have a positional probability close to 1, meaning that the transit event has a high chance of occurring on the target, rather than on a background or nearby star. The positional probability is accounted for in the prior probability. However, several of our highest ranked candidates have low positional probability. To ensure that our candidates are actually due to a transit on the target star, we removed 530 candidates with positional probability $<0.5$.

\item We also expect true planetary transits to not show high variability in their depth between different \tess sectors. 
While a few of the features used in the machine learning classifiers accounted for transit depth variability in the observations, some of our highest ranked candidates still had
significant changes in their depth between different sectors, mostly due to instrumental noise, stellar variability not correctly accounted for during detrending, detrending-introduced artifacts, dilution from a nearby source with changing contribution as the aperture changes in different sectors, incorrect periods, or unaccounted transit timing variations. Therefore, to easily remove these type of candidates, we applied another cut and discarded candidates with depth variations between sectors larger than 50\% of the total mean depth in at least one sector. 
We also removed candidates with transits shallower than 200~ppm in at least one sector, to avoid low S/N candidates.
We note that the depth was computed after flattening the light curves as described in Sect. \ref{sec:bls}. These criteria further removed 624 candidates.

\item Our BLS search accounted for signals at double and half a given period, however, we found that several of our candidates with high probability were actually higher harmonics of the real signal (mostly because the real signal is outside our period search limits). 
These candidates are not true recoveries, and including them in our samples would affect the sample completeness and introduce biases in future estimates of the underlying planet occurrence rate.
We do not expect our machine learning models to be able to correctly identify those candidates as non-transiting planets because our training sets do not include such harmonics. 
Therefore, we visually vetted the candidates by folding their light curves at various aliases of the detected period and visually discarding only those candidates that exhibit clear signs of period aliasing. 
This step further removed 79 candidates that mostly had clear, relatively deep transits.

\item In several cases, our pipeline identified as significant candidates with only two transits in which one or even both transits occur close to the start/end of the sector or close to intra-sector gaps, meaning that the full transit is not completely covered. These are regions subject to strong trends, and we expect several of the candidates found here to be false alarms. Therefore, we also discarded 34 candidates with less than two full transits (where we define a full transit as covering at least 90\% of its expected transit duration).
By definition, this two-transit constraint also removes any candidates showing only one transit. We do not want to include any candidate with a single transit in our samples because we cannot recover a reliable period with a single event. However, some of these single-transit candidates showed prominent transit depths and high posterior probabilities (despite having an incorrect period), making them interesting candidates for further follow-up. Hence, despite removing them from our list of candidates of interest, we make them available as a separate sample, see Sect. \ref{sec:single_transits}.

\item Some of our final candidates are signals with high \raven probability but very low transit S/N (small \rprstar with large uncertainties, also mostly with low SDE and MES values), for which our systematic application of \texttt{juliet} does not result in a reliable fit.
Therefore, we further removed 461 candidates with low transit S/N (defined as the \texttt{juliet} fit \rprstar divided by its lower uncertainty) by requiring them to have a transit S/N $\geq3$.
\end{itemize}


\section{Transiting planet candidate samples}\label{sec:res_cand}

\subsection{Vetted candidates}\label{sec:res_cand_vet}

\begin{figure}
\centering
\includegraphics[width=0.49\textwidth]{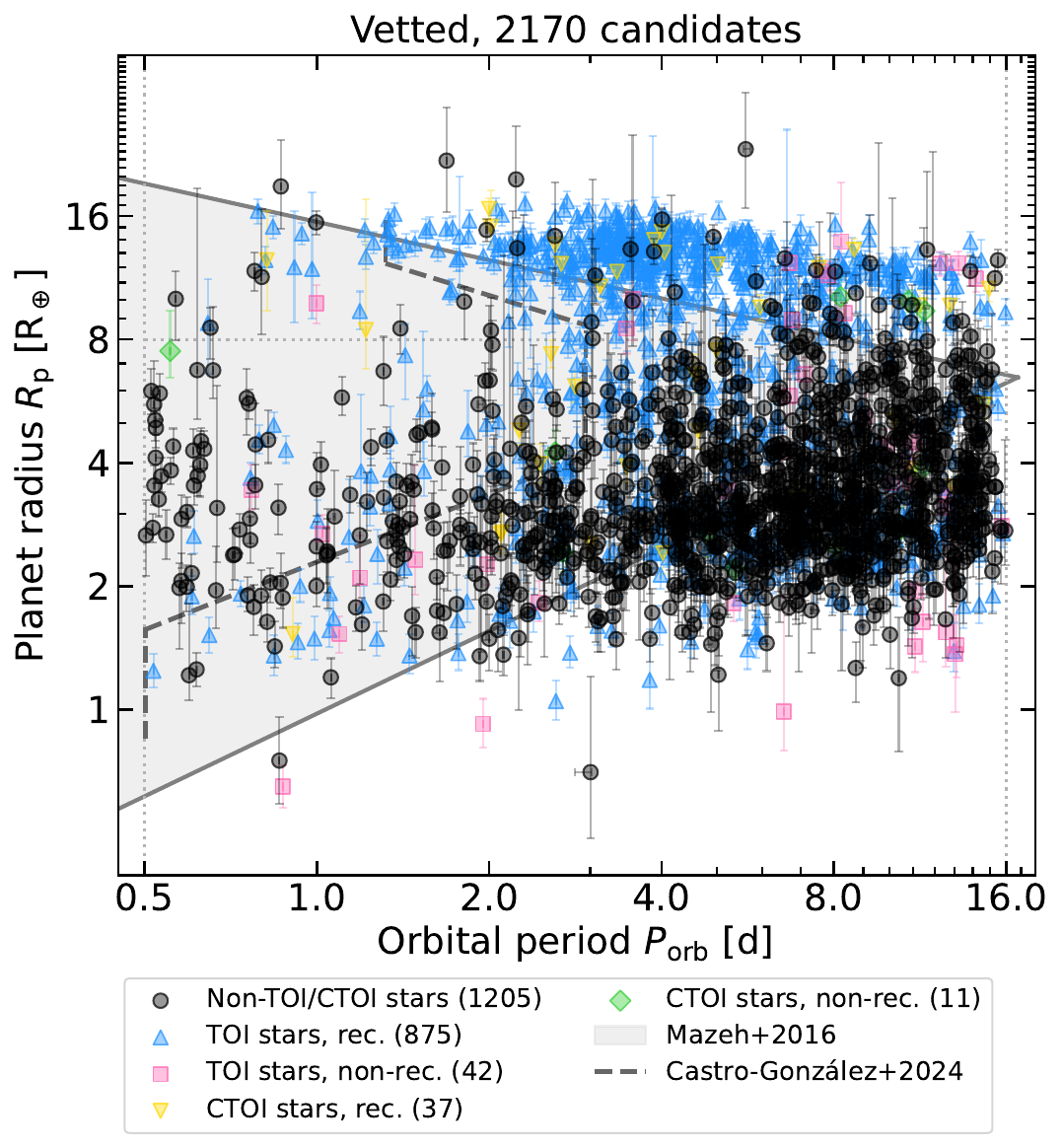}
\caption{Vetted sample of 2170 candidates in period-radius space. The values shown are the results of our \texttt{juliet} fits (see Sect. \ref{sec:juliet}). Black circles correspond to new candidates orbiting stars not known to host any TOI/CTOI, blue triangles show candidates matching known TOIs (i.e. recovered), pink squares show candidates on stars known to have a TOI that do not match the known TOI candidates (i.e. non-recovered, some of which could be new candidates), yellow down triangles show candidates matching known CTOIs (i.e. recovered), and green diamonds show candidates on stars known to have a CTOI that do not match the known CTOI candidates (i.e. non-recovered, some of which could be new candidates).
Solid grey lines and grey-shaded area show the Neptunian desert limits according to \citet{mazeh2016desert}, and dashed grey lines show the recently derived  limits between the Neptunian desert, ridge, and savannah from \citet{castrogonzalez2024neptune}.
Dotted grey vertical and horizontal lines show our pipeline's detection/validation limits: periods from 0.5 to 16~d and radii below $8~\Rearth$.
}
\label{fig:cand_per_rp_vet}
\end{figure}

After processing our sample of stars with \raven and applying all the constraints explained above (Sect. \ref{sec:raven}), we are left with a \emph{vetted} sample of candidates. To summarise, these are candidates that:
\begin{itemize}
\item have Planet-NSFP classifier probability $\geq0.9$,
\item have \raven probability $\geq0.9$, and
\item have been vetted with the further checks explained in Sect. \ref{sec:further_checks}.
\end{itemize}
This vetted sample has 2170 candidates from 2112 different stars. Over 1000 candidates are newly identified in this work (i.e. they are not previously known TOIs or CTOIs).
The properties of the stellar hosts of the vetted candidates are shown in Fig. \ref{fig:stellar_sample} in pink, and the main BLS parameters are shown in the third column of Fig. \ref{fig:cand_bls_hist}.
The distribution of vetted candidates in planetary radius and orbital period space is shown in Fig. \ref{fig:cand_per_rp_vet}.
The final \raven probability of most of these candidates is not high enough to statistically validate them, but they are significant TCEs identified by our pipeline.
Therefore, we make available the main properties of the vetted candidates in an online table (with columns described in Table \ref{tab:cand_vet}).
We also provide summary plots of all candidates, which show the FFI light curves used, light curves phase-folded to the candidate period with the best-fitting transit model obtained with \juliet, and per-sector light curves and phase-folded transits (see an example in Fig. \ref{fig:summary}).

The table also includes several candidates that we flag after a visual inspection of the light curves (see column Notes on the table). These include low S/N candidates that pass our S/N threshold but still result in an unreliable fit, candidates with correlated noise in the light curves, and candidates with transit-like features that occur close to data gaps, where the data is prone to systematics (in particular we find a large number of these in sector 18). We generally advise caution with candidates with less than three transits, correlated noise, transits occurring close to data gaps, and shallow depth (i.e. low S/N).

\begin{table*}
\caption{Properties of our vetted (see Sect. \ref{sec:res_cand_vet} for details) and validated (see Sect. \ref{sec:res_cand_val} for details) samples of candidates. The full tables are available online, and here we only describe the available columns.
The tables contain stellar properties from \gaia, 
BLS results (Sect. \ref{sec:bls}), 
Planet-FP posterior probabilities, positional probability, other vetting flags (Sect. \ref{sec:further_checks}), \raven probability, and \texttt{juliet} fitted and derived parameters (Sect. \ref{sec:juliet}). Note that for each of the \texttt{juliet} fitted and derived parameters, we also provide upper and lower uncertainties as separate columns, which are not explicitly mentioned here.
Candidates are sorted by TIC id and candidate number.}
\label{tab:cand_vet}
\begin{tabular}{p{0.26\linewidth}p{0.7\linewidth}}
\hline
Column & Description \\
\hline
TIC & TIC identifier. \\
Candidate & Candidate number, i.e., peak number in the BLS (from 1 to 5). \\
TOI & TOI identifier, if the star has been flagged as having any TOI. The specific candidate found in our search does not necessarily match any of the TOI candidates. \\
CTOI & CTOI identifier, if the star has been flagged as having any CTOI, not promoted to TOI. The specific candidate found in our search does not necessarily match any of the CTOI candidates. \\
\gaia DR3 & \gaia DR3 identifier. \\
\gaia DR2 & \gaia DR2 identifier. \\
RA & \gaia right ascension. \\
Dec & \gaia declination. \\
\hline
$G$ & Apparent \gaia $G$ magnitude. \\
$d$ & Distance derived from \gaia DR3 parallax ($10^3$/parallax) [pc]. \\
$R_\star$ & Stellar radius from \gaia DR2 [\Rsun]. \\
$T_\mathrm{eff}$ & Stellar effective temperature from \gaia DR2 [K]. \\
\logg & Stellar surface gravity from \gaia DR3 [dex]. \\
\hline
$P$ & BLS period [d]. \\
$t_0$ & BLS mid-transit time [BJD]. \\
$t_\mathrm{dur}$ & BLS transit duration [h]. \\
$\delta$ & Transit depth [ppm]. \\
SDE & Signal detection efficiency. \\
MES & Multiple event statistic. \\
\hline
Planet-NSFP mean classifier probability & Planet-NSFP classifiers mean probability (GBDT and GP). \\
Planet-EB posterior probability & Combination of the Planet-EB classifiers mean probability (GBDT and GP) and EB prior probability. \\
Planet-BEB posterior probability & Combination of Planet-BEB classifiers mean probability (GBDT and GP) and BEB prior probability. \\
Planet-HEB posterior probability & Combination of Planet-HEB classifiers mean probability (GBDT and GP) and HEB prior probability. \\
Planet-HTP posterior probability & Combination of Planet-HTP classifiers mean probability (GBDT and GP) and HTP prior probability. \\
Planet-NTP posterior probability & Combination of Planet-NTP classifiers mean probability (GBDT and GP) and NTP prior probability. \\
Planet-NEB posterior probability & Combination of Planet-NEB classifiers mean probability (GBDT and GP) and NEB prior probability. \\
Planet-NHEB posterior probability & Combination of Planet-NHEB classifiers mean probability (GBDT and GP) and NHEB prior probability. \\
Positional probability & Positional probability derived following \citet{hadjigeorghiou2024positional}. \\
\raven probability & Final \raven probability (minimum of the posterior probabilities of all scenarios). \\
\hline
Sector flag & True if transit depth significantly varies between sectors and/or is $<200$~ppm in at least one sector. \\
Number of full transits & Total number of full transits (at least 90\% of the expected transit duration) found in the light curves. \\
\hline
\texttt{juliet} $P$ & \texttt{juliet} fitted orbital period [d]. \\
\texttt{juliet} $t_0$ & \texttt{juliet} fitted mid-transit time [TBJD]. \\
\texttt{juliet} $\rprstar$ & \texttt{juliet} fitted planetary radius $[\Rstar]$. \\
\texttt{juliet} $b$ & \texttt{juliet} fitted impact parameter. \\
\texttt{juliet} $q1$, $q2$ & \texttt{juliet} fitted quadratic limb-darkening parameters \citep[parametrisation from][]{kipping2013ld}. \\
\texttt{juliet} $\rho_\star$ & \texttt{juliet} fitted stellar density [kg~m$^{-3}$]. \\
\texttt{juliet} $\sigma_\mathrm{GP}$ & \texttt{juliet} fitted GP amplitude [ppm]. \\
\texttt{juliet} $\rho_\mathrm{GP}$ & \texttt{juliet} fitted GP length scale [d]. \\
\texttt{juliet} $\sigma_w$ & \texttt{juliet} fitted jitter [ppm]. \\
\texttt{juliet} mflux & \texttt{juliet} fitted instrumental offset [ppm]. \\
\hline
\texttt{juliet} $a/R_\star$ & \texttt{juliet} derived orbital semi-major axis $[R_\star]$. \\
\texttt{juliet} $i_\mathrm{p}$ & \texttt{juliet} derived orbital inclination [deg]. \\
\texttt{juliet} $t_\mathrm{dur}$ & \texttt{juliet} derived transit duration [h]. \\
\texttt{juliet} $u_1$, $u_2$ & \texttt{juliet} derived  quadratic limb-darkening parameters. \\
\texttt{juliet} $\Rp$ & \texttt{juliet} derived planetary radius $[\Rearth$]. \\
\texttt{juliet} transit S/N & \texttt{juliet} transit S/N. \\
\hline
Notes & Visual vetting comments, including unreliable fits, harmonics, and extra candidates.\\
\hline
\end{tabular}
\begin{flushleft}
{\it Notes:} Full table available online, this table describes the available columns.
\end{flushleft}
\end{table*}


\subsection{Validated candidates}\label{sec:res_cand_val}

\begin{figure}
\centering
\includegraphics[width=0.49\textwidth]{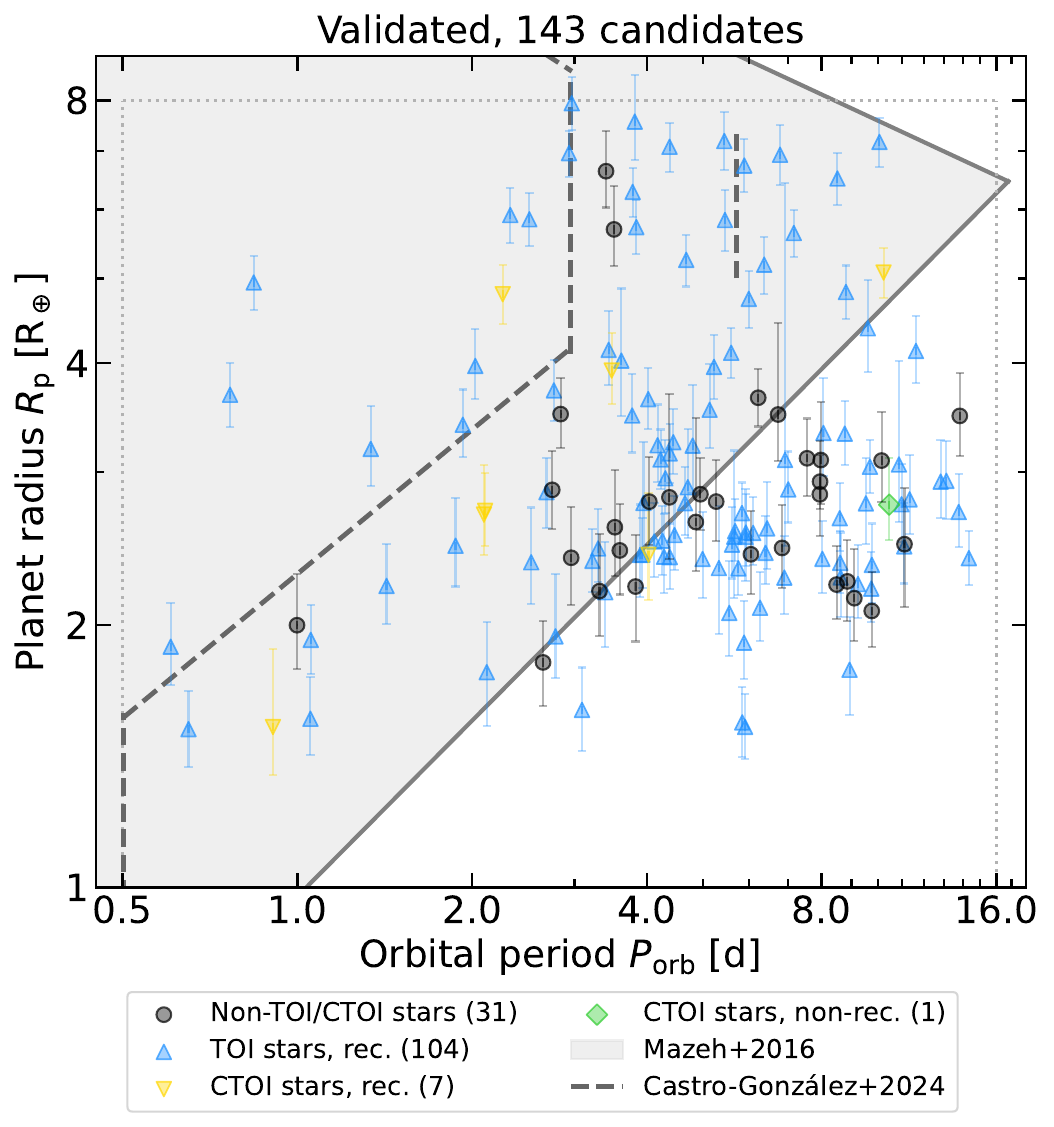}
\caption{Validated sample of 143 candidates in period-radius space. All symbols show the same as Fig. \ref{fig:cand_per_rp_vet} but for the validated sample rather than the vetted one. Note that the validated sample is included in the vetted one.
Note that the CTOI TIC~270471727 BLS peak 1 (green diamond) is actually a CTOI that we recover (TIC~270471727.01), despite the $t_0$ not matching (see Appendix \ref{sec:cand_vet_toi_other_recovery}).
}
\label{fig:cand_per_rp_val}
\end{figure}

To be able to validate our candidates, we require stronger constraints than those applied to the sample of vetted candidates presented in Sect. \ref{sec:res_cand_vet}.
We adopt a conservative approach to define a candidate as validated, as described below.

We initially applied \raven's machine learning classifiers to candidates defined by the ephemeris and orbital parameters ($P$, $t_0$, $t_\mathrm{dur}$, and $\delta$) given by our BLS search. However, the \juliet fits performed afterwards to highly-ranked candidates result in more precise orbital parameters than those from the BLS, which could result in slightly different features and hence different classification probabilities. Therefore, we re-computed the features and re-ran \raven's classifiers on all highly ranked candidates now using the parameters obtained from the \juliet fits instead of the BLS ones, and obtained updated NSFP and \raven probabilities. We then require the initial \raven probability and the Planet-NSFP classifier probability (based on the BLS parameters), as well as the updated \raven probability and the updated Planet-NSFP classifier probability (based on the \juliet fit parameters) all to be above the typical threshold of 0.99 for a candidate to be considered validated.

We only validate candidates with $\Rp \leq 8~\Rearth$ because, for larger radii, transits caused by planetary candidates look very similar to those caused by brown dwarfs and low-mass eclipsing binaries. This a common constraint when statistically validating planetary candidates from  that photometric transits \citep[see e.g.][]{shporer2017eb,giacalone2021triceratops,hadjigeorghiou2025raven}.

Additionally, we also performed a manual check of all highly-ranked candidates and avoid validating unclear cases that are relatively low S/N and/or show correlated noise in the light curves. This extra check removed 10 further candidates (TIC~39516274.01 or TOI-5997.01 in BLS peak 1, TIC~69369434.01 peak 3, TIC~95327577.01 in peak 1, TIC~117648031.01 in peak 1, TIC~198153540 BLS peak 2, TIC~356227008.01 in peak 1, TIC~365272555.01 in peak 1, TIC~390586021.01 in peak 2, TIC~457135306.01 in peak 3, and TIC~459215281.01 or TOI-3891.01 in peak 1). Note that we expect most of these to be very strong candidates and indeed likely true planets. The manual check is applied to be conservative in adding candidates to the sample of known planets.

In summary, we consider a candidate \emph{validated} if it:
\begin{itemize}
\item has an initial (based on the BLS) Planet-NSFP classifier probability $\geq0.99$,
\item has an initial (based on the BLS) \raven probability $\geq0.99$, 
\item has an updated (based on the \juliet fits) Planet-NSFP classifier probability $\geq0.99$,
\item has an updated (based on the \juliet fits) \raven probability $\geq0.99$, 
\item passes the further checks from section Sect. \ref{sec:further_checks}, with the difference that here we require that the candidates have 3 (instead of 2) full transits (i.e. covering at least 90\% of the expected transit duration),
\item has a planetary radius $\Rp \leq 8~\Rearth$,
\item and does not show correlated noise in the light curve and has sufficient S/N (as determined by a manual check).
\end{itemize}

Our validated sample has 143 candidates from 140 stars. Of these 143 candidates, 31 are newly identified in this work (i.e. they are not previously known TOIs or CTOIs).
Their stellar hosts properties are shown in Fig. \ref{fig:stellar_sample} in yellow, and the main BLS parameters are shown in the fourth column of Fig. \ref{fig:cand_bls_hist}.
The distribution of validated candidates in planetary radius and orbital period space is shown in Fig. \ref{fig:cand_per_rp_val}.
We also provide figures with the phase-folded transits of all validated candidates newly discovered here (31 candidates) in Appendix \ref{sec:lc_val_new}.
We make our validated candidates available in an online table (with columns described in Table \ref{tab:cand_vet}). Note that the validated candidates are also contained in our vetted sample, and therefore are also present in the corresponding table with the vetted sample candidates. We only provide them in a separate table for easy access. 

Without considering the updated NSFP and \raven probabilities obtained with the \juliet parameters, our validated sample would be of 220 candidates. In other words, requiring the updated NSFP and \raven probabilities to also be $>0.99$ removes 77 candidates with initial NSFP and \raven probabilities (from the BLS parameters) above 0.99.
The clear variation in final probability highlights that small ephemeris variation can cause scatter in the estimated probabilities of order of a percent or more. We guard against this by re-running validated candidates with different ephemeris, and applying stricter criteria for validation than a pure probability threshold. A potential variance in validation probabilities for the same candidates was also identified in \citet{armstrong2021kepler}, in that case between different validation methods on \textit{Kepler} data. Investigating the intrinsic precision and systemic biases of the validation process, for \raven and other methods, should be a key future goal for the field.


\subsection{Candidates with $\Rp>8~\Rearth$}

We find 207 high-probability (NSFP mean probability $\geq0.99$ and \raven probability $\geq0.99$) candidates that pass all further vetting checks explained in Sect. \ref{sec:further_checks} except having $\Rp>8~\Rearth$ (note that we have not re-ran \raven with the \juliet parameters as inputs for these candidates since we do not validate them). 
As mentioned in Sect. \ref{sec:res_cand_val}, we do not validate any candidates with $\Rp>8~\Rearth$ due to the degeneracy between planets, brown dwarfs, and low-mass eclipsing binaries.
However, we make them available here in a separate online Table (with columns described in Table \ref{tab:cand_vet}), as these are key targets for further follow-up such as multi-band photometry or reconnaissance spectroscopy.
Of the total 207 candidates, 6 of them are newly reported here, 196 are TOIs, and 5 are CTOIs. We recover all these TOIs and CTOIs. 
Following the TFOP Working Group (TFOPWG) dispositions,\footnote{KP: known planet, CP: confirmed planet, PC: planet candidate, APC: ambiguous planet candidate, FA: false alarm, FP: false positive.} the TOIs are 79 KPs, 29 CPs, 86 PCs, 1 APC, and only 1 FP.
Note that these candidates are already included in our vetted sample, but similarly to the validated sample, we make them available in their own individual table for ease of access.


\subsection{Mono-transit candidates}\label{sec:single_transits}

The BLS search identified a few significant signals (transit depth $\gtrsim1000$~ppm) with high \raven probability ($\geq0.9$) that correspond to single transits with an unreliable period.
These candidates show one clear transit, and further transits, according to the (incorrect) recovered period, fall on intra-sector gaps and/or just outside the sector time range, making the recovered period  unreliable. Despite that, these candidates initially made it into our vetted sample, and were discarded with the further checks described in Sect \ref{sec:further_checks}.
Since these are interesting mono-transit candidates, we provide newly identified (i.e. not previously flagged as TOI/CTOI and known planets) single-transit candidates in Table \ref{tab:mono} (4 candidates). 
We visually looked for further transits in sectors after 55 (the sector limit for our search) and examined the QLP and short-cadence data SPOC reductions,\footnote{using the \tess Light Curve Online Viewer \url{https://tess.cuikaiming.com/}} when available, since they can include sectors not reduced in the SPOC-FFI data. We detail this information in the same table.


\subsection{Visually identified multiple systems}\label{sec:multis_vis}

When manually vetting our candidates, we visually found several light curves showing clear extra transits that did not belong to the candidates identified with the BLS. We flag those stars in Table \ref{tab:multi} as multi-candidate systems (8 candidates). 
Finding multiple candidates in a given system reinforces the likelihood that these candidates are real transiting planets, since it has been observed that planet candidates in multi-candidate systems are likely to be true planets \citep{lissauer2012multis}.

When several clear transits are present in the FFI data, we performed a multi-planet fit with \texttt{juliet} to obtain the orbital parameters of the extra candidates. We used the same fit parameters and priors as described in Sect \ref{sec:juliet} and only added the orbital parameters for the second planet candidate (with central values for the priors for the period, mid-transit time, and depth visually identified from the light curves). We increased the number of live points to 3500 to account for the extra parameters. 

These second candidates have not been directly detected by our pipeline, and hence, we do not include them in our vetted or validated samples (moreover, several are outside our BLS search range). Likewise, for the original candidates, we do not update the parameters obtained from the single-planet \texttt{juliet} fit reported in the vetted and validated tables with those derived here from the multi-planet fit, since we want those samples to contain parameters derived uniformly for all candidates. In general, the orbital parameters of the initial candidates derived from the original single-planet \juliet fit agree within $1\sigma$ with the new parameters derived here with the two-planet fit.


\section{Discussion}\label{sec:discussion}

In this section we discuss our main results. Unless something different is specified, any parameters related to host stars are from \gaia (see Sect. \ref{sec:data} for details), and any parameters related to our candidates come from the \texttt{juliet} fits or, if the candidate is not included in our vetted/validated samples, from the BLS peaks.
Note that to refer to vetted candidates we use the convention TIC~id.01 (or TOI~id.01 if a TOI) for the first candidate, TIC~id.02 for the second one, and so on. For validated planets, we instead use TIC~id~b (or TOI~id~b if a TOI) for the first planet, TIC~id~c for the second one, and so on. For candidates that we identify as harmonics of true signals, we use TIC~id BLS peak number. We always specify the TIC id and the BLS peak in which a candidate is recovered to be able to identify candidates in our samples.


\subsection{TOI/CTOI candidates}\label{sec:res_toi}

There are several pipelines that scan \tess observations to detect transit-like candidates labelled as TCEs. The \tess Science Office performs a visual vetting of significant TCEs detected by the \tess SPOC and QLP pipelines, and promising candidates are classified as TOIs \citep[][]{guerrero2021toi}.
Community efforts to follow-up candidates for confirmation as planets (including ground-based photometry, spectroscopy, and imaging) are coordinated by the \tess follow-up observing program (TFOP) and available in the Exoplanet Follow-up Observing Program repository \citep[ExoFOP,][]{collins2018tfopwg,akeson2019tfopwg}. 
Other candidates not found by the main \tess pipelines but found by other searches are initially classified as CTOIs (also available in ExoFOP), which can be later on promoted to TOIs.

Hence, the TOI and CTOI samples are a list of well-vetted events that include known planets, known false positives, and a large number of planetary candidates not yet validated or confirmed.
In this section we study which of our candidates match known TOIs/CTOIs (we highlight new candidates found in stars known to host TOIs/CTOIs in Sect. \ref{sec:new_toi}).
For a detailed analysis of the application of \raven to TOIs classified as known/confirmed planets and false positives, we refer the reader to \citetalias{hadjigeorghiou2025raven}, where these pre-classified TOIs are used to test the performance of the pipeline.
The application of \raven to the full sample of TOIs with the goal of validating them will be subject of further work.

Our initial sample of stars contains a total of 3703 TOIs in 3492 different stars.\footnote{comparing with the ExoFOP databases, accessed on 13/08/2025, \url{https://exofop.ipac.caltech.edu}} 
Several of these TOIs have periods outside of our search range (from 0.5 to 16~d), and some have periods equal to 0, usually corresponding to mono-transits of long period candidates. Considering only TOIs with periods inside our search range, our initial sample has 3098 TOI in 2991 different stars.
Regarding the CTOIs not promoted to TOIs, in our initial sample there are 1330 CTOI candidates in 1256 different stars, which reduce to 1146 CTOI candidates in 1086 different stars when considering only candidates inside our period search range.


\subsubsection{TOI/CTOI BLS recovery}\label{sec:res_toi_bls}
We checked if the candidates found by our BLS search matched the periods and transit times of the known TOIs and CTOIs, that is, if we recovered these TOIs and CTOIs. We consider a TOI/CTOI as recovered in our BLS if a given candidate (any of the 5 BLS peaks) matches the reported TOI/CTOI period within 2\% and mid-transit time (adjusted by the recovered period) within 0.5~d. We use the TOI/CTOI period and mid-transit time values reported in ExoFOP and those derived from our BLS search.

Of the 3098 TOIs in our initial sample and with period within our search range, our BLS search was able to correctly recover 2633 TOIs in 2591 different stars (84.99\% recovery rate).
Following the TFOPWG dispositions, 
357 are KPs, 311 CPs, 1342 PCs, 175 APCs, 7 FAs, and 441 FPs.
Of the 1146 CTOIs, we recovered 615 candidates from 613 different stars (53.66\% recovery rate).

There are 465 TOIs in 447 different stars that we do not recover.
They are 27 KPs, 64 CPs, 209 PCs, 36 APCs, 25 FAs, and 104 FPs.
Of these non-recovered TOIs, we find a harmonic related to the period $P$ reported in ExoFOP for 80 of them (27 TOI with double the period $2P$, 48 with half the period $P/2$, 3 with triple the period $3P$, and 2 with a third of the period $P/3$), a match in period (within $1\%$) but not in mid-transit time for 40 of them, and no match for the remaining 345.
For the CTOIs, we do not recover 531 candidates in 499 different stars.
For 64 of them we find a harmonic related to the reported period $P$ (11 with $2P$, 31 with $P/2$, 11 with $3P$, and 11 with $P/3$), a match in period (within $1\%$) but not in mid-transit time for 153, and no match for the remaining 314.

We also looked at the recovery of multi-candidate systems. In our BLS sample, there are 92 TOI stars hosts of multi-candidate systems, with a total of 199 candidates (considering candidates within our period search range). Of these, our BLS recovered a total 121 candidates in 79 different hosts. These correspond to 39 host stars for which we recovered more than one candidate each (81 candidates in total), and for remaining 40 multi-candidate host stars, we only recovered one of the candidates. Therefore, our BLS only recovered about half of the multi-candidate TOI systems in our sample.
For CTOIs, our BLS recovered less candidates than for the TOI sample, which is also reflected in the low numbers of recovered multi-candidate systems. Our BLS sample has 59 CTOI stars with multi-candidate systems (119 candidates in total), of which we recovered 30 candidates in 28 stars. Hence, we only recovered 2 CTOI systems with more than one candidate.
{Typically, from our results we see that extra planets in multi-candidate systems are not recovered because they are not found by the BLS, due to the rest of the peaks being harmonics (other than half and double, which are initially removed by the BLS) of the first one, or because they have depths below $\sim200$~ppm, which also challenges their detection. In some cases, these other candidates are detected but have NSFP or \raven probabilities below the 0.99 or 0.9 cut, which removes them from our vetted and validated samples (see Section \ref{sec:new_multi_val}) for validated multi-candidate systems.}

Generally, we found that TOIs/CTOIs within our BLS search range are not recovered because their transits are shallow (\raven is reliable for depths above $\sim$200-300~ppm), the transits have low S/N (either because of the intrinsic small depth or because the light curve is noisy) or because they are part of multi-planet systems where other recovered candidates have transit depths significantly larger, which can challenge the detection of smaller transits.

We do not expect to recover all TOI/CTOIs for several reasons. Our BLS search is limited to 5 peaks, removes peaks with double and half the period of an identified signal, and it does not iteratively remove identified signals (e.g. by masking identified transits) when searching for the next ones.
Moreover, the TOI/CTOI candidates come from a range of heterogeneous sources and are identified with different methods. For instance, the \tess primary mission (2~min/20~s cadence), QLP, and the SPOC-FFI data (the one used here) have different cadences, object selection criteria (resulting in different sectors reduced by each pipeline), different photometry extraction, and use different candidate identification algorithms. CTOIs (and TOIs promoted from CTOIs) can include even further pipelines, selection methodologies, and data. Hence, more information than the FFI light curves and alternative methodologies might be needed to detect some TOIs/CTOIs.


\begin{figure}
\centering
\includegraphics[width=0.49\textwidth]{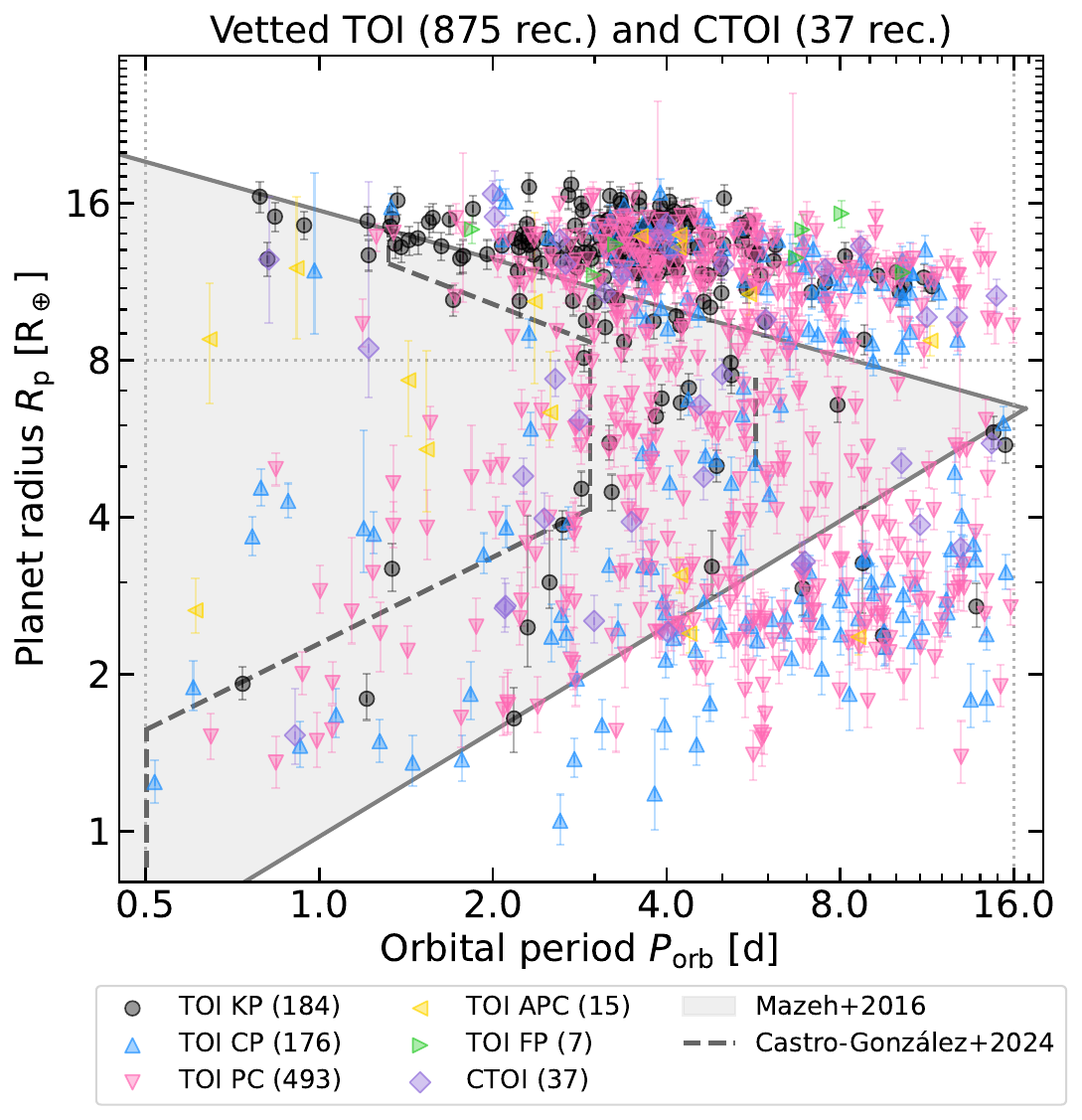}
\caption{TOI/CTOI candidates recovered in our vetted sample (875 TOIs and 37 CTOIs) in period-radius space. The values shown are the results of our \texttt{juliet} fits (see Sect. \ref{sec:juliet}). TOIs with ExoFOP disposition KP, CP, PC, APC, and FP are shown as black circles, blue up triangles, pink down triangles, yellow left triangles, and green right triangles, respectively. CTOIs are shown as purple diamonds.
Solid grey lines and grey-shaded area show the Neptunian desert limits according to \citet{mazeh2016desert}, and dashed grey lines show the recently derived  limits between the Neptunian desert, ridge, and savannah from \citet{castrogonzalez2024neptune}.
Dotted grey vertical and horizontal lines show our pipeline's detection/validation limits: periods from 0.5 to 16~d and radii below $8~\Rearth$.}
\label{fig:cand_per_rp_vet_toi}
\end{figure}

\begin{figure}
\centering
\includegraphics[width=0.49\textwidth]{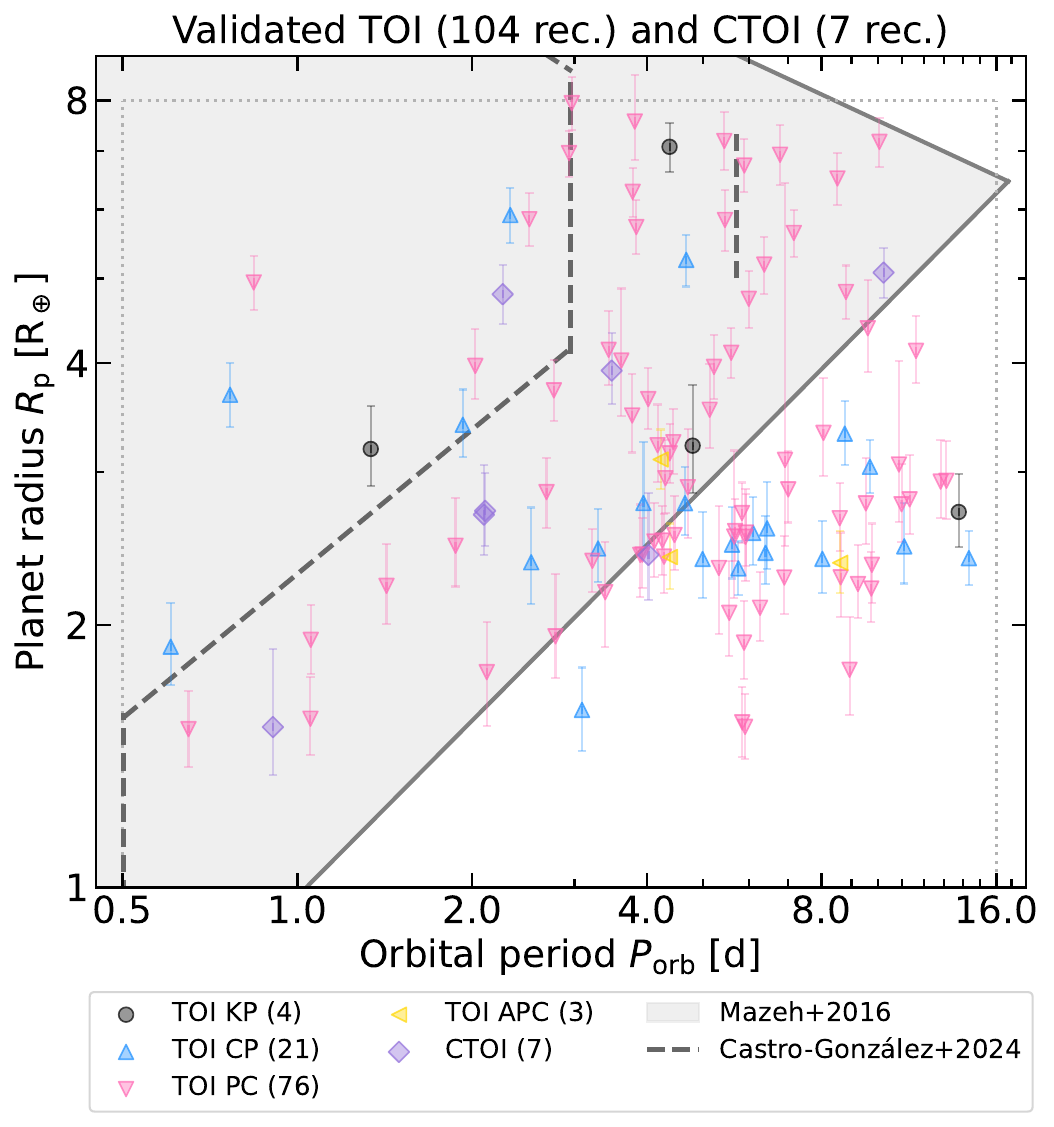}
\caption{TOI/CTOI candidates recovered in our validated sample (104 TOIs and 7 CTOIs) in period-radius space. All symbols show the same as Fig. \ref{fig:cand_per_rp_vet_toi} but for the validated sample rather than the vetted one.}
\label{fig:cand_per_rp_val_toi}
\end{figure}

\subsubsection{Vetted TOI/CTOI}
Our sample of vetted candidates includes 871 stars with 905 TOI identified in ExoFOP within our period range (878 stars with 925 TOI including those outside of our period search range), of which we recover 875 TOI in 863 stars. These 875 recovered TOIs are indicated as blue triangles in Fig. \ref{fig:cand_per_rp_vet}.
The breakdown of these recovered TOI into TFOP dispositions is 184 KPs, 176 CPs, 493 PCs, 15 APCs, and 7 FPs, see Fig. \ref{fig:cand_per_rp_vet_toi} for their distribution in period-radius space.
The 7 FPs are all known eclipsing binaries.
The vetted sample also contains 48 stars known to have 48 CTOIs (52 hosts and candidates outside of our period search range), of which we recover 37 candidates. These recovered CTOIs are indicated as yellow triangles in Fig. \ref{fig:cand_per_rp_vet}, and also appear in Fig. \ref{fig:cand_per_rp_vet_toi} as purple diamonds.

Amongst the stars known to have a TOI/CTOI in our vetted sample (including candidates with period outside our search range), there are some for which we do not recover exactly the reported candidates, and instead recover candidates that show multiples of the reported period (half $P/2$, double $2P$, one third $P/3$, and triple $3P$), only have a match in period and not in mid-transit time, or show no match at all. These are indicated as pink squares (TOI, 18 partially matched and 24 non-matched, 42 in total) and green diamonds (CTOI, 6 partially matched and 5 non-matched, 11 in total) in Fig. \ref{fig:cand_per_rp_vet}. Note that these do not appear in Fig. \ref{fig:cand_per_rp_vet_toi}, which shows only recovered TOIs/CTOIs, as they are not correctly recovered candidates.

Regarding the partially matched candidates, for the TOIs we recover $P/2$ for 11 TOI (10 PC and 1 FP, 7 within our period search), $2P$ for 1 (a PC inside our period search), $3P$ for 1 (a PC inside our period search), and for 2 (1 PC and 1 FP both inside our period search) we only have a match in period and not mid-transit time.
For the CTOIs, we recover $P/2$ for 3 candidates (all outside our period search), and match in period but not in mid-transit time for 3 candidates (all inside our period search).
We briefly describe these partially-matched TOI and CTOI candidates in Appendix \ref{sec:cand_vet_toi_other_recovery}. 
None of the 18 TOI candidates partially matched are in our validated sample.
Of the 5 CTOI partially matched, one is also in our validated sample, TIC~270471727 BLS peak 1 or now TIC~270471727~b. This candidate is actually correctly recovered, see next section \ref{sec:cand_val_toi}.

Regarding candidates without a match (i.e. non-recovered), we find 24 in stars hosting TOIs and 5 in stars hosting CTOIs.
Most of them correspond to recovered periods that are due to large gaps in the light curves, further aliases of periods outside of our search range that we did not consider above, or TOI parameters that are not correct. Two of them, TOI-6484.02 (TIC~294394558.02 in BLS peak 2 and TIC~142589416.02 (in BLS peak 2 of a CTOI) correspond to new candidates, see Sect. \ref{sec:new_toi} for details. 
We also describe all these non-matched candidates in detail in Appendix \ref{sec:cand_vet_toi_other_recovery}.

From the visual analysis presented in Appendix \ref{sec:cand_vet_toi_other_recovery}, we find that for 3 candidates (TIC~285094173 BLS peak 1, TIC~305048116 BLS peak 1, and TIC~346673534 BLS peak 1) the period reported in ExoFOP is a harmonic of the true period, and our BLS actually recovered the correct period, and for 5 candidates (TIC~155005217 BLS peak 1, TIC~270471727 BLS peak 1, TIC~435868942 BLS peak 1, TIC~458424950 BLS peak 1, and TIC~469775147 BLS peak 1) the match in mid-transit time $t_0$ failed because the $t_0$ reported in ExoFOP is out of phase by 0.5 and/or because ours and the ExoFOP's $t_0$ are separated in time and the $t_0$ precision (and/or the period precision) is not sufficient to result in a match when propagated in time.
Therefore, all these 9 candidates are actually correctly recovered signals.


\subsubsection{Validated TOI/CTOI}\label{sec:cand_val_toi}

In the sample of validated candidates (which is included in the vetted sample described above) there are 103 stars that have 107 TOI identified in ExoFOP (103 stars with 110 TOI including those outside of our period search range). We recovered 104 TOI in 103 stars. These recovered TOIs are indicated as blue triangles in Fig. \ref{fig:cand_per_rp_val}. According to the ExoFOP dispositions, these are 4 KPs, 21 CPs, 76 PCs, and 3 APCs, see Fig. \ref{fig:cand_per_rp_val_toi}. That is, we newly validate 79 TOIs with PC/APC disposition.
All the recovered TOIs in the validated sample have a match in period and mid-transit time (i.e. we do not find any TOI candidates at harmonics of the ExoFOP reported period).
There are no FPs or FAs amongst our validated TOIs.

Regarding CTOIs, our validated sample has 8 candidates in 8 stars (all within our period search range), of which we recover 7 candidates and partially recover 1 (TIC~270471727 BLS peak 1). The recovered candidates are indicated as yellow triangles in Fig. \ref{fig:cand_per_rp_val}, and also appear in Fig. \ref{fig:cand_per_rp_val_toi} as purple diamonds.
As mentioned above, the partially recovered candidate TIC~270471727 BLS peak 1 (green diamond in Fig.~\ref{fig:cand_per_rp_val}) failed the match in mid-transit time because our and the ExoFOP values are far apart in time, and the periods and/or mid-transit times are not precise enough to result in a match when propagated (see Appendix \ref{sec:cand_vet_toi_other_recovery} for more details). Hence, this is then a recovered candidate (TIC~270471727.01).
In total, we newly validate these 8 CTOI candidates.


\subsection{Newly validated candidates}

\begin{figure*}
\centering
\includegraphics[width=\textwidth]{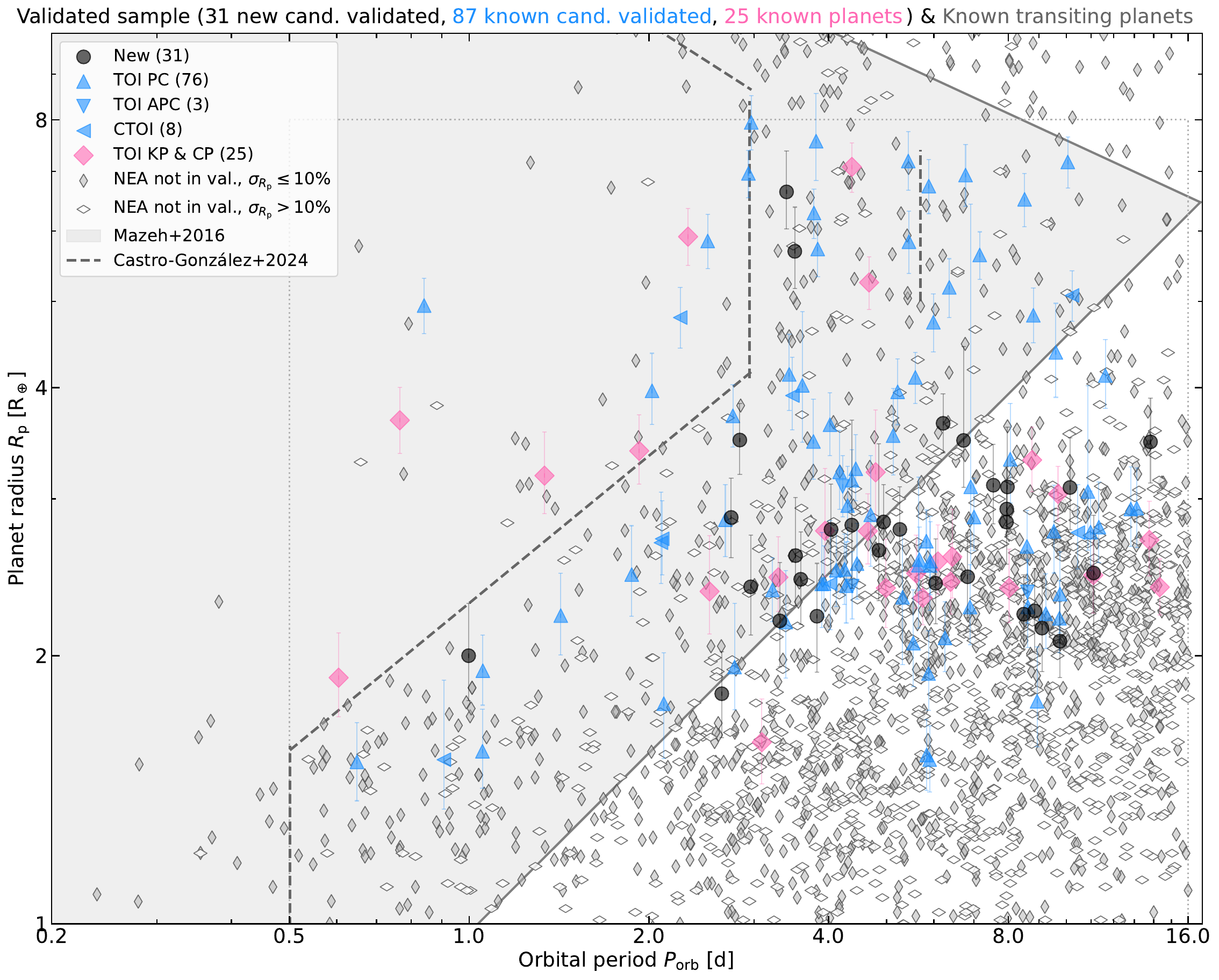}
\caption{Validated candidates in context. Validated sample of 143 candidates together with known transiting planets (with $\Rp\geq1~\Rearth$ and $\Porb<16$~d) from the NASA Exoplanet Archive \citep[NEA, Planetary Systems Composite Parameters Table, accessed on 13/08/2025][]{christiansen2025nea}, in period-radius space. The values shown for our candidates are the results of our \texttt{juliet} fits (see Sect. \ref{sec:juliet}), while the values from the rest of planets are taken from the NEA. Our validated candidates are divided into different groups: black circles show new validated planets identified here, blue triangles show previously known non-confirmed TOI/CTOI candidates which we now validate (upper triangles for TOI PCs, lower triangles for TOI APCs, and left triangles for CTOIs), and large pink diamonds show known and confirmed TOIs (KPs and CPs). Note that the CTOI TIC~270471727 BLS peak 1 or TIC~270471727.01 is included in `CTOI' category (left triangles), since this candidate is actually recovered, despite the $t_0$ not matching (see Appendix \ref{sec:cand_vet_toi_other_recovery}). Small diamonds show known transiting planets from the NEA that are not in our validated sample (flat, white diamonds for planets with relative radius uncertainty $>10\%$, and vertical, grey diamonds for planets with relative uncertainty $\leq10\%$). Solid grey lines and grey-shaded area show the Neptunian desert limits according to \citet{mazeh2016desert}, and dashed grey lines show the recently derived  limits between the Neptunian desert, ridge, and savannah from \citet{castrogonzalez2024neptune}. Dotted grey vertical and horizontal lines show our pipeline's detection/validation limits: periods from 0.5 to 16~d and radii below $8~\Rearth$.
}
\label{fig:cand_per_rp_val_nea}
\end{figure*}

Out of the total sample 143 validated planets, we newly validate 118 planets. These include 87 previously know TOIs/CTOIs candidates (76 TOI PCs, 3 TOI APCs, and 8 CTOIs) and 31 new candidates that we detect here.  We show all our validated sample in context with the currently known transiting exoplanet population in Fig. \ref{fig:cand_per_rp_val_nea}. 
In this section we highlight some interesting newly validated planets.


\subsubsection{Newly validated desert dwellers}

Within the Neptunian desert limits \citep[as defined by][]{castrogonzalez2024neptune}, we validate four TOI PCs: the ultra-short period TOI-7008~b (TOI-7008.01, TIC~335661164.01, $\Porb\simeq0.84$~d, $\Rp\simeq4.95\,\Rearth$), TOI-5486~b (TOI-5486.01, TIC~291109653.01, $\Porb\simeq2.02$~d, $\Rp\simeq3.97\,\Rearth$) and TOI-4030~b (TOI-4030.01, TIC~390021939.01, $\Porb\simeq2.51$~d, $\Rp\simeq5.85\,\Rearth$), both close to the desert edge, and TOI-2200~b (TOI-2200.01, TIC~142105158, $\Porb\simeq2.94$~d, $\Rp\simeq6.96\,\Rearth$), right at the limit with the Neptunian ridge (blue up triangles in Fig. \ref{fig:cand_per_rp_val_nea}). We also validate one CTOI, TIC~249022743~b (TIC~249022743.01, $\Porb\simeq2.26$~d, $\Rp\simeq4.80\,\Rearth$), identified in \citet{montalto2020diamante} and \citet{melton2024dtarps2} (blue left triangle in Fig. \ref{fig:cand_per_rp_val_nea}). All these planets are found in the first BLS peak.

TOI-7008~b ($\Porb\simeq0.84$~d, $\Rp\simeq4.95\,\Rearth$) is especially interesting because it is one of the shortest-period planets in the desert, together with Kepler-1520~b \citep[an evaporating planet with $\Porb\simeq0.65$~d, $\Rp\simeq5.77\,\Rearth$,][]{morton2016vespa}, 
LTT~9779~b \citep[a planet with extremely high albedo with $\Porb\simeq0.79$~d, $\Rp\simeq4.72\,\Rearth$,][]{jenkins2020ltt9779}, 
the extremely dense planets
TOI-3261~b \citep[$\Porb\simeq0.88$~d, $\Rp\simeq3.82\,\Rearth$,][]{nabbie2024toi3261}, 
TOI-849~b \citep[$\Porb\simeq0.77$~d, $\Rp\simeq3.62\,\Rearth$,][]{armstrong2020toi849}, and  
TOI-332~b \citep[$\Porb\simeq0.78$~d, $\Rp\simeq3.20\,\Rearth$,][]{osborn2023toi332}, and the compact multi-planet system K2-266~b \citep[$\Porb\simeq0.66$~d, $\Rp\simeq3.30\,\Rearth$,][]{rodriguez2018k2-266}.
The host TOI-7008 is a relatively faint star ($G\simeq13.44$~mag), slightly smaller and cooler than our Sun ($\Rstar\simeq0.95\,\Rsun$ and $\Teff\simeq5116$~K).

TOI-4030~b, TOI-2200~b, and TIC 249022743~b are also orbiting stars relatively faint and similar to our Sun ($G\simeq12.51-13.70$~mag, $\Rstar\simeq0.83-1.21\,\Rsun$ and $\Teff\simeq5036-5640$~K), while TOI-5486.01 is orbiting an M dwarf ($G\simeq13.35$~mag, $\Rstar\simeq0.55\,\Rsun$ and $\Teff\simeq3750$~K).

We note that the Neptunian desert is a region prone to FPs and generally has a low planet occurrence rate. \raven accounts for region-specific occurrence rates by constructing the simulated Planet training set based on current occurrence rates from \citet[][]{hsu2019occurrence} \citepalias[see details of our training sets in][]{hadjigeorghiou2025raven}. This means that Neptunian desert planets are more scarce in the training data, which the classifiers learn to recognise and take into account. On first order, the classifiers are less likely to attribute high probabilities to Neptunian desert planets (although the probability computation process is more complex and also depends on the specific FP scenario considered). Therefore, any validated candidates in the Neptunian desert region must have other convincing features as to their planetary nature to overcome the inherent bias against them, making them reliable validated candidates.


\subsubsection{Newly validated ultra-short-period super-Earths}\label{sec:usp}

Our validated sample has one new USP ($\Porb\lesssim1$~d) super-Earth, TIC~18942729~b (TIC~18942729.01, identified in the third BLS peak), just below the lower limit of the Neptunian desert \citep{castrogonzalez2024neptune}. It has $\Porb\simeq1.00$~d and $\Rp\simeq~2.00\,\Rearth$ and is orbiting a relatively faint star smaller and cooler than the Sun ($G\simeq12.27$~mag, $\Rstar\simeq0.82\,\Rsun$ and $\Teff\simeq4360$~K).
In the same star, we also visually identified a second candidate with $\Porb\simeq23$~d and $\Rp\simeq3.64\,\Rearth$ (TIC~18942729.02, see Sect. \ref{sec:multis_vis}).

In the same region, we also validate three TOI PCs, TOI-5736~b (TOI-5736.01, TIC~310380289.01, $P\simeq0.65$~d, $\Rp\simeq1.52\,\Rearth$), TOI-2345~b (TOI-2345.01, TIC~91555165.01, $P\simeq1.05$~d, $\Rp\simeq1.56\,\Rearth$), and TOI-6281~b (TOI-6281.01, TIC~152569268.01, $P\simeq1.05$~d, $\Rp\simeq1.92\,\Rearth$), and one CTOI, TIC~231949697~b (TIC~231949697.01, $P\simeq0.907204$~d, $\Rp\simeq1.528319\,\Rearth$) identified in \citet{eschen2024mdwarf}. We have detected all of them in the first BLS peak.
The TOIs TOI-6281~b and TOI-2345~b are orbiting K dwarf stars ($G\simeq10.65,\,11.03$~mag, $\Rstar\simeq0.87,\,0.76\,\Rsun$ and $\Teff\simeq4533,\,4584$~K, respectively).
TOI-5736~b has a late-K/early-M dwarf host ($G\simeq12.89$~mag, $\Rstar\simeq0.58\,\Rsun$, $\Teff\simeq4102$~K), and 
the CTOI TIC~231949697~b is orbiting a mid-M dwarf ($G\simeq12.36$~mag, $\Rstar\simeq0.49\,\Rsun$, $\Teff\simeq3451$~K).


\subsubsection{Newly validated multi-candidate systems}\label{sec:new_multi_val}

Our validated sample includes three systems for which we validate two planets each: TOI-1839 (TIC~381714186), TOI-4156 (TIC~462553953), and TIC~24750448. 
The general recovery of multi-candidate systems is further analysed in Sections \ref{sec:res_toi_bls} and \ref{sec:limit_multis}.

TOI-1839 (TIC~381714186) has two sub-Neptunes, an inner TOI PC ($\Porb\simeq1.42$~d and $\Rp\simeq2.22\,\Rearth$) and an outer CTOI identified in \citet{munavarhussain2025multis} ($\Porb\simeq4.02$~d and $\Rp\simeq\,2.41\Rearth$) that we detected in the first and second BLS peaks, respectively. They are orbiting a star slightly smaller and cooler than the Sun ($G\simeq10.73$~mag, $\Rstar\simeq0.85\,\Rsun$, $\Teff\simeq5315$~K).

Similarly, TOI-4156 (TIC~462553953) has two sub-Neptunes, both TOI PCs (with $\Porb\simeq4.46$~d and $\Rp\simeq2.54\,\Rearth$ for the inner planet and $\Porb\simeq12.82$~d and $\Rp\simeq2.92\,\Rearth$ for the outer) detected in the first and second BLS peaks. They are orbiting a star slightly larger and cooler than the Sun ($G\simeq11.12$~mag, $\Rstar\simeq1.04\,\Rsun$, $\Teff\simeq5971$~K).

We also detected two new validated sub-Neptunes in TIC~24750448, which does not have any known TOI/CTOI candidates. The inner planet has $\Porb\simeq3.59$~d and $\Rp\simeq2.44\,\Rearth$, detected in the third BLS peak, and the outer one, $\Porb\simeq10.14$~d and $\Rp\simeq3.09\,\Rearth$, detected in the second BLS peak. The periods are not clear harmonics of each other and the transits do not overlap. 
The host star is a faint K dwarf ($G\simeq13.21$~mag, $\Rstar\simeq0.71\,\Rsun$, $\Teff\simeq4545$~K).
Following the same process as explained in Section \ref{sec:multis_vis}, we performed a 2-planet fit with \juliet, which results in parameters consistent within 1$\sigma$ with those initially obtained from the 1-planet fits ($\Porb=3.591592^{+0.000038}_{-0.000051}$~d, $\Rp=2.41^{+0.27}_{-0.25}\,\Rearth$, and $t_0=2058.3725^{+0.0037}_{-0.0074}$~TBJD for the inner planet and $\Porb=10.143847^{+0.00012}_{-0.00012}$~d, $\Rp=3.03^{+0.29}_{-0.27}\,\Rearth$, and  $t_0=2085.0386^{+0.0053}_{-0.0046}$~TBJD for the outer one).


\subsection{New vetted candidates in TOI/CTOI host stars}\label{sec:new_toi}

In the vetted sample (and not in the validated one), we find one new candidate each in two TOI/CTOI-host stars, TIC~294394558 (TOI-6484) and TIC~142589416. As mentioned before in Section \ref{sec:multis_vis}, the fact that these vetted candidates are in stars already known to host other planet candidates reinforces their potential planetary nature.

TIC~294394558  has a known PC (TOI-6484.01) with $P\simeq18.25$~d and $\Rp\simeq3.87\,\Rearth$. This is outside of our BLS search range, but we recover the first harmonic in the first BLS peak (not in our vetted or validated samples). The second BLS peak corresponds to a new vetted candidate (TOI-6484.02) with $P\simeq10.90$~d (close to 5/3 the known PC period) and $\Rp\simeq3.61\,\Rearth$.
The transits of the two candidates do not overlap, although they occur close together in sector 35. The reported depth for the TOI is $\simeq3100$~ppm, while for the new candidate with $P\simeq10.90$~d, we find a depth of $\simeq1800$~ppm. 
Sector 34 further shows an extra transit-like feature that could be due to an outer candidate.
The stellar host is a relatively faint K dwarf ($G\simeq12.66$~mag, $\Rstar\simeq0.78\,\Rsun$, $\Teff\simeq4934$~K).
We performed a two-planet fit with \juliet and the orbital parameters agree within $1\sigma$ with those from the single-planet fit ($\Porb=10.900490^{+0.000072}_{-0.000075}$~d, $\Rp=3.51^{+0.25}_{-0.24}$, and $t_0=2205.3299^{+0.0029}_{-0.0024}$~TBJD for the new inner planet).
An ASTEP \citep{guillot2015astep,mekarnia2016astep} transit observation of the known PC on 02/06/2024 reported on ExoFOP showed that the transit started 17~min earlier than expected from the \tess ephemerides, which could indicate TTVs caused by the inner candidate we identified here. 

We also identified a new candidate in the CTOI host TIC~142589416. This star has a known CTOI, TIC~142589416.01 \citep{montalto2023diamante} with $P\simeq2.10$~d and $\Rp\simeq2.57\,\Rearth$, recovered in our first BLS peak and in our validated sample. The second BLS candidate of this star is in our vetted sample. This second candidate has a period of $\simeq4.80$~d, which is not a clear alias of the known CTOI, and a radius of $\Rp\simeq2.70\,\Rearth$. The transits of the two candidates do not coincide. 
They are orbiting faint late-type K dwarf ($G\simeq13.56$~mag, $\Rstar\simeq0.71\,\Rsun$, $\Teff\simeq4396$~K).
A two-planet fit with \juliet results in parameters that agree within $1\sigma$ with those from the two separate single-planet fits ($\Porb=2.095183^{+0.000010}_{-0.000010}$, $\Rp=2.67^{+0.35}_{-0.27}$, $t_0=2614.12^{+0.0020}_{-0.0021}$ for the known inner CTOI and $\Porb=4.804919^{+0.000048}_{-0.000055}$, $\Rp=2.88^{+0.49}_{-0.33}$, $t_0=2618.7575^{+0.0038}_{-0.0038}$ for the new outer candidate).


\subsection{Limitations}

Here we highlight some of the shortcomings of our work. We refer the reader to \citetalias{hadjigeorghiou2025raven} for an in-depth discussion of the pipeline and its main limitations.

\subsubsection{Harmonics}

Our region of interest for this work is the Neptunian desert and its surrounding regions, that is, short-period ($<16$~d), $\sim$Neptune-sized planets. Hence, the simulations used by \raven's machine learning models contained only transiting planets with radius within 1 and 16~\Rearth and periods from 0.5 to 16~d. Accordingly, our BLS search looked for signals within 0.5 and 16~d.
This allowed us to recover several known planets and planet candidates occupying this parameters space, as well as detect new candidates. 

Our methodology is in principle not sensitive to planets outside this period range, however, we still recovered harmonics of signals outside our range.
We performed visual vetting of our high-probability candidates to remove those corresponding to harmonics. However, it was not always possible to correctly identify a candidate as a harmonic of the true period, especially in cases of low-S/N, shallow transits, and in cases where transits would fall on data gaps.
Indeed, we observed that several TOI with period outside of our range were partially-recovered by our pipeline (i.e. our signal matched one of the true period harmonics rather than the actual period) after the our visual vetting step. This especially occurred when gaps in the data allowed for period shorter than the true one (i.e. the periods needed for the shorter harmonic to be the true period fell on data gaps).

In some cases we also recovered harmonics of a true signal that was within our BLS search range.
To try to mitigate selecting harmonics, once a signal is found, the BLS discarded half and double the period of that signal. This procedure is performed on the five highest peaks, in order of decreasing power (SDE). Generally, we expect the true period to produce a peak with higher SDE than its harmonics, but this might not be the case for all light curves. The BLS cannot assess if the original signal found is the true period or a harmonic, which implies that, if a harmonic is selected first, the true signal will be discarded. Moreover, harmonics different than two and half times the period of a signal can still be included in the recovered BLS peaks. Our vetted sample included several TOIs with true period within our range, for which we recover a harmonic (see Appendix \ref{sec:cand_vet_toi_other_recovery}). This occurred mostly for candidates with numerous gaps in the light curves (due to data being flagged as low quality) and/or for signals whose transits could fall in intra-sector gaps and/or just outside the start/end of a sector.

\subsubsection{Multi-candidate systems} \label{sec:limit_multis}

In general, the number of recovered candidates in multiple systems in our BLS, vetted, and validated samples is low, as explained in Sect. \ref{sec:res_toi}.
This is not an exclusive characteristic of \raven, since other approaches such as the \textit{Kepler} pipeline are also prone to missing multi-candidate systems \citep[e.g.][]{thompson2018kepler25,hedges2019k2}. Here we briefly discuss some of the possible reasons behind this low detection of multi-planet systems for \raven.

As shown in \citetalias{hadjigeorghiou2025raven}, \raven's performance when it comes to recover and validate candidates in multi-planet systems is complex.
Aside from the BLS search being limited to periods from 0.5 up to 16~d and to 5 candidates per star, multi-planet systems are not present in the simulations used to train the machine learning models, and the pipeline does not account for effects such as transit timing variations, all of which could result in a lower performance for multi-planet systems.
As reported in \citetalias{hadjigeorghiou2025raven}, for a sample of 709 confirmed planet TOIs (i.e. KP and CP) \raven obtained a mean final probability of 0.78 and median of 0.96. When considering only those in multi-planet systems (95 planets), the performance decreased slightly, with a mean final probability of 0.68 and median of 0.84.
However, the final probabilities for planets in multi-planet systems still span the whole possible range, and 37 of them had probability $>0.9$. The fact that a planet in a multiple system ends up with a low probability is not simply because of the system having more than one planet, but also depends on the transit S/N, impact parameter, ephemeris, and/or dilution level.
Finally, as shown in \citetalias{hadjigeorghiou2025raven}, \raven’s performance for multi-candidate systems can be significantly improved when the transits of the other candidates are masked (when this information is previously known). Note that such masking was not possible here as harmonics of the candidates were also present in our candidate list.

\subsubsection{Depth}

The simulations of transiting planets included systems down to 1~\Rearth.
In general, to correctly compute reliable features, \raven requires transits to have relatively high S/N, which generally implies relatively large depths.
Our BLS removed signals with low SDE and low MES, which generally correspond to shallow transits.
Moreover, when vetting our candidates, we also imposed a minimum S/N threshold and cuts in the transit depth.
All of these conditions increase the reliability of our candidates, but at the same time, they limit the power of the pipeline to detect small planets.
Indeed, the bulk of our candidates in the vetted sample is generally above $\sim1.2\,\Rearth$.

In our validated sample we only included candidates with $\Rp\leq8\,\Rearth$. As mentioned in Sect. \ref{sec:res_cand_val}, we impose this cut because, for larger radii, planets, brown dwarfs, and low-mass stars produce very similar transits and are very challenging to differentiate from each other. Again, this increases the reliability of our validated candidates as true transiting planets, while at the same time decreasing the detection of large Saturn- and Jupiter-like planets.

\subsubsection{Crowded regions}

When vetting transits, we discarded candidates with a positional probability (the probability that the event occurs on target rather than on a known nearby source) below 0.5. This cut, together with the nearby scenario classifiers, ensures that our candidates occur on the target star. At the same time, this means that our sensitivity in crowded regions of the sky might decrease, since the positional probability (even for candidates found on-target) tends to decrease for targets in crowded regions \citep[see Fig. 12 in][]{hadjigeorghiou2024positional}. Despite that, other factors than the crowding fraction can affect the positional probability of a target, such as the S/N of the transit (i.e. on-target, low-S/N transits might also result in a low positional probability). Hence, the sensitivity of the pipeline to candidates in crowded regions is complex and depends on multiple properties of the observed light curves.

\subsubsection{Orbital parameters}

\begin{figure}
\centering
\includegraphics[width=0.5\textwidth]{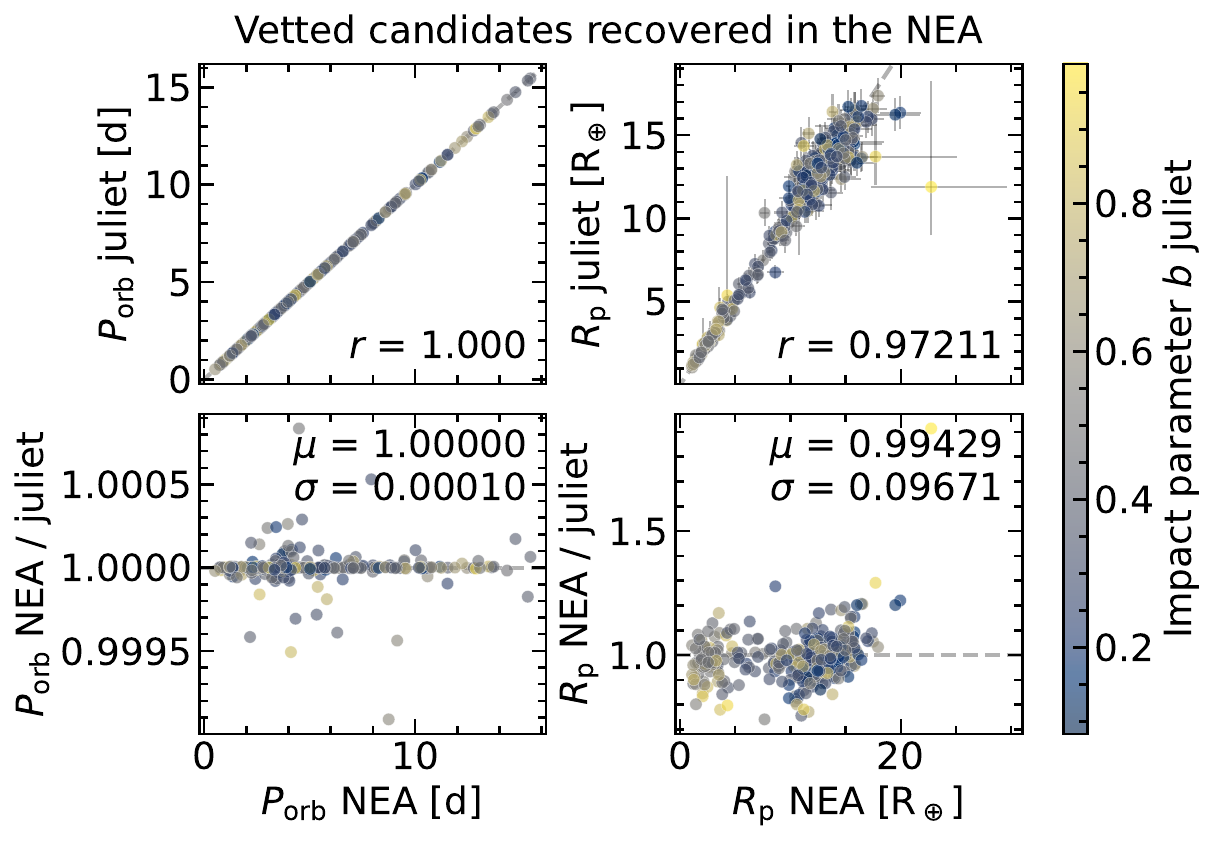}
\caption{Comparison between our \texttt{juliet} fits and values from the NEA, with periods on the left panels and radii on the right panels. Top panels show \texttt{juliet} vs NEA values and Pearson's correlation coefficient ($r$) in the inset text, and bottom panels show the ratio of NEA to \texttt{juliet} values as a function of the NEA values, with the inset text showing the ratio mean ($\mu$) and standard deviation ($\sigma$).
Note that the top left panel contains the period uncertainties, but they are too small to be seen (the typical period uncertainty is of the order of $10^{-5}-10^{-6}$).}
\label{fig:juliet_nea_compare}
\end{figure}

For the candidates in our vetted and validated samples, we provide a uniform set of system parameters: stellar parameters from \gaia and planetary parameters from our \texttt{juliet} fits to the \tess sectors 1 to 55. 
For recovered planets in our vetted sample confirmed in the NEA, we compared our \texttt{juliet} fit values to the values reported in the NEA (Fig. \ref{fig:juliet_nea_compare}).
Generally, we find a good match between our \texttt{juliet} periods and radii and those in the NEA.
The \texttt{juliet} and NEA periods show a Pearson's correlation coefficient of 1.0 and the ratio of the two values has a standard deviation of $\sim0.0001$.
When comparing the radii, we find a correlation of 0.97, and the radius ratio shows a standard deviation of $<0.1$.
We note again that any dubious fits (such as those with high impact parameter, low S/N transit, or the presence of noise in the light curve) have been flagged in Table \ref{tab:cand_vet}, under the \emph{Notes} column.

Despite being generally reliable and useful for statistical analyses, we advise caution when using the orbital parameters provided here. Obtaining a uniform sample of parameters required us to use the same priors, fit procedure, and assumptions for all candidates (e.g. circular orbit, single-planet with no TTVs, stellar/instrumental variations well represented by a GP with a Matérn-3/2 kernel), which might result in less accurate and/or less precise parameters than performing a tailored fit for each target. 
Moreover, we only used \tess FFI data from sectors 1 to 55, but for several of our targets there are further sectors available, as well as other photometric and/or spectroscopic data that can constrain orbital parameters better than only using limited \tess data.

\section{Conclusion}\label{sec:conclusion}

We have applied the \raven pipeline to $\sim$4 years of \tess SPOC FFI light curves (sectors 1-55) of a sample of $\sim$2.26 million main sequence stars well-characterised by \gaia. \raven identifies transiting planet candidates with a BLS and classifies these candidates between likely transiting planets and several FP scenarios using machine learning models trained with realistic simulations. The final output of the pipeline is the minimum probability of all the planet-FP classifiers combined with the priors for each scenario, the \raven probability. We used this \raven probability to rank our candidates and performed further vetting of the most promising ones. 

We provide several samples or lists of candidates with their main properties.
We provide a list of statistically validated planets (143 planets), containing candidates that have passed all of our vetting thresholds and have probability $\geq0.99$.
We also provide a sample of vetted candidates (2170 candidates), containing vetted, highly ranked candidates with transit-like events (probability $\geq0.9$) whose probability is generally not high enough to be statistically validated (note that the validated candidates are included in this sample).
Complementing the previous tables, we also provide summary plots for all candidates in the vetted and validated samples, showing the FFI light curves used and the light curves phase-folded to the candidate period showing the detected transit.
\raven also detected significant single-transit events (4 candidates, probability~$>0.9$) with an incorrect period, and significant candidates with large radius ($\Rp\geq8\,\Rearth$, 207 candidates, probability~$>0.99$) not validated due to the degeneracy in radius between planets, brown dwarfs and low-mass stars. We also make these two extra samples available here.
New candidates will be reported in the ExoFOP portal as CTOIs.

Finally, we also make available all candidates with an initial vetting performed with our Planet-NSFP classifiers, which trace instrumental and stellar variability (mean NSFP classifier probability $\geq0.9$, 14\,815 candidates). This sample contains significant TCEs, most of which are FPs (which are further vetted with the rest of \raven classifiers in the aforementioned samples).

Regarding our sample of validated planets, we would like to advise the community to exercise caution and to in general not treat validated planets, from any validation method, in the same way as confirmed planets with known mass are treated. Instead, validated planets are planets that have a very high planet probability, better than planet candidates, but are not equivalent to confirmed planets. Some of our validated planets (including all non-TOI candidates newly discovered here) do not yet have any follow-up observations, which should help confirm that the transit events occur on the host star, show no transit chromaticity, and eventually should be able to measure the planet's mass, allowing their confirmation.

As a summary, we provide several samples of transiting planet candidates (with orbital periods between 0.5 and 16~d) derived from a large, uniform sample of well-characterised main sequence stars. These candidates include over 2000 vetted candidates, $\sim$1000 of which are newly identified here. We statistically validate as planets 87 known TOI/CTOI with planet candidate status and 31 new candidates identified here. 
As shown in \citetalias{hadjigeorghiou2025raven}, the different steps performed by the \raven pipeline have been extensively well characterised, which implies that the different biases present in our samples of candidates can be well understood. In particular, our vetted sample is suitable for statistical analyses (note that, in our validated sample, we have performed a manual removal of unreliable candidates, which biases the sample). This characterisation of our pipeline and vetted sample allows for the inference of underlying exoplanet occurrence rates, which is the subject of further work presented in \citet{cui2026occurrence}.


\section*{Acknowledgements}

%
This research was in part funded by the UKRI (Grant EP/X027562/1).
Computing facilities were provided by the Scientific Computing Research Technology Platform of the University of Warwick.
Funding for the \tess mission is provided by NASA's Science Mission directorate.
This paper includes data collected by the \tess mission, which are publicly available from the Mikulski Archive for Space Telescopes (MAST).
We acknowledge the use of public TOI Release data from pipelines at the \tess Science Office and at the \tess Science Processing Operations Center.
This research has made use of the Exoplanet Follow-up Observation Program (ExoFOP; DOI: 10.26134/ExoFOP5) website, which is operated by the California Institute of Technology, under contract with the National Aeronautics and Space Administration under the Exoplanet Exploration Program.
This research has made use of the NASA Exoplanet Archive, which is operated by the California Institute of Technology, under contract with the National Aeronautics and Space Administration under the Exoplanet Exploration Program.
This work has made use of data from the European Space Agency (ESA) mission \gaia (\url{https://www.cosmos.esa.int/gaia}), processed by the \gaia Data Processing and Analysis Consortium (DPAC, \url{https://www.cosmos.esa.int/web/gaia/dpac/consortium}). Funding for the DPAC has been provided by national institutions, in particular the institutions participating in the \gaia Multilateral Agreement.
This work made use of \texttt{numpy} \citep{harris2020numpy}, \texttt{scipy} \citep{virtanen2020scipy}, \texttt{matplotlib} \citep{hunter2007matplotlib}, \texttt{astropy} \citep{astropycollaboration2013astropy,astropycollaboration2018astropy}, \texttt{astroquery} \citep{ginsburg2019astroquery}, \texttt{cuvarbase} \citep{hoffman2022cuvarbase}, \texttt{Scikit-learn} \citep{pedregosa2011sklearn}, \texttt{skorch} \citep{tietz2017skorch}, \texttt{XGBoost} \citep{chen2016xgboost}, \texttt{Lightkurve} \citep{lightkurve2018}, \texttt{juliet} \citep{espinoza2019juliet}, \texttt{dynesty} \citep{speagle2020dynesty,koposov2024dynesty}, \texttt{batman} \citep{kreidberg2015batman}, \texttt{celerite} \citep{foremanmackey2017celerite}, \raven \citep{hadjigeorghiou2024positional,hadjigeorghiou2025raven}, \texttt{PASTIS} \citep{diaz2014pastis,santerne2015pastis}, and the \tess Light Curve Online Viewer \url{https://tess.cuikaiming.com/}.
We thank the anonymous referee for their review of our draft, which helped improve the results presented in this work.

\section*{Data Availability}

We make our samples of candidates and relevant parameters publicly available as online material in \url{https://zenodo.org/records/19661443}.
This paper includes data collected by the \tess mission, which are publicly available from the Mikulski Archive for Space Telescopes (MAST), DOI \href{https://dx.doi.org/10.17909/t9-wpz1-8s54}{10.17909/t9-wpz1-8s54}.
This paper uses data from the Planetary Systems Composite Parameters Table available in the NASA Exoplanet Archive, DOI \href{https://doi.org/10.26133/NEA13}{10.26133/NEA13}, accessed on 13/08/2025.
This paper uses data from the TOIs and CTOIs Tables in ExoFOP, DOI \href{https://exofop.ipac.caltech.edu/}{10.26134/ExoFOP5}, accessed on 13/08/2025.
This work has made use of data from the European Space Agency (ESA) mission \gaia (\url{https://www.cosmos.esa.int/gaia}), processed by the \gaia Data Processing and Analysis Consortium (DPAC, \url{https://www.cosmos.esa.int/web/gaia/dpac/consortium}).




\bibliographystyle{mnras}
\bibliography{raven_candidates} 

@ARTICLE{hadjigeorghiou2025raven,
       author = {{Hadjigeorghiou}, Andreas and {Armstrong}, David J. and {Cui}, Kaiming and {Lafarga Magro}, Marina and {Agust{\'\i}n Nieto}, Luis and {D{\'\i}az}, Rodrigo F. and {Doyle}, Lauren and {Kunovac}, Vedad},
        title = "{RAVEN: RAnking and Validation of ExoplaNets}",
      journal = {arXiv e-prints},
         year = 2025,
        month = sep,
          eid = {arXiv:2509.17645},
        pages = {arXiv:2509.17645},
          doi = {10.48550/arXiv.2509.17645},
archivePrefix = {arXiv},
       eprint = {2509.17645},
 primaryClass = {astro-ph.EP},
       adsurl = {https://ui.adsabs.harvard.edu/abs/2025arXiv250917645H}
}

@ARTICLE{cui2026occurrence,
       author = {{Cui}, Kaiming and {Armstrong}, David J. and {Hadjigeorghiou}, Andreas and {Lafarga}, Marina and {Kunovac}, Vedad and {Doyle}, Lauren and {Nieto}, Luis Agust{\'\i}n and {D{\'\i}az}, Rodrigo F.},
        title = "{Demographics of close-in TESS exoplanets orbiting FGK main-sequence stars}",
      journal = {\mnras},
         year = 2026,
        month = feb,
       volume = {546},
       number = {2},
          eid = {stag022},
        pages = {stag022},
          doi = {10.1093/mnras/stag022},
archivePrefix = {arXiv},
       eprint = {2601.09492},
 primaryClass = {astro-ph.EP},
       adsurl = {https://ui.adsabs.harvard.edu/abs/2026MNRAS.546ag022C}
}

@manual{tietz2017skorch,
  author       = {Marian Tietz and Thomas J. Fan and Daniel Nouri and Benjamin Bossan and {skorch Developers}},
  title        = {skorch: A scikit-learn compatible neural network library that wraps PyTorch},
  month        = jul,
  year         = 2017,
  url          = {https://skorch.readthedocs.io/en/stable/}
}

@inproceedings{akeson2019tfopwg,
  title = {{{TESS Follow-up Observing Program Working Group}} ({{TFOP WG}}): {{The ExoFOP-TESS Website}}},
  shorttitle = {{{TESS Follow-up Observing Program Working Group}} ({{TFOP WG}})},
  booktitle = {American {{Astronomical Society Meeting Abstracts}} \#233},
  author = {Akeson, Rachel and Christiansen, Jessie},
  year = {2019},
  month = jan,
  volume = {233},
  pages = {140.09},
  urldate = {2025-08-20}
}

@article{andrae2018gaia,
  title = {Gaia {{Data Release}} 2 - {{First}} Stellar Parameters from {{Apsis}}},
  author = {Andrae, Ren{\'e} and Fouesneau, Morgan and Creevey, Orlagh and Ordenovic, Christophe and Mary, Nicolas and Burlacu, Alexandru and Chaoul, Laurence and {Jean-Antoine-Piccolo}, Anne and Kordopatis, Georges and Korn, Andreas and Lebreton, Yveline and Panem, Chantal and Pichon, Bernard and Th{\'e}venin, Fr{\'e}d{\'e}ric and Walmsley, Gavin and {Bailer-Jones}, Coryn A. L.},
  year = {2018},
  month = aug,
  journal = {A\&A},
  volume = {616},
  pages = {A8},
  issn = {0004-6361, 1432-0746},
  doi = {10.1051/0004-6361/201732516},
  urldate = {2025-09-08},
  copyright = {{\copyright} ESO 2018},
  langid = {english}
}

@article{ansdell2018exonet,
  title = {Scientific {{Domain Knowledge Improves Exoplanet Transit Classification}} with {{Deep Learning}}},
  author = {Ansdell, Megan and Ioannou, Yani and Osborn, Hugh P. and Sasdelli, Michele and {(2018 NASA Frontier Development Lab Exoplanet Team)} and Smith, Jeffrey C. and Caldwell, Douglas and Jenkins, Jon M. and R{\"a}issi, Chedy and Angerhausen, Daniel and {(2018 NASA Frontier Development Lab Exoplanet Mentors)}},
  year = {2018},
  month = dec,
  journal = {ApJL},
  volume = {869},
  number = {1},
  pages = {L7},
  issn = {2041-8205},
  doi = {10.3847/2041-8213/aaf23b},
  urldate = {2025-09-02},
  langid = {english}
}

@article{armstrong2017som,
  title = {Transit Shapes and Self-Organizing Maps as a Tool for Ranking Planetary Candidates: Application to {{Kepler}} and {{K2}}},
  shorttitle = {Transit Shapes and Self-Organizing Maps as a Tool for Ranking Planetary Candidates},
  author = {Armstrong, D. J. and Pollacco, D. and Santerne, A.},
  year = {2017},
  month = mar,
  journal = {\mnras},
  volume = {465},
  pages = {2634--2642},
  issn = {0035-8711},
  doi = {10.1093/mnras/stw2881},
  urldate = {2023-12-07}
}

@article{armstrong2020toi849,
  title = {A Remnant Planetary Core in the Hot-{{Neptune}} Desert},
  author = {Armstrong, David J. and Lopez, Th{\'e}o A. and Adibekyan, Vardan and Booth, Richard A. and Bryant, Edward M. and Collins, Karen A. and Deleuil, Magali and Emsenhuber, Alexandre and Huang, Chelsea X. and King, George W. and {Lillo-Box}, Jorge and Lissauer, Jack J. and Matthews, Elisabeth and Mousis, Olivier and Nielsen, Louise D. and Osborn, Hugh and Otegi, Jon and Santos, Nuno C. and Sousa, S{\'e}rgio G. and Stassun, Keivan G. and Veras, Dimitri and Ziegler, Carl and Acton, Jack S. and Almenara, Jose M. and Anderson, David R. and Barrado, David and Barros, Susana C. C. and Bayliss, Daniel and Belardi, Claudia and Bouchy, Francois and Brice{\~n}o, C{\'e}sar and Brogi, Matteo and Brown, David J. A. and Burleigh, Matthew R. and Casewell, Sarah L. and Chaushev, Alexander and Ciardi, David R. and Collins, Kevin I. and Col{\'o}n, Knicole D. and Cooke, Benjamin F. and Crossfield, Ian J. M. and D{\'i}az, Rodrigo F. and Delgado Mena, Elisa and Demangeon, Olivier D. S. and Dorn, Caroline and Dumusque, Xavier and Eigm{\"u}ller, Philipp and Fausnaugh, Michael and Figueira, Pedro and Gan, Tianjun and Gandhi, Siddharth and Gill, Samuel and Gonzales, Erica J. and Goad, Michael R. and G{\"u}nther, Maximilian N. and Helled, Ravit and Hojjatpanah, Saeed and Howell, Steve B. and Jackman, James and Jenkins, James S. and Jenkins, Jon M. and Jensen, Eric L. N. and Kennedy, Grant M. and Latham, David W. and Law, Nicholas and Lendl, Monika and Lozovsky, Michael and Mann, Andrew W. and Moyano, Maximiliano and McCormac, James and Meru, Farzana and Mordasini, Christoph and Osborn, Ares and Pollacco, Don and Queloz, Didier and Raynard, Liam and Ricker, George R. and Rowden, Pamela and Santerne, Alexandre and Schlieder, Joshua E. and Seager, Sara and Sha, Lizhou and Tan, Thiam-Guan and Tilbrook, Rosanna H. and Ting, Eric and Udry, St{\'e}phane and Vanderspek, Roland and Watson, Christopher A. and West, Richard G. and Wilson, Paul A. and Winn, Joshua N. and Wheatley, Peter and Villasenor, Jesus Noel and Vines, Jose I. and Zhan, Zhuchang},
  year = {2020},
  month = jul,
  journal = {Nature},
  volume = {583},
  pages = {39--42},
  issn = {0028-0836},
  doi = {10.1038/s41586-020-2421-7},
  urldate = {2022-09-20}
}

@article{armstrong2021kepler,
  title = {Exoplanet Validation with Machine Learning: 50 New Validated {{Kepler}} Planets},
  shorttitle = {Exoplanet Validation with Machine Learning},
  author = {Armstrong, David J. and Gamper, Jevgenij and Damoulas, Theodoros},
  year = {2021},
  month = jul,
  journal = {\mnras},
  volume = {504},
  pages = {5327--5344},
  issn = {0035-8711},
  doi = {10.1093/mnras/staa2498},
  urldate = {2023-07-15}
}

@article{astropycollaboration2013astropy,
  title = {Astropy: {{A}} Community {{Python}} Package for Astronomy},
  shorttitle = {Astropy},
  author = {{Astropy Collaboration} and Robitaille, Thomas P. and Tollerud, Erik J. and Greenfield, Perry and Droettboom, Michael and Bray, Erik and Aldcroft, Tom and Davis, Matt and Ginsburg, Adam and {Price-Whelan}, Adrian M. and Kerzendorf, Wolfgang E. and Conley, Alexander and Crighton, Neil and Barbary, Kyle and Muna, Demitri and Ferguson, Henry and Grollier, Fr{\'e}d{\'e}ric and Parikh, Madhura M. and Nair, Prasanth H. and Unther, Hans M. and Deil, Christoph and Woillez, Julien and Conseil, Simon and Kramer, Roban and Turner, James E. H. and Singer, Leo and Fox, Ryan and Weaver, Benjamin A. and Zabalza, Victor and Edwards, Zachary I. and Azalee Bostroem, K. and Burke, D. J. and Casey, Andrew R. and Crawford, Steven M. and Dencheva, Nadia and Ely, Justin and Jenness, Tim and Labrie, Kathleen and Lim, Pey Lian and Pierfederici, Francesco and Pontzen, Andrew and Ptak, Andy and Refsdal, Brian and Servillat, Mathieu and Streicher, Ole},
  year = {2013},
  month = oct,
  journal = {A\&A},
  volume = {558},
  pages = {A33},
  issn = {0004-6361},
  doi = {10.1051/0004-6361/201322068},
  urldate = {2020-01-23}
}

@article{astropycollaboration2018astropy,
  title = {The {{Astropy Project}}: {{Building}} an {{Open-science Project}} and {{Status}} of the v2.0 {{Core Package}}},
  shorttitle = {The {{Astropy Project}}},
  author = {{Astropy Collaboration} and {Price-Whelan}, A. M. and Sip{\H o}cz, B. M. and G{\"u}nther, H. M. and Lim, P. L. and Crawford, S. M. and Conseil, S. and Shupe, D. L. and Craig, M. W. and Dencheva, N. and Ginsburg, A. and VanderPlas, J. T. and Bradley, L. D. and {P{\'e}rez-Su{\'a}rez}, D. and {de Val-Borro}, M. and Aldcroft, T. L. and Cruz, K. L. and Robitaille, T. P. and Tollerud, E. J. and Ardelean, C. and Babej, T. and Bach, Y. P. and Bachetti, M. and Bakanov, A. V. and Bamford, S. P. and Barentsen, G. and Barmby, P. and Baumbach, A. and Berry, K. L. and Biscani, F. and Boquien, M. and Bostroem, K. A. and Bouma, L. G. and Brammer, G. B. and Bray, E. M. and Breytenbach, H. and Buddelmeijer, H. and Burke, D. J. and Calderone, G. and Cano Rodr{\'i}guez, J. L. and Cara, M. and Cardoso, J. V. M. and Cheedella, S. and Copin, Y. and Corrales, L. and Crichton, D. and D'Avella, D. and Deil, C. and Depagne, {\'E}. and Dietrich, J. P. and Donath, A. and Droettboom, M. and Earl, N. and Erben, T. and Fabbro, S. and Ferreira, L. A. and Finethy, T. and Fox, R. T. and Garrison, L. H. and Gibbons, S. L. J. and Goldstein, D. A. and Gommers, R. and Greco, J. P. and Greenfield, P. and Groener, A. M. and Grollier, F. and Hagen, A. and Hirst, P. and Homeier, D. and Horton, A. J. and Hosseinzadeh, G. and Hu, L. and Hunkeler, J. S. and Ivezi{\'c}, {\v Z}. and Jain, A. and Jenness, T. and Kanarek, G. and Kendrew, S. and Kern, N. S. and Kerzendorf, W. E. and Khvalko, A. and King, J. and Kirkby, D. and Kulkarni, A. M. and Kumar, A. and Lee, A. and Lenz, D. and Littlefair, S. P. and Ma, Z. and Macleod, D. M. and Mastropietro, M. and McCully, C. and Montagnac, S. and Morris, B. M. and Mueller, M. and Mumford, S. J. and Muna, D. and Murphy, N. A. and Nelson, S. and Nguyen, G. H. and Ninan, J. P. and N{\"o}the, M. and Ogaz, S. and Oh, S. and Parejko, J. K. and Parley, N. and Pascual, S. and Patil, R. and Patil, A. A. and Plunkett, A. L. and Prochaska, J. X. and Rastogi, T. and Reddy Janga, V. and Sabater, J. and Sakurikar, P. and Seifert, M. and Sherbert, L. E. and {Sherwood-Taylor}, H. and Shih, A. Y. and Sick, J. and Silbiger, M. T. and Singanamalla, S. and Singer, L. P. and Sladen, P. H. and Sooley, K. A. and Sornarajah, S. and Streicher, O. and Teuben, P. and Thomas, S. W. and Tremblay, G. R. and Turner, J. E. H. and Terr{\'o}n, V. and {van Kerkwijk}, M. H. and {de la Vega}, A. and Watkins, L. L. and Weaver, B. A. and Whitmore, J. B. and Woillez, J. and Zabalza, V. and {Astropy Contributors}},
  year = {2018},
  month = sep,
  journal = {AJ},
  volume = {156},
  pages = {123},
  issn = {0004-6256},
  doi = {10.3847/1538-3881/aabc4f},
  urldate = {2020-01-23}
}

@article{borucki2010kepler,
  title = {Kepler {{Planet-Detection Mission}}: {{Introduction}} and {{First Results}}},
  shorttitle = {Kepler {{Planet-Detection Mission}}},
  author = {Borucki, William J. and Koch, David and Basri, Gibor and Batalha, Natalie and Brown, Timothy and Caldwell, Douglas and Caldwell, John and {Christensen-Dalsgaard}, J{\o}rgen and Cochran, William D. and DeVore, Edna and Dunham, Edward W. and Dupree, Andrea K. and Gautier, Thomas N. and Geary, John C. and Gilliland, Ronald and Gould, Alan and Howell, Steve B. and Jenkins, Jon M. and Kondo, Yoji and Latham, David W. and Marcy, Geoffrey W. and Meibom, S{\o}ren and Kjeldsen, Hans and Lissauer, Jack J. and Monet, David G. and Morrison, David and Sasselov, Dimitar and Tarter, Jill and Boss, Alan and Brownlee, Don and Owen, Toby and Buzasi, Derek and Charbonneau, David and Doyle, Laurance and Fortney, Jonathan and Ford, Eric B. and Holman, Matthew J. and Seager, Sara and Steffen, Jason H. and Welsh, William F. and Rowe, Jason and Anderson, Howard and Buchhave, Lars and Ciardi, David and Walkowicz, Lucianne and Sherry, William and Horch, Elliott and Isaacson, Howard and Everett, Mark E. and Fischer, Debra and Torres, Guillermo and Johnson, John Asher and Endl, Michael and MacQueen, Phillip and Bryson, Stephen T. and Dotson, Jessie and Haas, Michael and Kolodziejczak, Jeffrey and Van Cleve, Jeffrey and Chandrasekaran, Hema and Twicken, Joseph D. and Quintana, Elisa V. and Clarke, Bruce D. and Allen, Christopher and Li, Jie and Wu, Haley and Tenenbaum, Peter and Verner, Ekaterina and Bruhweiler, Frederick and Barnes, Jason and Prsa, Andrej},
  year = {2010},
  month = feb,
  journal = {Science},
  volume = {327},
  pages = {977},
  issn = {0036-8075},
  doi = {10.1126/science.1185402},
  urldate = {2025-09-01}
}

@article{caceres2019arps,
  title = {Autoregressive {{Planet Search}}: {{Methodology}}},
  shorttitle = {Autoregressive {{Planet Search}}},
  author = {Caceres, Gabriel A. and Feigelson, Eric D. and Babu, G. Jogesh and Bahamonde, Natalia and Christen, Alejandra and Bertin, Karine and Meza, Cristian and Cur{\'e}, Michel},
  year = {2019},
  month = jul,
  journal = {AJ},
  volume = {158},
  number = {2},
  pages = {57},
  issn = {1538-3881},
  doi = {10.3847/1538-3881/ab26b8},
  urldate = {2025-08-15},
  langid = {english}
}

@article{caceres2019kepler,
  title = {Autoregressive {{Planet Search}}: {{Application}} to the {{Kepler Mission}}},
  shorttitle = {Autoregressive {{Planet Search}}},
  author = {Caceres, Gabriel A. and Feigelson, Eric D. and Babu, G. Jogesh and Bahamonde, Natalia and Christen, Alejandra and Bertin, Karine and Meza, Cristian and Cur{\'e}, Michel},
  year = {2019},
  month = jul,
  journal = {AJ},
  volume = {158},
  number = {2},
  pages = {58},
  issn = {1538-3881},
  doi = {10.3847/1538-3881/ab26ba},
  urldate = {2025-08-15},
  langid = {english}
}

@article{caldwell2020spocffi,
  title = {{{TESS Science Processing Operations Center FFI Target List Products}}},
  author = {Caldwell, Douglas A. and Tenenbaum, Peter and Twicken, Joseph D. and Jenkins, Jon M. and Ting, Eric and Smith, Jeffrey C. and Hedges, Christina and Fausnaugh, Michael M. and Rose, Mark and Burke, Christopher},
  year = {2020},
  month = nov,
  journal = {Res. Notes AAS},
  volume = {4},
  number = {11},
  pages = {201},
  issn = {2515-5172},
  doi = {10.3847/2515-5172/abc9b3},
  urldate = {2024-12-06},
  langid = {english}
}

@article{castrogonzalez2024neptune,
  title = {Mapping the Exo-{{Neptunian}} Landscape - {{A}} Ridge between the Desert and Savanna},
  author = {{Castro-Gonz{\'a}lez}, A. and Bourrier, V. and {Lillo-Box}, J. and Delisle, J.-B. and Armstrong, D. J. and Barrado, D. and Correia, A. C. M.},
  year = {2024},
  month = sep,
  journal = {A\&A},
  volume = {689},
  pages = {A250},
  issn = {0004-6361, 1432-0746},
  doi = {10.1051/0004-6361/202450957},
  urldate = {2025-03-13},
  copyright = {{\copyright} The Authors 2024},
  langid = {english}
}

@inproceedings{chen2016xgboost,
  title = {{{XGBoost}}: {{A Scalable Tree Boosting System}}},
  shorttitle = {{{XGBoost}}},
  booktitle = {Proceedings of the 22nd {{ACM SIGKDD International Conference}} on {{Knowledge Discovery}} and {{Data Mining}}},
  author = {Chen, Tianqi and Guestrin, Carlos},
  year = {2016},
  month = aug,
  series = {{{KDD}} '16},
  pages = {785--794},
  publisher = {Association for Computing Machinery},
  address = {New York, NY, USA},
  doi = {10.1145/2939672.2939785},
  urldate = {2025-08-19},
  isbn = {978-1-4503-4232-2}
}

@article{christiansen2025nea,
  title = {The {{NASA Exoplanet Archive}} and {{Exoplanet Follow-up Observing Program}}: {{Data}}, {{Tools}}, and {{Usage}}},
  shorttitle = {The {{NASA Exoplanet Archive}} and {{Exoplanet Follow-up Observing Program}}},
  author = {Christiansen, Jessie L. and McElroy, Douglas L. and Harbut, Marcy and Ciardi, David R. and Crane, Megan and Good, John and {Hardegree-Ullman}, Kevin K. and Kesseli, Aurora Y. and Lund, Michael B. and Lynn, Meca and Muthiar, Ananda and Nilsson, Ricky and Oluyide, Toba and Papin, Michael and Rivera, Amalia and Swain, Melanie and Susemiehl, Nicholas D. and Tam, Raymond and van Eyken, Julian and Beichman, Charles},
  year = {2025},
  month = aug,
  journal = {Planet. Sci. J.},
  volume = {6},
  number = {8},
  pages = {186},
  issn = {2632-3338},
  doi = {10.3847/PSJ/ade3c2},
  urldate = {2025-08-20},
  langid = {english}
}

@article{cloutier2019orion,
  title = {The {{Independent Discovery}} of {{Planet Candidates}} around {{Low-mass Stars}} and {{Astrophysical False Positives}} from the {{First Two TESS Sectors}}},
  author = {Cloutier, Ryan},
  year = {2019},
  month = jul,
  journal = {AJ},
  volume = {158},
  number = {2},
  pages = {81},
  issn = {1538-3881},
  doi = {10.3847/1538-3881/ab27be},
  urldate = {2025-09-02},
  langid = {english}
}

@inproceedings{collins2018tfopwg,
  title = {{{TESS Follow-up Observing Program}} ({{TFOP}}) {{Working Group}}:{{A Mission-led Effort}} to {{Coordinate Community Resources}} to {{Confirm TESS Planets}}},
  shorttitle = {{{TESS Follow-up Observing Program}} ({{TFOP}}) {{Working Group}}},
  booktitle = {American {{Astronomical Society Meeting Abstracts}} \#231},
  author = {Collins, Karen and Quinn, Samuel N. and Latham, David W. and Christiansen, Jessie and Ciardi, David and Dragomir, Diana and Crossfield, Ian and Seager, Sara},
  year = {2018},
  month = jan,
  volume = {231},
  pages = {439.08},
  urldate = {2025-08-20}
}

@article{coughlin2016robovetter,
  title = {{{PLANETARY CANDIDATES OBSERVED BY KEPLER}}. {{VII}}. {{THE FIRST FULLY UNIFORM CATALOG BASED ON THE ENTIRE}} 48-{{MONTH DATA SET}} ({{Q1}}--{{Q17 DR24}})},
  author = {Coughlin, Jeffrey L. and Mullally, F. and Thompson, Susan E. and Rowe, Jason F. and Burke, Christopher J. and Latham, David W. and Batalha, Natalie M. and Ofir, Aviv and Quarles, Billy L. and Henze, Christopher E. and Wolfgang, Angie and Caldwell, Douglas A. and Bryson, Stephen T. and Shporer, Avi and Catanzarite, Joseph and Akeson, Rachel and Barclay, Thomas and Borucki, William J. and Boyajian, Tabetha S. and Campbell, Jennifer R. and Christiansen, Jessie L. and Girouard, Forrest R. and Haas, Michael R. and Howell, Steve B. and Huber, Daniel and Jenkins, Jon M. and Li, Jie and {Patil-Sabale}, Anima and Quintana, Elisa V. and Ramirez, Solange and Seader, Shawn and Smith, Jeffrey C. and Tenenbaum, Peter and Twicken, Joseph D. and Zamudio, Khadeejah A.},
  year = {2016},
  month = may,
  journal = {ApJS},
  volume = {224},
  number = {1},
  pages = {12},
  issn = {0067-0049},
  doi = {10.3847/0067-0049/224/1/12},
  urldate = {2025-09-03},
  langid = {english}
}

@article{cumming2008distribution,
  title = {The {{Keck Planet Search}}: {{Detectability}} and the {{Minimum Mass}} and {{Orbital Period Distribution}} of {{Extrasolar Planets}}},
  shorttitle = {The {{Keck Planet Search}}},
  author = {Cumming, Andrew and Butler, R. Paul and Marcy, Geoffrey W. and Vogt, Steven S. and Wright, Jason T. and Fischer, Debra A.},
  year = {2008},
  month = may,
  journal = {PASP},
  volume = {120},
  number = {867},
  pages = {531},
  issn = {1538-3873},
  doi = {10.1086/588487},
  urldate = {2025-09-09},
  langid = {english}
}

@article{dattilo2019astronetK2,
  title = {Identifying {{Exoplanets}} with {{Deep Learning}}. {{II}}. {{Two New Super-Earths Uncovered}} by a {{Neural Network}} in {{K2 Data}}},
  author = {Dattilo, Anne and Vanderburg, Andrew and Shallue, Christopher J. and Mayo, Andrew W. and Berlind, Perry and Bieryla, Allyson and Calkins, Michael L. and Esquerdo, Gilbert A. and Everett, Mark E. and Howell, Steve B. and Latham, David W. and Scott, Nicholas J. and Yu, Liang},
  year = {2019},
  month = apr,
  journal = {AJ},
  volume = {157},
  number = {5},
  pages = {169},
  issn = {1538-3881},
  doi = {10.3847/1538-3881/ab0e12},
  urldate = {2025-09-02},
  langid = {english}
}

@article{devorapajares2024sherlock,
  title = {The Sherlock Pipeline: New Exoplanet Candidates in the {{WASP-16}}, {{HAT-P-27}}, {{HAT-P-26}}, and {{TOI-2411}} Systems},
  shorttitle = {The Sherlock Pipeline},
  author = {{D{\'e}vora-Pajares}, Mart{\'i}n and Pozuelos, Francisco J and Thuillier, Antoine and Timmermans, Mathilde and Van~Grootel, Val{\'e}rie and Bonidie, Victoria and Mota, Luis Cerde{\~n}o and Su{\'a}rez, Juan C},
  year = {2024},
  month = aug,
  journal = {\mnras},
  volume = {532},
  number = {4},
  pages = {4752--4773},
  issn = {0035-8711},
  doi = {10.1093/mnras/stae1740},
  urldate = {2025-08-21}
}

@article{diaz2014pastis,
  title = {Pastis: {{Bayesian}} Extrasolar Planet Validation -- {{I}}. {{General}} Framework, Models, and Performance},
  shorttitle = {Pastis},
  author = {D{\'i}az, R. F. and Almenara, J. M. and Santerne, A. and Moutou, C. and Lethuillier, A. and Deleuil, M.},
  year = {2014},
  month = jun,
  journal = {\mnras},
  volume = {441},
  number = {2},
  pages = {983--1004},
  issn = {0035-8711},
  doi = {10.1093/mnras/stu601},
  urldate = {2025-06-13}
}

@article{doyle2024tessgaia,
  title = {The {{TESS-SPOC FFI}} Target Sample Explored with {{Gaia}}},
  author = {Doyle, Lauren and Armstrong, David J and Bayliss, Daniel and Rodel, Toby and Kunovac, Vedad},
  year = {2024},
  month = apr,
  journal = {\mnras},
  volume = {529},
  number = {2},
  pages = {1802--1813},
  issn = {0035-8711},
  doi = {10.1093/mnras/stae616},
  urldate = {2024-08-21}
}

@article{eisner2021pht,
  title = {Planet {{Hunters TESS II}}: Findings from the First Two Years of {{TESS}}},
  shorttitle = {Planet {{Hunters TESS II}}},
  author = {Eisner, N L and Barrag{\'a}n, O and Lintott, C and Aigrain, S and Nicholson, B and Boyajian, T S and Howell, S and Johnston, C and Lakeland, B and Miller, G and McMaster, A and Parviainen, H and Safron, E J and Schwamb, M E and Trouille, L and Vaughan, S and Zicher, N and Allen, C and Allen, S and Bouslog, M and Johnson, C and Simon, M N and Wolfenbarger, Z and Baeten, E M L and Bundy, D M and Hoffman, T},
  year = {2021},
  month = mar,
  journal = {\mnras},
  volume = {501},
  number = {4},
  pages = {4669--4690},
  issn = {0035-8711},
  doi = {10.1093/mnras/staa3739},
  urldate = {2025-09-03}
}

@article{eschen2024mdwarf,
  title = {Nine New {{M}} Dwarf Planet Candidates from {{TESS}} Including Five Gas Giants},
  author = {Eschen, Yoshi Nike Emilia and Kunimoto, Michelle},
  year = {2024},
  month = jul,
  journal = {\mnras},
  volume = {531},
  number = {4},
  pages = {5053--5060},
  issn = {0035-8711},
  doi = {10.1093/mnras/stae1496},
  urldate = {2025-08-28}
}

@ARTICLE{espinoza2019juliet,
       author = {{Espinoza}, N{\'e}stor and {Kossakowski}, Diana and {Brahm}, Rafael},
        title = "{juliet: a versatile modelling tool for transiting and non-transiting exoplanetary systems}",
      journal = {\mnras},
         year = 2019,
        month = dec,
       volume = {490},
       number = {2},
        pages = {2262-2283},
          doi = {10.1093/mnras/stz2688},
archivePrefix = {arXiv},
       eprint = {1812.08549},
 primaryClass = {astro-ph.EP},
       adsurl = {https://ui.adsabs.harvard.edu/abs/2019MNRAS.490.2262E}
}

@article{feinstein2019eleanor,
  title = {Eleanor: {{An Open-source Tool}} for {{Extracting Light Curves}} from the {{TESS Full-frame Images}}},
  shorttitle = {Eleanor},
  author = {Feinstein, Adina D. and Montet, Benjamin T. and {Foreman-Mackey}, Daniel and Bedell, Megan E. and Saunders, Nicholas and Bean, Jacob L. and Christiansen, Jessie L. and Hedges, Christina and Luger, Rodrigo and Scolnic, Daniel and Cardoso, Jos{\'e} Vin{\'i}cius de Miranda},
  year = {2019},
  month = jul,
  journal = {PASP},
  volume = {131},
  number = {1003},
  pages = {094502},
  issn = {1538-3873},
  doi = {10.1088/1538-3873/ab291c},
  urldate = {2025-09-09},
  langid = {english}
}

@article{feliz2021NEMESIS,
  title = {{{NEMESIS}}: {{Exoplanet Transit Survey}} of {{Nearby M-dwarfs}} in {{TESS FFIs}}. {{I}}.},
  shorttitle = {{{NEMESIS}}},
  author = {Feliz, Dax L. and Plavchan, Peter and Bianco, Samantha N. and Jimenez, Mary and Collins, Kevin I. and Villarreal Alvarado, Bryan and Stassun, Keivan G.},
  year = {2021},
  month = apr,
  journal = {AJ},
  volume = {161},
  number = {5},
  pages = {247},
  publisher = {The American Astronomical Society},
  issn = {1538-3881},
  doi = {10.3847/1538-3881/abedb3},
  urldate = {2025-08-29},
  langid = {english}
}

@article{fernandes2022pterodactyls,
  title = {Pterodactyls: {{A Tool}} to {{Uniformly Search}} and {{Vet}} for {{Young Transiting Planets}} in {{TESS Primary Mission Photometry}}},
  shorttitle = {Pterodactyls},
  author = {Fernandes, Rachel B. and Mulders, Gijs D. and Pascucci, Ilaria and Bergsten, Galen J. and Koskinen, Tommi T. and {Hardegree-Ullman}, Kevin K. and Pearson, Kyle A. and Giacalone, Steven and Zink, Jon and Ciardi, David R. and O'Brien, Patrick},
  year = {2022},
  month = aug,
  journal = {AJ},
  volume = {164},
  number = {3},
  pages = {78},
  issn = {1538-3881},
  doi = {10.3847/1538-3881/ac7b29},
  urldate = {2025-09-02},
  langid = {english}
}

@article{fiscale2025dartvetter,
  title = {{{DART-Vetter}}: {{A Deep Learning Tool}} for {{Automatic Triage}} of {{Exoplanet Candidates}}},
  shorttitle = {{{DART-Vetter}}},
  author = {Fiscale, Stefano and Inno, Laura and Rotundi, Alessandra and Ciaramella, Angelo and Ferone, Alessio and Magliano, Christian and Cacciapuoti, Luca and Kostov, Veselin and Quintana, Elisa V. and Covone, Giovanni and Muscari Tomajoli, Maria Teresa and Saggese, Vito and Tonietti, Luca and Vanzanella, Antonio and Della Corte, Vincenzo},
  year = {2025},
  month = jul,
  journal = {AJ},
  volume = {170},
  number = {2},
  pages = {73},
  issn = {1538-3881},
  doi = {10.3847/1538-3881/addf4d},
  urldate = {2025-09-02},
  langid = {english}
}

@article{foremanmackey2017celerite,
  title = {Fast and {{Scalable Gaussian Process Modeling}} with {{Applications}} to {{Astronomical Time Series}}},
  author = {{Foreman-Mackey}, Daniel and Agol, Eric and Ambikasaran, Sivaram and Angus, Ruth},
  year = {2017},
  month = nov,
  journal = {AJ},
  volume = {154},
  number = {6},
  pages = {220},
  issn = {1538-3881},
  doi = {10.3847/1538-3881/aa9332},
  urldate = {2025-08-20},
  langid = {english}
}

@article{fulton2017radiusvalley,
  title = {The {{California-Kepler Survey}}. {{III}}. {{A Gap}} in the {{Radius Distribution}} of {{Small Planets}}},
  author = {Fulton, Benjamin J. and Petigura, Erik A. and Howard, Andrew W. and Isaacson, Howard and Marcy, Geoffrey W. and Cargile, Phillip A. and Hebb, Leslie and Weiss, Lauren M. and Johnson, John Asher and Morton, Timothy D. and Sinukoff, Evan and Crossfield, Ian J. M. and Hirsch, Lea A.},
  year = {2017},
  month = sep,
  journal = {AJ},
  volume = {154},
  pages = {109},
  issn = {0004-6256},
  doi = {10.3847/1538-3881/aa80eb},
  urldate = {2021-10-19}
}

@article{gaia2016mission,
  title = {The {{Gaia}} Mission},
  author = {Gaia Collaboration and Prusti, T. and de Bruijne, J. H. J. and Brown, A. G. A. and Vallenari, A. and Babusiaux, C. and {Bailer-Jones}, C. a. L. and Bastian, U. and Biermann, M. and Evans, D. W. and Eyer, L. and Jansen, F. and Jordi, C. and Klioner, S. A. and Lammers, U. and Lindegren, L. and Luri, X. and Mignard, F. and Milligan, D. J. and Panem, C. and Poinsignon, V. and Pourbaix, D. and Randich, S. and Sarri, G. and Sartoretti, P. and Siddiqui, H. I. and Soubiran, C. and Valette, V. and van Leeuwen, F. and Walton, N. A. and Aerts, C. and Arenou, F. and Cropper, M. and Drimmel, R. and H{\o}g, E. and Katz, D. and Lattanzi, M. G. and O'Mullane, W. and Grebel, E. K. and Holland, A. D. and Huc, C. and Passot, X. and Bramante, L. and Cacciari, C. and Casta{\~n}eda, J. and Chaoul, L. and Cheek, N. and Angeli, F. De and Fabricius, C. and Guerra, R. and Hern{\'a}ndez, J. and {Jean-Antoine-Piccolo}, A. and Masana, E. and Messineo, R. and Mowlavi, N. and Nienartowicz, K. and {Ord{\'o}{\~n}ez-Blanco}, D. and Panuzzo, P. and Portell, J. and Richards, P. J. and Riello, M. and Seabroke, G. M. and Tanga, P. and Th{\'e}venin, F. and Torra, J. and Els, S. G. and {Gracia-Abril}, G. and Comoretto, G. and {Garcia-Reinaldos}, M. and Lock, T. and Mercier, E. and Altmann, M. and Andrae, R. and Astraatmadja, T. L. and {Bellas-Velidis}, I. and Benson, K. and Berthier, J. and Blomme, R. and Busso, G. and Carry, B. and Cellino, A. and Clementini, G. and Cowell, S. and Creevey, O. and Cuypers, J. and Davidson, M. and Ridder, J. De and de Torres, A. and Delchambre, L. and Dell'Oro, A. and Ducourant, C. and Fr{\'e}mat, Y. and {Garc{\'i}a-Torres}, M. and Gosset, E. and Halbwachs, J.-L. and Hambly, N. C. and Harrison, D. L. and Hauser, M. and Hestroffer, D. and Hodgkin, S. T. and Huckle, H. E. and Hutton, A. and Jasniewicz, G. and Jordan, S. and Kontizas, M. and Korn, A. J. and Lanzafame, A. C. and Manteiga, M. and Moitinho, A. and Muinonen, K. and Osinde, J. and Pancino, E. and Pauwels, T. and Petit, J.-M. and {Recio-Blanco}, A. and Robin, A. C. and Sarro, L. M. and Siopis, C. and Smith, M. and Smith, K. W. and Sozzetti, A. and Thuillot, W. and van Reeven, W. and Viala, Y. and Abbas, U. and Aramburu, A. Abreu and Accart, S. and Aguado, J. J. and Allan, P. M. and Allasia, W. and Altavilla, G. and {\'A}lvarez, M. A. and Alves, J. and Anderson, R. I. and Andrei, A. H. and Varela, E. Anglada and Antiche, E. and Antoja, T. and Ant{\'o}n, S. and Arcay, B. and Atzei, A. and Ayache, L. and Bach, N. and Baker, S. G. and {Balaguer-N{\'u}{\~n}ez}, L. and Barache, C. and Barata, C. and Barbier, A. and Barblan, F. and Baroni, M. and y Navascu{\'e}s, D. Barrado and Barros, M. and Barstow, M. A. and Becciani, U. and Bellazzini, M. and Bellei, G. and Garc{\'i}a, A. Bello and Belokurov, V. and Bendjoya, P. and Berihuete, A. and Bianchi, L. and Bienaym{\'e}, O. and Billebaud, F. and Blagorodnova, N. and {Blanco-Cuaresma}, S. and Boch, T. and Bombrun, A. and Borrachero, R. and Bouquillon, S. and Bourda, G. and Bouy, H. and Bragaglia, A. and Breddels, M. A. and Brouillet, N. and Br{\"u}semeister, T. and Bucciarelli, B. and Budnik, F. and Burgess, P. and Burgon, R. and Burlacu, A. and Busonero, D. and Buzzi, R. and Caffau, E. and Cambras, J. and Campbell, H. and Cancelliere, R. and {Cantat-Gaudin}, T. and Carlucci, T. and Carrasco, J. M. and Castellani, M. and Charlot, P. and Charnas, J. and Charvet, P. and Chassat, F. and Chiavassa, A. and Clotet, M. and Cocozza, G. and Collins, R. S. and Collins, P. and Costigan, G. and Crifo, F. and Cross, N. J. G. and Crosta, M. and Crowley, C. and Dafonte, C. and Damerdji, Y. and Dapergolas, A. and David, P. and David, M. and Cat, P. De and de Felice, F. and de Laverny, P. and Luise, F. De and March, R. De and de Martino, D. and de Souza, R. and Debosscher, J. and del Pozo, E. and Delbo, M. and Delgado, A. and Delgado, H. E. and di Marco, F. and Matteo, P. Di and Diakite, S. and Distefano, E. and Dolding, C. and Anjos, S. Dos and Drazinos, P. and Dur{\'a}n, J. and Dzigan, Y. and Ecale, E. and Edvardsson, B. and Enke, H. and Erdmann, M. and Escolar, D. and Espina, M. and Evans, N. W. and Bontemps, G. Eynard and Fabre, C. and Fabrizio, M. and Faigler, S. and Falc{\~a}o, A. J. and Casas, M. Farr{\`a}s and Faye, F. and Federici, L. and Fedorets, G. and {Fern{\'a}ndez-Hern{\'a}ndez}, J. and Fernique, P. and Fienga, A. and Figueras, F. and Filippi, F. and Findeisen, K. and Fonti, A. and Fouesneau, M. and Fraile, E. and Fraser, M. and Fuchs, J. and Furnell, R. and Gai, M. and Galleti, S. and Galluccio, L. and Garabato, D. and {Garc{\'i}a-Sedano}, F. and Gar{\'e}, P. and Garofalo, A. and Garralda, N. and Gavras, P. and Gerssen, J. and Geyer, R. and Gilmore, G. and Girona, S. and Giuffrida, G. and Gomes, M. and {Gonz{\'a}lez-Marcos}, A. and {Gonz{\'a}lez-N{\'u}{\~n}ez}, J. and {Gonz{\'a}lez-Vidal}, J. J. and Granvik, M. and Guerrier, A. and Guillout, P. and Guiraud, J. and G{\'u}rpide, A. and {Guti{\'e}rrez-S{\'a}nchez}, R. and Guy, L. P. and Haigron, R. and Hatzidimitriou, D. and Haywood, M. and Heiter, U. and Helmi, A. and Hobbs, D. and Hofmann, W. and Holl, B. and Holland, G. and Hunt, J. a. S. and Hypki, A. and Icardi, V. and Irwin, M. and de Fombelle, G. Jevardat and Jofr{\'e}, P. and Jonker, P. G. and Jorissen, A. and Julbe, F. and Karampelas, A. and Kochoska, A. and Kohley, R. and Kolenberg, K. and Kontizas, E. and Koposov, S. E. and Kordopatis, G. and Koubsky, P. and Kowalczyk, A. and {Krone-Martins}, A. and Kudryashova, M. and Kull, I. and Bachchan, R. K. and {Lacoste-Seris}, F. and Lanza, A. F. and Lavigne, J.-B. and {Poncin-Lafitte}, C. Le and Lebreton, Y. and Lebzelter, T. and Leccia, S. and Leclerc, N. and {Lecoeur-Taibi}, I. and Lemaitre, V. and Lenhardt, H. and Leroux, F. and Liao, S. and Licata, E. and Lindstr{\o}m, H. E. P. and Lister, T. A. and Livanou, E. and Lobel, A. and L{\"o}ffler, W. and L{\'o}pez, M. and {Lopez-Lozano}, A. and Lorenz, D. and Loureiro, T. and MacDonald, I. and Fernandes, T. Magalh{\~a}es and Managau, S. and Mann, R. G. and Mantelet, G. and Marchal, O. and Marchant, J. M. and Marconi, M. and Marie, J. and Marinoni, S. and Marrese, P. M. and Marschalk{\'o}, G. and Marshall, D. J. and {Mart{\'i}n-Fleitas}, J. M. and Martino, M. and Mary, N. and Matijevi{\v c}, G. and Mazeh, T. and McMillan, P. J. and Messina, S. and Mestre, A. and Michalik, D. and Millar, N. R. and Miranda, B. M. H. and Molina, D. and Molinaro, R. and Molinaro, M. and Moln{\'a}r, L. and Moniez, M. and Montegriffo, P. and Monteiro, D. and Mor, R. and Mora, A. and Morbidelli, R. and Morel, T. and Morgenthaler, S. and Morley, T. and Morris, D. and Mulone, A. F. and Muraveva, T. and Musella, I. and Narbonne, J. and Nelemans, G. and Nicastro, L. and Noval, L. and Ord{\'e}novic, C. and {Ordieres-Mer{\'e}}, J. and Osborne, P. and Pagani, C. and Pagano, I. and Pailler, F. and Palacin, H. and Palaversa, L. and Parsons, P. and Paulsen, T. and Pecoraro, M. and Pedrosa, R. and Pentik{\"a}inen, H. and Pereira, J. and Pichon, B. and Piersimoni, A. M. and Pineau, F.-X. and Plachy, E. and Plum, G. and Poujoulet, E. and Pr{\v s}a, A. and Pulone, L. and Ragaini, S. and Rago, S. and Rambaux, N. and {Ramos-Lerate}, M. and Ranalli, P. and Rauw, G. and Read, A. and Regibo, S. and Renk, F. and Reyl{\'e}, C. and Ribeiro, R. A. and Rimoldini, L. and Ripepi, V. and Riva, A. and Rixon, G. and Roelens, M. and {Romero-G{\'o}mez}, M. and Rowell, N. and Royer, F. and Rudolph, A. and {Ruiz-Dern}, L. and Sadowski, G. and Sell{\'e}s, T. Sagrist{\`a} and Sahlmann, J. and Salgado, J. and Salguero, E. and Sarasso, M. and Savietto, H. and Schnorhk, A. and Schultheis, M. and Sciacca, E. and Segol, M. and Segovia, J. C. and Segransan, D. and Serpell, E. and Shih, I.-C. and Smareglia, R. and Smart, R. L. and Smith, C. and Solano, E. and Solitro, F. and Sordo, R. and Nieto, S. Soria and Souchay, J. and Spagna, A. and Spoto, F. and Stampa, U. and Steele, I. A. and Steidelm{\"u}ller, H. and Stephenson, C. A. and Stoev, H. and Suess, F. F. and S{\"u}veges, M. and Surdej, J. and Szabados, L. and {Szegedi-Elek}, E. and Tapiador, D. and Taris, F. and Tauran, G. and Taylor, M. B. and Teixeira, R. and Terrett, D. and Tingley, B. and Trager, S. C. and Turon, C. and Ulla, A. and Utrilla, E. and Valentini, G. and van Elteren, A. and Hemelryck, E. Van and van Leeuwen, M. and Varadi, M. and Vecchiato, A. and Veljanoski, J. and Via, T. and Vicente, D. and Vogt, S. and Voss, H. and Votruba, V. and Voutsinas, S. and Walmsley, G. and Weiler, M. and Weingrill, K. and Werner, D. and Wevers, T. and Whitehead, G. and Wyrzykowski, {\L} and Yoldas, A. and {\v Z}erjal, M. and Zucker, S. and Zurbach, C. and Zwitter, T. and Alecu, A. and Allen, M. and Prieto, C. Allende and Amorim, A. and {Anglada-Escud{\'e}}, G. and Arsenijevic, V. and Azaz, S. and Balm, P. and Beck, M. and Bernstein, H.-H. and Bigot, L. and Bijaoui, A. and Blasco, C. and Bonfigli, M. and Bono, G. and Boudreault, S. and Bressan, A. and Brown, S. and Brunet, P.-M. and Bunclark, P. and Buonanno, R. and Butkevich, A. G. and Carret, C. and Carrion, C. and Chemin, L. and Ch{\'e}reau, F. and Corcione, L. and Darmigny, E. and de Boer, K. S. and de Teodoro, P. and de Zeeuw, P. T. and Luche, C. Delle and Domingues, C. D. and Dubath, P. and Fodor, F. and Fr{\'e}zouls, B. and Fries, A. and Fustes, D. and Fyfe, D. and Gallardo, E. and Gallegos, J. and Gardiol, D. and Gebran, M. and Gomboc, A. and G{\'o}mez, A. and Grux, E. and Gueguen, A. and Heyrovsky, A. and Hoar, J. and Iannicola, G. and Parache, Y. Isasi and Janotto, A.-M. and Joliet, E. and Jonckheere, A. and Keil, R. and Kim, D.-W. and Klagyivik, P. and Klar, J. and Knude, J. and Kochukhov, O. and Kolka, I. and Kos, J. and Kutka, A. and Lainey, V. and LeBouquin, D. and Liu, C. and Loreggia, D. and Makarov, V. V. and Marseille, M. G. and Martayan, C. and {Martinez-Rubi}, O. and Massart, B. and Meynadier, F. and Mignot, S. and Munari, U. and Nguyen, A.-T. and Nordlander, T. and Ocvirk, P. and O'Flaherty, K. S. and Sanz, A. Olias and Ortiz, P. and Osorio, J. and Oszkiewicz, D. and Ouzounis, A. and Palmer, M. and Park, P. and Pasquato, E. and Peltzer, C. and Peralta, J. and P{\'e}turaud, F. and Pieniluoma, T. and Pigozzi, E. and Poels, J. and Prat, G. and Prod'homme, T. and Raison, F. and Rebordao, J. M. and Risquez, D. and {Rocca-Volmerange}, B. and Rosen, S. and {Ruiz-Fuertes}, M. I. and Russo, F. and Sembay, S. and Vizcaino, I. Serraller and Short, A. and Siebert, A. and Silva, H. and Sinachopoulos, D. and Slezak, E. and Soffel, M. and Sosnowska, D. and Strai{\v z}ys, V. and ter Linden, M. and Terrell, D. and Theil, S. and Tiede, C. and Troisi, L. and Tsalmantza, P. and Tur, D. and Vaccari, M. and Vachier, F. and Valles, P. and Hamme, W. Van and Veltz, L. and Virtanen, J. and Wallut, J.-M. and Wichmann, R. and Wilkinson, M. I. and Ziaeepour, H. and Zschocke, S.},
  year = {2016},
  month = nov,
  journal = {A\&A},
  volume = {595},
  pages = {A1},
  issn = {0004-6361, 1432-0746},
  doi = {10.1051/0004-6361/201629272},
  urldate = {2025-10-06},
  copyright = {{\copyright} ESO, 2016},
  langid = {english}
}

@article{gaia2023dr3,
  title = {Gaia {{Data Release}} 3. {{Summary}} of the Content and Survey Properties},
  author = {{Gaia Collaboration} and Vallenari, A. and Brown, A. G. A. and Prusti, T. and {de Bruijne}, J. H. J. and Arenou, F. and Babusiaux, C. and Biermann, M. and Creevey, O. L. and Ducourant, C. and Evans, D. W. and Eyer, L. and Guerra, R. and Hutton, A. and Jordi, C. and Klioner, S. A. and Lammers, U. L. and Lindegren, L. and Luri, X. and Mignard, F. and Panem, C. and Pourbaix, D. and Randich, S. and Sartoretti, P. and Soubiran, C. and Tanga, P. and Walton, N. A. and {Bailer-Jones}, C. A. L. and Bastian, U. and Drimmel, R. and Jansen, F. and Katz, D. and Lattanzi, M. G. and {van Leeuwen}, F. and Bakker, J. and Cacciari, C. and Casta{\~n}eda, J. and De Angeli, F. and Fabricius, C. and Fouesneau, M. and Fr{\'e}mat, Y. and Galluccio, L. and Guerrier, A. and Heiter, U. and Masana, E. and Messineo, R. and Mowlavi, N. and Nicolas, C. and Nienartowicz, K. and Pailler, F. and Panuzzo, P. and Riclet, F. and Roux, W. and Seabroke, G. M. and Sordo, R. and Th{\'e}venin, F. and {Gracia-Abril}, G. and Portell, J. and Teyssier, D. and Altmann, M. and Andrae, R. and Audard, M. and {Bellas-Velidis}, I. and Benson, K. and Berthier, J. and Blomme, R. and Burgess, P. W. and Busonero, D. and Busso, G. and C{\'a}novas, H. and Carry, B. and Cellino, A. and Cheek, N. and Clementini, G. and Damerdji, Y. and Davidson, M. and {de Teodoro}, P. and Nu{\~n}ez Campos, M. and Delchambre, L. and Dell'Oro, A. and Esquej, P. and {Fern{\'a}ndez-Hern{\'a}ndez}, J. and Fraile, E. and Garabato, D. and {Garc{\'i}a-Lario}, P. and Gosset, E. and Haigron, R. and Halbwachs, J. -L. and Hambly, N. C. and Harrison, D. L. and Hern{\'a}ndez, J. and Hestroffer, D. and Hodgkin, S. T. and Holl, B. and Jan{\ss}en, K. and {Jevardat de Fombelle}, G. and Jordan, S. and {Krone-Martins}, A. and Lanzafame, A. C. and L{\"o}ffler, W. and Marchal, O. and Marrese, P. M. and Moitinho, A. and Muinonen, K. and Osborne, P. and Pancino, E. and Pauwels, T. and {Recio-Blanco}, A. and Reyl{\'e}, C. and Riello, M. and Rimoldini, L. and Roegiers, T. and Rybizki, J. and Sarro, L. M. and Siopis, C. and Smith, M. and Sozzetti, A. and Utrilla, E. and {van Leeuwen}, M. and Abbas, U. and {\'A}brah{\'a}m, P. and Abreu Aramburu, A. and Aerts, C. and Aguado, J. J. and Ajaj, M. and {Aldea-Montero}, F. and Altavilla, G. and {\'A}lvarez, M. A. and Alves, J. and Anders, F. and Anderson, R. I. and Anglada Varela, E. and Antoja, T. and Baines, D. and Baker, S. G. and {Balaguer-N{\'u}{\~n}ez}, L. and Balbinot, E. and Balog, Z. and Barache, C. and Barbato, D. and Barros, M. and Barstow, M. A. and Bartolom{\'e}, S. and Bassilana, J. -L. and Bauchet, N. and Becciani, U. and Bellazzini, M. and Berihuete, A. and Bernet, M. and Bertone, S. and Bianchi, L. and Binnenfeld, A. and {Blanco-Cuaresma}, S. and Blazere, A. and Boch, T. and Bombrun, A. and Bossini, D. and Bouquillon, S. and Bragaglia, A. and Bramante, L. and Breedt, E. and Bressan, A. and Brouillet, N. and Brugaletta, E. and Bucciarelli, B. and Burlacu, A. and Butkevich, A. G. and Buzzi, R. and Caffau, E. and Cancelliere, R. and {Cantat-Gaudin}, T. and Carballo, R. and Carlucci, T. and Carnerero, M. I. and Carrasco, J. M. and Casamiquela, L. and Castellani, M. and {Castro-Ginard}, A. and Chaoul, L. and Charlot, P. and Chemin, L. and Chiaramida, V. and Chiavassa, A. and Chornay, N. and Comoretto, G. and Contursi, G. and Cooper, W. J. and Cornez, T. and Cowell, S. and Crifo, F. and Cropper, M. and Crosta, M. and Crowley, C. and Dafonte, C. and Dapergolas, A. and David, M. and David, P. and {de Laverny}, P. and De Luise, F. and De March, R.},
  year = {2023},
  month = jun,
  journal = {A\&A},
  volume = {674},
  pages = {A1},
  issn = {0004-6361},
  doi = {10.1051/0004-6361/202243940},
  urldate = {2025-06-11}
}

@article{giacalone2021triceratops,
  title = {Vetting of 384 {{TESS Objects}} of {{Interest}} with {{TRICERATOPS}} and {{Statistical Validation}} of 12 {{Planet Candidates}}},
  author = {Giacalone, Steven and Dressing, Courtney D. and Jensen, Eric L. N. and Collins, Karen A. and Ricker, George R. and Vanderspek, Roland and Seager, S. and Winn, Joshua N. and Jenkins, Jon M. and Barclay, Thomas and Barkaoui, Khalid and Cadieux, Charles and Charbonneau, David and Collins, Kevin I. and Conti, Dennis M. and Doyon, Ren{\'e} and Evans, Phil and Ghachoui, Mourad and Gillon, Micha{\"e}l and Guerrero, Natalia M. and Hart, Rhodes and Jehin, Emmanu{\"e}l and Kielkopf, John F. and McLean, Brian and Murgas, Felipe and Palle, Enric and Parviainen, Hannu and Pozuelos, Francisco J. and Relles, Howard M. and Shporer, Avi and Socia, Quentin and Stockdale, Chris and Tan, Thiam-Guan and Torres, Guillermo and Twicken, Joseph D. and Waalkes, William C. and Waite, Ian A.},
  year = {2021},
  month = jan,
  journal = {AJ},
  volume = {161},
  pages = {24},
  issn = {0004-6256},
  doi = {10.3847/1538-3881/abc6af},
  urldate = {2024-12-06}
}

@article{giacalone2022terrestrial,
  title = {Validation of 13 {{Hot}} and {{Potentially Terrestrial TESS Planets}}},
  author = {Giacalone, Steven and Dressing, Courtney D. and Hedges, Christina and Kostov, Veselin B. and Collins, Karen A. and Jensen, Eric L. N. and Yahalomi, Daniel A. and Bieryla, Allyson and Ciardi, David R. and Howell, Steve B. and {Lillo-Box}, Jorge and Barkaoui, Khalid and Winters, Jennifer G. and Matthews, Elisabeth and Livingston, John H. and Quinn, Samuel N. and Safonov, Boris S. and Cadieux, Charles and Furlan, E. and Crossfield, Ian J. M. and Mandell, Avi M. and Gilbert, Emily A. and Kruse, Ethan and Quintana, Elisa V. and Ricker, George R. and Seager, S. and Winn, Joshua N. and Jenkins, Jon M. and Duffy Adkins, Britt and Baker, David and Barclay, Thomas and Barrado, David and Batalha, Natalie M. and Belinski, Alexander A. and Benkhaldoun, Zouhair and Buchhave, Lars A. and Cacciapuoti, Luca and Charbonneau, David and Chontos, Ashley and Christiansen, Jessie L. and Cloutier, Ryan and Collins, Kevin I. and Conti, Dennis M. and Cutting, Neil and Dixon, Scott and Doyon, Ren{\'e} and Mufti, Mohammed El and {Esparza-Borges}, Emma and Essack, Zahra and Fukui, Akihiko and Gan, Tianjun and Gary, Kaz and Ghachoui, Mourad and Gillon, Micha{\"e}l and Girardin, Eric and Glidden, Ana and Gonzales, Erica J. and Guerra, Pere and Horch, Elliott P. and He{\l}miniak, Krzysztof G. and Howard, Andrew W. and Huber, Daniel and Irwin, Jonathan M. and Isopi, Giovanni and Jehin, Emmanu{\"e}l and Kagetani, Taiki and Kane, Stephen R. and Kawauchi, Kiyoe and Kielkopf, John F. and Lewin, Pablo and Luker, Lindy and Lund, Michael B. and Mallia, Franco and Mao, Shude and Massey, Bob and Matson, Rachel A. and Mireles, Ismael and Mori, Mayuko and Murgas, Felipe and Narita, Norio and O'Dwyer, Tanner and Petigura, Erik A. and Polanski, Alex S. and Pozuelos, Francisco J. and Palle, Enric and Parviainen, Hannu and Plavchan, Peter P. and Relles, Howard M. and Robertson, Paul and Rose, Mark E. and Rowden, Pamela and Roy, Arpita and Savel, Arjun B. and Schlieder, Joshua E. and Schnaible, Chloe and Schwarz, Richard P. and Sefako, Ramatholo and Selezneva, Aleksandra and Skinner, Brett and Stockdale, Chris and Strakhov, Ivan A. and Tan, Thiam-Guan and Torres, Guillermo and Tronsgaard, Ren{\'e} and Twicken, Joseph D. and Vermilion, David and Waite, Ian A. and Walter, Bradley and Wang, Gavin and Ziegler, Carl and Zou, Yujie},
  year = {2022},
  month = jan,
  journal = {AJ},
  volume = {163},
  number = {2},
  pages = {99},
  issn = {1538-3881},
  doi = {10.3847/1538-3881/ac4334},
  urldate = {2025-10-10},
  langid = {english}
}

@article{ginsburg2019astroquery,
  title = {Astroquery: {{An Astronomical Web-querying Package}} in {{Python}}},
  shorttitle = {Astroquery},
  author = {Ginsburg, Adam and Sip{\H o}cz, Brigitta M. and Brasseur, C. E. and Cowperthwaite, Philip S. and Craig, Matthew W. and Deil, Christoph and Groener, Austen M. and Guillochon, James and Guzman, Giannina and Liedtke, Simon and Lim, Pey Lian and Lockhart, Kelly E. and Mommert, Michael and Morris, Brett M. and Norman, Henrik and Parikh, Madhura and Persson, Magnus V. and Robitaille, Thomas P. and Segovia, Juan-Carlos and Singer, Leo P. and Tollerud, Erik J. and {de Val-Borro}, Miguel and Valtchanov, Ivan and Woillez, Julien and subset of the astropy collaboration {The Astroquery collaboration}, a},
  year = {2019},
  month = feb,
  journal = {AJ},
  volume = {157},
  number = {3},
  pages = {98},
  issn = {1538-3881},
  doi = {10.3847/1538-3881/aafc33},
  urldate = {2025-10-07},
  langid = {english}
}

@article{gomezbarrientos2025triceratops,
  title = {Validation of {{TESS Planet Candidates}} with {{Multicolor Transit Photometry}} and {{TRICERATOPS}}+},
  author = {Gomez Barrientos, Jonathan and {Greklek-McKeon}, Michael and Knutson, Heather A. and Giacalone, Steven and Levine, W. Garrett and Saidel, Morgan and Vissapragada, Shreyas and Ciardi, David R. and Collins, Karen A. and Latham, David W. and Watkins, Cristilyn N. and Budnikova, Polina A. and Cheryasov, Dmitry V. and Fukui, Akihiko and Bieryla, Allyson and Shporer, Avi and Tofflemire, Benjamin M. and Clark, Catherine A. and Stockdale, Chris and Littlefield, Colin and Gilbert, Emily and Palle, Enric and Girardin, Eric and Murgas, Felipe and Bergsten, Galen J. and Osborn, Hugh P. and Crossfield, Ian J. M. and {de Leon}, Jerome and Higuera, Jesus and Isogai, Keisuke and Everett, Mark E. and Lund, Michael B. and Narita, Norio and Schwarz, Richard P. and Zambelli, Roberto and Howell, Steve B.},
  year = {2025},
  month = aug,
  journal = {AJ},
  volume = {170},
  number = {3},
  pages = {148},
  issn = {1538-3881},
  doi = {10.3847/1538-3881/ade68b},
  urldate = {2025-09-02},
  langid = {english}
}

@article{guerrero2021toi,
  title = {The {{TESS Objects}} of {{Interest Catalog}} from the {{TESS Prime Mission}}},
  author = {Guerrero, Natalia M. and Seager, S. and Huang, Chelsea X. and Vanderburg, Andrew and Soto, Aylin Garcia and Mireles, Ismael and Hesse, Katharine and Fong, William and Glidden, Ana and Shporer, Avi and Latham, David W. and Collins, Karen A. and Quinn, Samuel N. and Burt, Jennifer and Dragomir, Diana and Crossfield, Ian and Vanderspek, Roland and Fausnaugh, Michael and Burke, Christopher J. and Ricker, George and Daylan, Tansu and Essack, Zahra and G{\"u}nther, Maximilian N. and Osborn, Hugh P. and Pepper, Joshua and Rowden, Pamela and Sha, Lizhou and Jr, Steven Villanueva and Yahalomi, Daniel A. and Yu, Liang and Ballard, Sarah and Batalha, Natalie M. and Berardo, David and Chontos, Ashley and Dittmann, Jason A. and Esquerdo, Gilbert A. and {Mikal-Evans}, Thomas and Jayaraman, Rahul and Krishnamurthy, Akshata and Louie, Dana R. and Mehrle, Nicholas and Niraula, Prajwal and Rackham, Benjamin V. and Rodriguez, Joseph E. and Rowden, Stephen J. L. and {Sousa-Silva}, Clara and Watanabe, David and Wong, Ian and Zhan, Zhuchang and Zivanovic, Goran and Christiansen, Jessie L. and Ciardi, David R. and Swain, Melanie A. and Lund, Michael B. and Mullally, Susan E. and Fleming, Scott W. and Rodriguez, David R. and Boyd, Patricia T. and Quintana, Elisa V. and Barclay, Thomas and Col{\'o}n, Knicole D. and Rinehart, S. A. and Schlieder, Joshua E. and Clampin, Mark and Jenkins, Jon M. and Twicken, Joseph D. and Caldwell, Douglas A. and Coughlin, Jeffrey L. and Henze, Chris and Lissauer, Jack J. and Morris, Robert L. and Rose, Mark E. and Smith, Jeffrey C. and Tenenbaum, Peter and Ting, Eric B. and Wohler, Bill and Bakos, G. {\'A} and Bean, Jacob L. and {Berta-Thompson}, Zachory K. and Bieryla, Allyson and Bouma, Luke G. and Buchhave, Lars A. and Butler, Nathaniel and Charbonneau, David and Doty, John P. and Ge, Jian and Holman, Matthew J. and Howard, Andrew W. and Kaltenegger, Lisa and Kane, Stephen R. and Kjeldsen, Hans and Kreidberg, Laura and Lin, Douglas N. C. and Minsky, Charlotte and Narita, Norio and Paegert, Martin and P{\'a}l, Andr{\'a}s and Palle, Enric and Sasselov, Dimitar D. and Spencer, Alton and Sozzetti, Alessandro and Stassun, Keivan G. and Torres, Guillermo and Udry, Stephane and Winn, Joshua N.},
  year = {2021},
  month = jun,
  journal = {ApJS},
  volume = {254},
  number = {2},
  pages = {39},
  issn = {0067-0049},
  doi = {10.3847/1538-4365/abefe1},
  urldate = {2023-09-11},
  langid = {english}
}

@article{guillot2015astep,
  title = {Thermalizing a Telescope in {{Antarctica}} -- Analysis of {{ASTEP}} Observations},
  author = {Guillot, T. and Abe, L. and Agabi, A. and Rivet, J.-P. and Daban, J.-B. and M{\'e}karnia, D. and Aristidi, E. and Schmider, F.-X. and Crouzet, N. and Gon{\c c}alves, I. and Gouvret, C. and Ottogalli, S. and Faradji, H. and Blanc, P.-E. and Bondoux, E. and Valbousquet, F.},
  year = {2015},
  journal = {Astrono. Nachr.},
  volume = {336},
  number = {7},
  pages = {638--656},
  issn = {1521-3994},
  doi = {10.1002/asna.201512174},
  urldate = {2025-10-20},
  copyright = {Copyright {\copyright} 2015 WILEY-VCH Verlag GmbH \& Co. KGaA, Weinheim},
  langid = {english}
}

@article{hadjigeorghiou2024positional,
  title = {The Positional Probability and True Host Star Identification of {{TESS}} Exoplanet Candidates},
  author = {Hadjigeorghiou, Andreas and Armstrong, David J.},
  year = {2024},
  month = jan,
  journal = {\mnras},
  volume = {527},
  pages = {4018--4030},
  issn = {0035-8711},
  doi = {10.1093/mnras/stad3286},
  urldate = {2024-01-15}
}

@article{harris2020numpy,
  title = {Array Programming with {{NumPy}}},
  author = {Harris, Charles R. and Millman, K. Jarrod and {van der Walt}, St{\'e}fan J. and Gommers, Ralf and Virtanen, Pauli and Cournapeau, David and Wieser, Eric and Taylor, Julian and Berg, Sebastian and Smith, Nathaniel J. and Kern, Robert and Picus, Matti and Hoyer, Stephan and {van Kerkwijk}, Marten H. and Brett, Matthew and Haldane, Allan and {del R{\'i}o}, Jaime Fern{\'a}ndez and Wiebe, Mark and Peterson, Pearu and {G{\'e}rard-Marchant}, Pierre and Sheppard, Kevin and Reddy, Tyler and Weckesser, Warren and Abbasi, Hameer and Gohlke, Christoph and Oliphant, Travis E.},
  year = {2020},
  month = sep,
  journal = {Nature},
  volume = {585},
  pages = {357--362},
  issn = {0028-0836},
  doi = {10.1038/s41586-020-2649-2},
  urldate = {2022-08-26}
}

@article{hedges2019k2,
  title = {Four {{Small Planets Buried}} in {{K2 Systems}}: {{What Can We Learn}} for {{TESS}}?},
  shorttitle = {Four {{Small Planets Buried}} in {{K2 Systems}}},
  author = {Hedges, Christina and Saunders, Nicholas and Barentsen, Geert and Coughlin, Jeffrey L. and Cardoso, Jos{\`e} Vin{\'i}cius de Miranda and Kostov, Veselin B. and Dotson, Jessie and Cody, Ann Marie},
  year = {2019},
  month = jul,
  journal = {ApJ},
  volume = {880},
  pages = {L5},
  issn = {0004-637X},
  doi = {10.3847/2041-8213/ab2a74},
  urldate = {2025-12-22}
}

@article{hernandezcarnerero2025attention,
  title = {Additive {{Attention}} for {{Vetting Transiting Exoplanet Candidates}}},
  author = {{Hern{\`a}ndez-Carnerero}, {\`A}lvar and {S{\`a}nchez-Marr{\`e}}, Miquel and Morales, Juan Carlos},
  year = {2025},
  month = jun,
  journal = {AJ},
  volume = {170},
  number = {1},
  pages = {21},
  issn = {1538-3881},
  doi = {10.3847/1538-3881/add2f1},
  urldate = {2025-09-02},
  langid = {english}
}

@article{hippke2019TLS,
  title = {Optimized Transit Detection Algorithm to Search for Periodic Transits of Small Planets},
  author = {Hippke, Michael and Heller, Ren{\'e}},
  year = {2019},
  month = mar,
  journal = {A\&A},
  volume = {623},
  pages = {A39},
  issn = {0004-6361},
  doi = {10.1051/0004-6361/201834672},
  urldate = {2023-12-07}
}

@article{hoffman2022cuvarbase,
  title = {Cuvarbase: Fast Period Finding Utilities for {{GPUs}}},
  shorttitle = {Cuvarbase},
  author = {Hoffman, John},
  year = {2022},
  month = oct,
  journal = {Astrophysics Source Code Library},
  pages = {ascl:2210.030},
  urldate = {2025-08-19}
}

@article{howell2014k2,
  title = {The {{K2 Mission}}: {{Characterization}} and {{Early Results}}},
  shorttitle = {The {{K2 Mission}}},
  author = {Howell, Steve B. and Sobeck, Charlie and Haas, Michael and Still, Martin and Barclay, Thomas and Mullally, Fergal and Troeltzsch, John and Aigrain, Suzanne and Bryson, Stephen T. and Caldwell, Doug and Chaplin, William J. and Cochran, William D. and Huber, Daniel and Marcy, Geoffrey W. and Miglio, Andrea and Najita, Joan R. and Smith, Marcie and Twicken, J. D. and Fortney, Jonathan J.},
  year = {2014},
  month = apr,
  journal = {PASP},
  volume = {126},
  pages = {398},
  issn = {0004-6280},
  doi = {10.1086/676406},
  urldate = {2025-09-01}
}

@article{hsu2019occurrence,
  title = {Occurrence {{Rates}} of {{Planets Orbiting FGK Stars}}: {{Combining Kepler DR25}}, {{Gaia DR2}}, and {{Bayesian Inference}}},
  shorttitle = {Occurrence {{Rates}} of {{Planets Orbiting FGK Stars}}},
  author = {Hsu, Danley C. and Ford, Eric B. and Ragozzine, Darin and Ashby, Keir},
  year = {2019},
  month = aug,
  journal = {AJ},
  volume = {158},
  number = {3},
  pages = {109},
  issn = {1538-3881},
  doi = {10.3847/1538-3881/ab31ab},
  urldate = {2025-12-10},
  langid = {english}
}

@article{huang2020qlp1,
  title = {Photometry of 10 {{Million Stars}} from the {{First Two Years}} of {{TESS Full Frame Images}}: {{Part I}}},
  shorttitle = {Photometry of 10 {{Million Stars}} from the {{First Two Years}} of {{TESS Full Frame Images}}},
  author = {Huang, Chelsea X. and Vanderburg, Andrew and P{\'a}l, Andras and Sha, Lizhou and Yu, Liang and Fong, Willie and Fausnaugh, Michael and Shporer, Avi and Guerrero, Natalia and Vanderspek, Roland and Ricker, George},
  year = {2020},
  month = nov,
  journal = {Res. Notes AAS},
  volume = {4},
  number = {11},
  pages = {204},
  issn = {2515-5172},
  doi = {10.3847/2515-5172/abca2e},
  urldate = {2025-08-22},
  langid = {english}
}

@article{huang2020qlp2,
  title = {Photometry of 10 {{Million Stars}} from the {{First Two Years}} of {{TESS Full Frame Images}}: {{Part II}}},
  shorttitle = {Photometry of 10 {{Million Stars}} from the {{First Two Years}} of {{TESS Full Frame Images}}},
  author = {Huang, Chelsea X. and Vanderburg, Andrew and P{\'a}l, Andras and Sha, Lizhou and Yu, Liang and Fong, Willie and Fausnaugh, Michael and Shporer, Avi and Guerrero, Natalia and Vanderspek, Roland and Ricker, George},
  year = {2020},
  month = nov,
  journal = {Res. Notes AAS},
  volume = {4},
  number = {11},
  pages = {206},
  issn = {2515-5172},
  doi = {10.3847/2515-5172/abca2d},
  urldate = {2025-08-22},
  langid = {english}
}

@article{hunter2007matplotlib,
  title = {Matplotlib: {{A 2D Graphics Environment}}},
  shorttitle = {Matplotlib},
  author = {Hunter, John D.},
  year = {2007},
  month = may,
  journal = {Comput. Sci. Eng.},
  volume = {9},
  pages = {90--95},
  doi = {10.1109/MCSE.2007.55},
  urldate = {2022-08-26}
}

@inproceedings{jenkins2016tessspoc,
  title = {The {{TESS}} Science Processing Operations Center},
  booktitle = {Software and {{Cyberinfrastructure}} for {{Astronomy IV}}},
  author = {Jenkins, Jon M. and Twicken, Joseph D. and McCauliff, Sean and Campbell, Jennifer and Sanderfer, Dwight and Lung, David and {Mansouri-Samani}, Masoud and Girouard, Forrest and Tenenbaum, Peter and Klaus, Todd and Smith, Jeffrey C. and Caldwell, Douglas A. and Chacon, A. Dean and Henze, Christopher and Heiges, Cory and Latham, David W. and Morgan, Edward and Swade, Daryl and Rinehart, Stephen and Vanderspek, Roland},
  year = {2016},
  month = aug,
  volume = {9913},
  pages = {1232--1251},
  publisher = {SPIE},
  doi = {10.1117/12.2233418},
  urldate = {2025-06-11}
}

@article{jenkins2020ltt9779,
  title = {An Ultrahot {{Neptune}} in the {{Neptune}} Desert},
  author = {Jenkins, James S. and D{\'i}az, Mat{\'i}as R. and Kurtovic, Nicol{\'a}s T. and Espinoza, N{\'e}stor and Vines, Jose I. and Rojas, Pablo A. Pe{\~n}a and Brahm, Rafael and Torres, Pascal and {Cort{\'e}s-Zuleta}, P{\'i}a and Soto, Maritza G. and Lopez, Eric D. and King, George W. and Wheatley, Peter J. and Winn, Joshua N. and Ciardi, David R. and Ricker, George and Vanderspek, Roland and Latham, David W. and Seager, Sara and Jenkins, Jon M. and Beichman, Charles A. and Bieryla, Allyson and Burke, Christopher J. and Christiansen, Jessie L. and Henze, Christopher E. and Klaus, Todd C. and McCauliff, Sean and Mori, Mayuko and Narita, Norio and Nishiumi, Taku and Tamura, Motohide and {de Leon}, Jerome Pitogo and Quinn, Samuel N. and Villase{\~n}or, Jesus Noel and Vezie, Michael and Lissauer, Jack J. and Collins, Karen A. and Collins, Kevin I. and Isopi, Giovanni and Mallia, Franco and Ercolino, Andrea and Petrovich, Cristobal and Jord{\'a}n, Andr{\'e}s and Acton, Jack S. and Armstrong, David J. and Bayliss, Daniel and Bouchy, Fran{\c c}ois and Belardi, Claudia and Bryant, Edward M. and Burleigh, Matthew R. and Cabrera, Juan and Casewell, Sarah L. and Chaushev, Alexander and Cooke, Benjamin F. and Eigm{\"u}ller, Philipp and Erikson, Anders and Foxell, Emma and G{\"a}nsicke, Boris T. and Gill, Samuel and Gillen, Edward and G{\"u}nther, Maximilian N. and Goad, Michael R. and Hooton, Matthew J. and Jackman, James A. G. and Louden, Tom and McCormac, James and Moyano, Maximiliano and Nielsen, Louise D. and Pollacco, Don and Queloz, Didier and Rauer, Heike and Raynard, Liam and Smith, Alexis M. S. and Tilbrook, Rosanna H. and {Titz-Weider}, Ruth and Turner, Oliver and Udry, St{\'e}phane and Walker, Simon R. and Watson, Christopher A. and West, Richard G. and Palle, Enric and Ziegler, Carl and Law, Nicholas and Mann, Andrew W.},
  year = {2020},
  month = dec,
  journal = {Nat Astron},
  volume = {4},
  number = {12},
  pages = {1148--1157},
  issn = {2397-3366},
  doi = {10.1038/s41550-020-1142-z},
  urldate = {2025-10-20},
  copyright = {2020 The Author(s), under exclusive licence to Springer Nature Limited},
  langid = {english}
}

@article{kipping2010binning,
  title = {Binning Is Sinning: Morphological Light-Curve Distortions Due to Finite Integration Time},
  shorttitle = {Binning Is Sinning},
  author = {Kipping, David M.},
  year = {2010},
  month = nov,
  journal = {\mnras},
  volume = {408},
  number = {3},
  pages = {1758--1769},
  issn = {0035-8711},
  doi = {10.1111/j.1365-2966.2010.17242.x},
  urldate = {2025-09-29}
}

@article{kipping2013ld,
  title = {Efficient, Uninformative Sampling of Limb Darkening Coefficients for Two-Parameter Laws},
  author = {Kipping, David M.},
  year = {2013},
  month = nov,
  journal = {\mnras},
  volume = {435},
  number = {3},
  pages = {2152--2160},
  issn = {0035-8711},
  doi = {10.1093/mnras/stt1435},
  urldate = {2025-02-11}
}

@article{kohonen1982som,
  title = {Self-Organized Formation of Topologically Correct Feature Maps},
  author = {Kohonen, Teuvo},
  year = {1982},
  month = jan,
  journal = {Biol. Cybern.},
  volume = {43},
  number = {1},
  pages = {59--69},
  issn = {1432-0770},
  doi = {10.1007/BF00337288},
  urldate = {2025-10-01},
  langid = {english}
}

@misc{koposov2024dynesty,
  title = {Joshspeagle/Dynesty: V2.1.4},
  shorttitle = {Joshspeagle/Dynesty},
  author = {Koposov, Sergey and Speagle, Josh and Barbary, Kyle and Ashton, Gregory and Bennett, Ed and Buchner, Johannes and Scheffler, Carl and Cook, Ben and Talbot, Colm and Guillochon, James and Cubillos, Patricio and Ramos, Andr{\'e}s Asensio and Dartiailh, Matthieu and Ilya and Tollerud, Erik and Lang, Dustin and Johnson, Ben and {jtmendel} and Higson, Edward and Vandal, Thomas and Daylan, Tansu and Angus, Ruth and {patelR} and Cargile, Phillip and Sheehan, Patrick and Pitkin, Matt and Kirk, Matthew and Leja, Joel and {joezuntz} and Goldstein, Danny},
  year = {2024},
  month = jun,
  urldate = {2025-08-20},
  howpublished = {Zenodo}
}

@article{kostov2019dave,
  title = {Discovery and {{Vetting}} of {{Exoplanets}}. {{I}}. {{Benchmarking K2 Vetting Tools}}},
  author = {Kostov, Veselin B. and Mullally, Susan E. and Quintana, Elisa V. and Coughlin, Jeffrey L. and Mullally, Fergal and Barclay, Thomas and Col{\'o}n, Knicole D. and Schlieder, Joshua E. and Barentsen, Geert and Burke, Christopher J.},
  year = {2019},
  month = feb,
  journal = {AJ},
  volume = {157},
  number = {3},
  pages = {124},
  issn = {1538-3881},
  doi = {10.3847/1538-3881/ab0110},
  urldate = {2025-09-02},
  langid = {english}
}

@article{kovacs2002bls,
  title = {A Box-Fitting Algorithm in the Search for Periodic Transits},
  author = {Kov{\'a}cs, G. and Zucker, S. and Mazeh, T.},
  year = {2002},
  month = aug,
  journal = {A\&A},
  volume = {391},
  pages = {369--377},
  issn = {0004-6361},
  doi = {10.1051/0004-6361:20020802},
  urldate = {2023-09-11}
}

@article{kreidberg2015batman,
  title = {Batman: {{BAsic Transit Model cAlculatioN}} in {{Python}}},
  shorttitle = {Batman},
  author = {Kreidberg, Laura},
  year = {2015},
  month = nov,
  journal = {PASP},
  volume = {127},
  pages = {1161},
  issn = {0004-6280},
  doi = {10.1086/683602},
  urldate = {2019-05-24}
}

@article{kunimoto2022qlpfaint,
  title = {The {{TESS Faint-star Search}}: 1617 {{TOIs}} from the {{TESS Primary Mission}}},
  shorttitle = {The {{TESS Faint-star Search}}},
  author = {Kunimoto, Michelle and Daylan, Tansu and Guerrero, Natalia and Fong, William and Bryson, Steve and Ricker, George R. and Fausnaugh, Michael and Huang, Chelsea X. and Sha, Lizhou and Shporer, Avi and Vanderburg, Andrew and Vanderspek, Roland K. and Yu, Liang},
  year = {2022},
  month = mar,
  journal = {ApJS},
  volume = {259},
  number = {2},
  pages = {33},
  issn = {0067-0049},
  doi = {10.3847/1538-4365/ac5688},
  urldate = {2025-08-22},
  langid = {english}
}

@article{lightkurve2018,
  title = {Lightkurve: {{Kepler}} and {{TESS}} Time Series Analysis in {{Python}}},
  shorttitle = {Lightkurve},
  author = {{Lightkurve Collaboration} and Cardoso, Jos{\'e} Vin{\'i}cius de Miranda and Hedges, Christina and {Gully-Santiago}, Michael and Saunders, Nicholas and Cody, Ann Marie and Barclay, Thomas and Hall, Oliver and Sagear, Sheila and Turtelboom, Emma and Zhang, Johnny and Tzanidakis, Andy and Mighell, Ken and Coughlin, Jeff and Bell, Keaton and {Berta-Thompson}, Zach and Williams, Peter and Dotson, Jessie and Barentsen, Geert},
  year = {2018},
  month = dec,
  journal = {Astrophysics Source Code Library},
  pages = {ascl:1812.013},
  urldate = {2025-10-07}
}

@article{lissauer2012multis,
  title = {{{ALMOST ALL OF KEPLER}}'{{S MULTIPLE-PLANET CANDIDATES ARE PLANETS}}},
  author = {Lissauer, Jack J. and Marcy, Geoffrey W. and Rowe, Jason F. and Bryson, Stephen T. and Adams, Elisabeth and Buchhave, Lars A. and Ciardi, David R. and Cochran, William D. and Fabrycky, Daniel C. and Ford, Eric B. and Fressin, Francois and Geary, John and Gilliland, Ronald L. and Holman, Matthew J. and Howell, Steve B. and Jenkins, Jon M. and Kinemuchi, Karen and Koch, David G. and Morehead, Robert C. and Ragozzine, Darin and Seader, Shawn E. and Tanenbaum, Peter G. and Torres, Guillermo and Twicken, Joseph D.},
  year = {2012},
  month = apr,
  journal = {ApJ},
  volume = {750},
  number = {2},
  pages = {112},
  issn = {0004-637X},
  doi = {10.1088/0004-637X/750/2/112},
  urldate = {2025-10-20},
  langid = {english}
}

@article{lundkvist2016desert,
  title = {Hot Super-{{Earths}} Stripped by Their Host Stars},
  author = {Lundkvist, M. S. and Kjeldsen, H. and Albrecht, S. and Davies, G. R. and Basu, S. and Huber, D. and Justesen, A. B. and Karoff, C. and Silva Aguirre, V. and Van Eylen, V. and Vang, C. and Arentoft, T. and Barclay, T. and Bedding, T. R. and Campante, T. L. and Chaplin, W. J. and {Christensen-Dalsgaard}, J. and Elsworth, Y. P. and Gilliland, R. L. and Handberg, R. and Hekker, S. and Kawaler, S. D. and Lund, M. N. and Metcalfe, T. S. and Miglio, A. and Rowe, J. F. and Stello, D. and Tingley, B. and White, T. R.},
  year = {2016},
  month = apr,
  journal = {Nat. Commun.},
  volume = {7},
  number = {1},
  pages = {11201},
  issn = {2041-1723},
  doi = {10.1038/ncomms11201},
  urldate = {2025-09-09},
  copyright = {2016 The Author(s)},
  langid = {english}
}

@article{mazeh2016desert,
  title = {Dearth of Short-Period {{Neptunian}} Exoplanets: {{A}} Desert in Period-Mass and Period-Radius Planes},
  shorttitle = {Dearth of Short-Period {{Neptunian}} Exoplanets},
  author = {Mazeh, T. and Holczer, T. and Faigler, S.},
  year = {2016},
  month = may,
  journal = {A\&A},
  volume = {589},
  pages = {A75},
  issn = {0004-6361},
  doi = {10.1051/0004-6361/201528065},
  urldate = {2022-09-20},
  langid = {english}
}

@article{mccauliff2015autovetter,
  title = {{{AUTOMATIC CLASSIFICATION OF KEPLER PLANETARY TRANSIT CANDIDATES}}},
  author = {McCauliff, Sean D. and Jenkins, Jon M. and Catanzarite, Joseph and Burke, Christopher J. and Coughlin, Jeffrey L. and Twicken, Joseph D. and Tenenbaum, Peter and Seader, Shawn and Li, Jie and Cote, Miles},
  year = {2015},
  month = jun,
  journal = {ApJ},
  volume = {806},
  number = {1},
  pages = {6},
  issn = {0004-637X},
  doi = {10.1088/0004-637X/806/1/6},
  urldate = {2025-09-03},
  langid = {english}
}

@article{mekarnia2016astep,
  title = {Transiting Planet Candidates with {{ASTEP}}~400 at {{Dome~C}}, {{Antarctica}}},
  author = {M{\'e}karnia, D. and Guillot, T. and Rivet, J.-P. and Schmider, F.-X. and Abe, L. and Gon{\c c}alves, I. and Agabi, A. and Crouzet, N. and Fruth, T. and Barbieri, M. and Bayliss, D. D. R. and Zhou, G. and Aristidi, E. and Szulagyi, J. and Daban, J.-B. and {Fante{\"i}-Caujolle}, Y. and Gouvret, C. and Erikson, A. and Rauer, H. and Bouchy, F. and Gerakis, J. and Bouchez, G.},
  year = {2016},
  month = nov,
  journal = {\mnras},
  volume = {463},
  number = {1},
  pages = {45--62},
  issn = {0035-8711},
  doi = {10.1093/mnras/stw1934},
  urldate = {2025-10-20}
}

@article{melton2024dtarps1,
  title = {{{DIAmante TESS AutoRegressive Planet Search}} ({{DTARPS}}). {{I}}. {{Analysis}} of 0.9 {{Million Light Curves}}},
  author = {Melton, Elizabeth J. and Feigelson, Eric D. and Montalto, Marco and Caceres, Gabriel A. and Rosenswie, Andrew W. and Abelson, Cullen S.},
  year = {2024},
  month = apr,
  journal = {AJ},
  volume = {167},
  number = {5},
  pages = {202},
  issn = {1538-3881},
  doi = {10.3847/1538-3881/ad29f0},
  urldate = {2024-08-30},
  langid = {english}
}

@article{melton2024dtarps2,
  title = {{{DIAmante TESS AutoRegressive Planet Search}} ({{DTARPS}}). {{II}}. {{Hundreds}} of {{New TESS Candidate Exoplanets}}},
  author = {Melton, Elizabeth J. and Feigelson, Eric D. and Montalto, Marco and Caceres, Gabriel A. and Rosenswie, Andrew W. and Abelson, Cullen S.},
  year = {2024},
  month = apr,
  journal = {AJ},
  volume = {167},
  number = {5},
  pages = {203},
  issn = {1538-3881},
  doi = {10.3847/1538-3881/ad29f1},
  urldate = {2024-08-30},
  langid = {english}
}

@article{melton2024dtarps3,
  title = {{{DIAmante TESS AutoRegressive Planet Search}} ({{DTARPS}}). {{III}}. {{Understanding}} the {{DTARPS-S Candidate Transiting Planet Catalogs}}},
  author = {Melton, Elizabeth J. and Feigelson, Eric D. and Montalto, Marco and Caceres, Gabriel A. and Rosenswie, Andrew W. and Abelson, Cullen S.},
  year = {2024},
  month = dec,
  journal = {AJ},
  volume = {168},
  pages = {271},
  issn = {0004-6256},
  doi = {10.3847/1538-3881/ad835510.1134/S1063772908070081},
  urldate = {2025-01-20}
}

@article{montalto2020diamante,
  title = {A Search for Transiting Planets around {{FGKM}} Dwarfs and Subgiants in the {{TESS}} Full Frame Images of the {{Southern}} Ecliptic Hemisphere},
  author = {Montalto, M and Borsato, L and Granata, V and Lacedelli, G and Malavolta, L and Manthopoulou, E E and Nardiello, D and Nascimbeni, V and Piotto, G},
  year = {2020},
  month = sep,
  journal = {\mnras},
  volume = {498},
  number = {2},
  pages = {1726--1749},
  issn = {0035-8711},
  doi = {10.1093/mnras/staa2438},
  urldate = {2025-08-12}
}

@article{montalto2023diamante,
  title = {A Search for Planetary Transits on a Set of 1.4 Million Multisector {{DIAmante}} Light Curves},
  author = {Montalto, M},
  year = {2023},
  month = jan,
  journal = {\mnras},
  volume = {518},
  number = {1},
  pages = {L31-L35},
  issn = {1745-3925},
  doi = {10.1093/mnrasl/slac131},
  urldate = {2025-08-26}
}

@article{morton2012validation,
  title = {{{AN EFFICIENT AUTOMATED VALIDATION PROCEDURE FOR EXOPLANET TRANSIT CANDIDATES}}},
  author = {Morton, Timothy D.},
  year = {2012},
  month = nov,
  journal = {ApJ},
  volume = {761},
  number = {1},
  pages = {6},
  issn = {0004-637X},
  doi = {10.1088/0004-637X/761/1/6},
  urldate = {2025-09-02},
  langid = {english}
}

@article{morton2016vespa,
  title = {False {{Positive Probabilities}} for All {{Kepler Objects}} of {{Interest}}: 1284 {{Newly Validated Planets}} and 428 {{Likely False Positives}}},
  shorttitle = {False {{Positive Probabilities}} for All {{Kepler Objects}} of {{Interest}}},
  author = {Morton, Timothy D. and Bryson, Stephen T. and Coughlin, Jeffrey L. and Rowe, Jason F. and Ravichandran, Ganesh and Petigura, Erik A. and Haas, Michael R. and Batalha, Natalie M.},
  year = {2016},
  month = may,
  journal = {ApJ},
  volume = {822},
  pages = {86},
  issn = {0004-637X},
  doi = {10.3847/0004-637X/822/2/86},
  urldate = {2023-09-11}
}

@article{mullally2015robovetter,
  title = {{{PLANETARY CANDIDATES OBSERVED BY KEPLER}}. {{VI}}. {{PLANET SAMPLE FROM Q1}}--{{Q16}} (47 {{MONTHS}})},
  author = {Mullally, F. and Coughlin, Jeffrey L. and Thompson, Susan E. and Rowe, Jason and Burke, Christopher and Latham, David W. and Batalha, Natalie M. and Bryson, Stephen T. and Christiansen, Jessie and Henze, Christopher E. and Ofir, Aviv and Quarles, Billy and Shporer, Avi and Eylen, Vincent Van and Laerhoven, Christa Van and Shah, Yash and Wolfgang, Angie and Chaplin, W. J. and Xie, Ji-Wei and Akeson, Rachel and Argabright, Vic and Bachtell, Eric and Barclay, Thomas and Borucki, William J. and Caldwell, Douglas A. and Campbell, Jennifer R. and Catanzarite, Joseph H. and Cochran, William D. and Duren, Riley M. and Fleming, Scott W. and Fraquelli, Dorothy and Girouard, Forrest R. and Haas, Michael R. and He{\l}miniak, Krzysztof G. and Howell, Steve B. and Huber, Daniel and Larson, Kipp and III, Thomas N. Gautier and Jenkins, Jon M. and Li, Jie and Lissauer, Jack J. and McArthur, Scot and Miller, Chris and Morris, Robert L. and {Patil-Sabale}, Anima and Plavchan, Peter and Putnam, Dustin and Quintana, Elisa V. and Ramirez, Solange and Aguirre, V. Silva and Seader, Shawn and Smith, Jeffrey C. and Steffen, Jason H. and Stewart, Chris and Stober, Jeremy and Still, Martin and Tenenbaum, Peter and Troeltzsch, John and Twicken, Joseph D. and Zamudio, Khadeejah A.},
  year = {2015},
  month = apr,
  journal = {ApJS},
  volume = {217},
  number = {2},
  pages = {31},
  issn = {0067-0049},
  doi = {10.1088/0067-0049/217/2/31},
  urldate = {2025-09-03},
  langid = {english}
}

@article{munavarhussain2025multis,
  title = {Eight {{New Candidate Multiplanet Systems}} among {{TESS Objects}} of {{Interest}}},
  author = {Munavar Hussain, Mohamed Jassim and Kunimoto, Michelle},
  year = {2025},
  month = jan,
  journal = {AJ},
  volume = {169},
  number = {2},
  pages = {91},
  issn = {1538-3881},
  doi = {10.3847/1538-3881/ad9a6c},
  urldate = {2025-12-19},
  langid = {english}
}

@article{nabbie2024toi3261,
  title = {Surviving in the {{Hot-Neptune Desert}}: {{The Discovery}} of the {{Ultrahot Neptune TOI-3261b}}},
  shorttitle = {Surviving in the {{Hot-Neptune Desert}}},
  author = {Nabbie, Emma and Huang, Chelsea X. and Burt, Jennifer A. and Armstrong, David J. and Mamajek, Eric E. and Adibekyan, Vardan and Sousa, S{\'e}rgio G. and Lopez, Eric D. and Thorngren, Daniel and Fern{\'a}ndez Fern{\'a}ndez, Jorge and Li, Gongjie and Jenkins, James S. and Vines, Jose I. and {Gomes da Silva}, Jo{\~a}o and Wittenmyer, Robert A. and Bayliss, Daniel and Brice{\~n}o, C{\'e}sar and Collins, Karen A. and Dumusque, Xavier and Horne, Keith and F. Keniger, Marcelo Aron and Law, Nicholas and {Lillo-Box}, Jorge and Liu, Shang-Fei and Mann, Andrew W. and D. Nielsen, Louise and Osborn, Ares and Relles, Howard M. and Rodrigues, Jos{\'e} J. and Serrano Bell, Juan and Srdoc, Gregor and Stockdale, Chris and Str{\o}m, Paul A. and Watkins, Cristilyn N. and Wheatley, Peter J. and Wright, Duncan J. and Zhou, George and Ziegler, Carl and Ricker, George and Seager, Sara and Vanderspek, Roland and Winn, Joshua N. and Jenkins, Jon M. and Fausnaugh, Michael and Kunimoto, Michelle and Osborn, Hugh P. and Quinn, Samuel N. and Wohler, Bill},
  year = {2024},
  month = aug,
  journal = {AJ},
  volume = {168},
  number = {3},
  pages = {132},
  issn = {1538-3881},
  doi = {10.3847/1538-3881/ad60be},
  urldate = {2025-10-20},
  langid = {english}
}

@article{nardiello2020pathos2,
  title = {A {{PSF-based Approach}} to {{TESS High}} Quality Data {{Of Stellar}} Clusters ({{PATHOS}}) -- {{II}}. {{Search}} for Exoplanets in Open Clusters of the {{Southern}} Ecliptic Hemisphere and Their Frequency},
  author = {Nardiello, D and Piotto, G and Deleuil, M and Malavolta, L and Montalto, M and Bedin, L R and Borsato, L and Granata, V and Libralato, M and Manthopoulou, E E},
  year = {2020},
  month = jul,
  journal = {\mnras},
  volume = {495},
  number = {4},
  pages = {4924--4942},
  issn = {0035-8711},
  doi = {10.1093/mnras/staa1465},
  urldate = {2025-09-03}
}

@article{olmschenk2021nn,
  title = {Identifying {{Planetary Transit Candidates}} in {{TESS Full-frame Image Light Curves}} via {{Convolutional Neural Networks}}},
  author = {Olmschenk, Greg and Ishitani Silva, Stela and Rau, Gioia and Barry, Richard K. and Kruse, Ethan and Cacciapuoti, Luca and Kostov, Veselin and Powell, Brian P. and Wyrwas, Edward and Schnittman, Jeremy D. and Barclay, Thomas},
  year = {2021},
  month = may,
  journal = {AJ},
  volume = {161},
  number = {6},
  pages = {273},
  issn = {1538-3881},
  doi = {10.3847/1538-3881/abf4c6},
  urldate = {2025-08-29},
  langid = {english}
}

@article{osborn2020nn,
  title = {Rapid Classification of {{TESS}} Planet Candidates with Convolutional Neural Networks},
  author = {Osborn, H. P. and Ansdell, M. and Ioannou, Y. and Sasdelli, M. and Angerhausen, D. and Caldwell, D. and Jenkins, J. M. and R{\"a}issi, C. and Smith, J. C.},
  year = {2020},
  month = jan,
  journal = {A\&A},
  volume = {633},
  pages = {A53},
  issn = {0004-6361, 1432-0746},
  doi = {10.1051/0004-6361/201935345},
  urldate = {2025-09-02},
  copyright = {{\copyright} H. P. Osborn et al. 2020},
  langid = {english}
}

@article{osborn2023toi332,
  title = {{{TOI-332}}\,b: A Super Dense {{Neptune}} Found Deep within the {{Neptunian}} Desert},
  shorttitle = {{{TOI-332}}\,b},
  author = {Osborn, Ares and Armstrong, David J and Fern{\'a}ndez~Fern{\'a}ndez, Jorge and Knierim, Henrik and Adibekyan, Vardan and Collins, Karen A and {Delgado-Mena}, Elisa and Fridlund, Malcolm and {Gomes~da~Silva}, Jo{\~a}o and Hellier, Coel and Jackson, David G and King, George W and {Lillo-Box}, Jorge and Matson, Rachel A and Matthews, Elisabeth C and Santos, Nuno C and Sousa, S{\'e}rgio G and Stassun, Keivan G and Tan, Thiam-Guan and Ricker, George R and Vanderspek, Roland and Latham, David W and Seager, Sara and Winn, Joshua N and Jenkins, Jon M and Bayliss, Daniel and Bouma, Luke G and Ciardi, David R and Collins, Kevin I and Col{\'o}n, Knicole D and Crossfield, Ian J M and Demangeon, Olivier D S and D{\'i}az, Rodrigo F and Dorn, Caroline and Dumusque, Xavier and Keniger, Marcelo Aron Fetzner and Figueira, Pedro and Gan, Tianjun and Goeke, Robert F and Hadjigeorghiou, Andreas and Hawthorn, Faith and Helled, Ravit and Howell, Steve B and Nielsen, Louise D and Osborn, Hugh P and Quinn, Samuel N and Sefako, Ramotholo and Shporer, Avi and Str{\o}m, Paul A and Twicken, Joseph D and Vanderburg, Andrew and Wheatley, Peter J},
  year = {2023},
  month = nov,
  journal = {\mnras},
  volume = {526},
  number = {1},
  pages = {548--566},
  issn = {0035-8711},
  doi = {10.1093/mnras/stad2575},
  urldate = {2025-09-14}
}

@article{pedregosa2011sklearn,
  title = {Scikit-Learn: {{Machine Learning}} in {{Python}}},
  shorttitle = {Scikit-Learn},
  author = {Pedregosa, Fabian and Varoquaux, Ga{\"e}l and Gramfort, Alexandre and Michel, Vincent and Thirion, Bertrand and Grisel, Olivier and Blondel, Mathieu and Prettenhofer, Peter and Weiss, Ron and Dubourg, Vincent and Vanderplas, Jake and Passos, Alexandre and Cournapeau, David and Brucher, Matthieu and Perrot, Matthieu and Duchesnay, {\'E}douard},
  year = {2011},
  journal = {Journal of Machine Learning Research},
  volume = {12},
  number = {85},
  pages = {2825--2830},
  issn = {1533-7928},
  urldate = {2025-10-24}
}

@article{rao2021nigraha,
  title = {Nigraha: {{Machine-learning-based}} Pipeline to Identify and Evaluate Planet Candidates from {{TESS}}},
  shorttitle = {Nigraha},
  author = {Rao, Sriram and Mahabal, Ashish and Rao, Niyanth and Raghavendra, Cauligi},
  year = {2021},
  month = apr,
  journal = {\mnras},
  volume = {502},
  number = {2},
  pages = {2845--2858},
  issn = {0035-8711},
  doi = {10.1093/mnras/stab203},
  urldate = {2025-09-03}
}

@book{rasmussen2006gp,
  title = {Gaussian {{Processes}} for {{Machine Learning}}},
  author = {Rasmussen, Carl Edward and Williams, Christopher K. I.},
  year = {2006},
  month = jan,
  urldate = {2025-10-01}
  
}

@article{ricker2015tess,
  title = {Transiting {{Exoplanet Survey Satellite}} ({{TESS}})},
  author = {Ricker, George R. and Winn, Joshua N. and Vanderspek, Roland and Latham, David W. and Bakos, G{\'a}sp{\'a}r {\'A}. and Bean, Jacob L. and {Berta-Thompson}, Zachory K. and Brown, Timothy M. and Buchhave, Lars and Butler, Nathaniel R. and Butler, R. Paul and Chaplin, William J. and Charbonneau, David and {Christensen-Dalsgaard}, J{\o}rgen and Clampin, Mark and Deming, Drake and Doty, John and De Lee, Nathan and Dressing, Courtney and Dunham, Edward W. and Endl, Michael and Fressin, Francois and Ge, Jian and Henning, Thomas and Holman, Matthew J. and Howard, Andrew W. and Ida, Shigeru and Jenkins, Jon M. and Jernigan, Garrett and Johnson, John Asher and Kaltenegger, Lisa and Kawai, Nobuyuki and Kjeldsen, Hans and Laughlin, Gregory and Levine, Alan M. and Lin, Douglas and Lissauer, Jack J. and MacQueen, Phillip and Marcy, Geoffrey and McCullough, Peter R. and Morton, Timothy D. and Narita, Norio and Paegert, Martin and Palle, Enric and Pepe, Francesco and Pepper, Joshua and Quirrenbach, Andreas and Rinehart, Stephen A. and Sasselov, Dimitar and Sato, Bun'ei and Seager, Sara and Sozzetti, Alessandro and Stassun, Keivan G. and Sullivan, Peter and Szentgyorgyi, Andrew and Torres, Guillermo and Udry, Stephane and Villasenor, Joel},
  year = {2015},
  month = jan,
  journal = {J. Astron. Tel. Inst. Syst.},
  volume = {1},
  pages = {014003},
  doi = {10.1117/1.JATIS.1.1.014003},
  urldate = {2022-10-11}
}

@article{rodriguez2018k2-266,
  title = {A {{Compact Multi-planet System}} with a {{Significantly Misaligned Ultra Short Period Planet}}},
  author = {Rodriguez, Joseph E. and Becker, Juliette C. and Eastman, Jason D. and Hadden, Sam and Vanderburg, Andrew and Khain, Tali and Quinn, Samuel N. and Mayo, Andrew and Dressing, Courtney D. and Schlieder, Joshua E. and Ciardi, David R. and Latham, David W. and Rappaport, Saul and Adams, Fred C. and Berlind, Perry and Bieryla, Allyson and Calkins, Michael L. and Esquerdo, Gilbert A. and Kristiansen, Martti H. and Omohundro, Mark and Schwengeler, Hans Martin and Stassun, Keivan G. and Terentev, Ivan},
  year = {2018},
  month = nov,
  journal = {AJ},
  volume = {156},
  number = {5},
  pages = {245},
  issn = {1538-3881},
  doi = {10.3847/1538-3881/aae530},
  urldate = {2025-10-20},
  langid = {english}
}

@article{salinas2023nn,
  title = {Distinguishing a Planetary Transit from False Positives: A {{Transformer-based}} Classification for Planetary Transit Signals},
  shorttitle = {Distinguishing a Planetary Transit from False Positives},
  author = {Salinas, Helem and Pichara, Karim and Brahm, Rafael and {P{\'e}rez-Galarce}, Francisco and Mery, Domingo},
  year = {2023},
  month = jul,
  journal = {\mnras},
  volume = {522},
  number = {3},
  pages = {3201--3216},
  issn = {0035-8711},
  doi = {10.1093/mnras/stad1173},
  urldate = {2025-09-02}
}

@article{salinas2025transformer,
  title = {Exoplanet Transit Candidate Identification in {{TESS}} Full-Frame Images via a Transformer-Based Algorithm},
  author = {Salinas, Helem and Brahm, Rafael and Olmschenk, Greg and Barry, Richard K and Pichara, Karim and Ishitani~Silva, Stela and Araujo, Vladimir},
  year = {2025},
  month = apr,
  journal = {\mnras},
  volume = {538},
  number = {3},
  pages = {2031--2049},
  issn = {0035-8711},
  doi = {10.1093/mnras/staf347},
  urldate = {2025-08-26}
}

@article{santerne2015pastis,
  title = {{{PASTIS}}: {{Bayesian}} Extrasolar Planet Validation - {{II}}. {{Constraining}} Exoplanet Blend Scenarios Using Spectroscopic Diagnoses},
  shorttitle = {{{PASTIS}}},
  author = {Santerne, A. and D{\'i}az, R. F. and Almenara, J.-M. and Bouchy, F. and Deleuil, M. and Figueira, P. and H{\'e}brard, G. and Moutou, C. and Rodionov, S. and Santos, N. C.},
  year = {2015},
  month = aug,
  journal = {\mnras},
  volume = {451},
  pages = {2337--2351},
  issn = {0035-8711},
  doi = {10.1093/mnras/stv1080},
  urldate = {2018-08-31}
}

@article{savitzky1964smooth,
  title = {Smoothing and {{Differentiation}} of {{Data}} by {{Simplified Least Squares Procedures}}.},
  author = {Savitzky, {\relax Abraham}. and Golay, M. J. E.},
  year = {1964},
  month = jul,
  journal = {Anal. Chem.},
  volume = {36},
  number = {8},
  pages = {1627--1639},
  issn = {0003-2700},
  doi = {10.1021/ac60214a047},
  urldate = {2025-08-19}
}

@article{shallue2018astronet,
  title = {Identifying {{Exoplanets}} with {{Deep Learning}}: {{A Five-planet Resonant Chain}} around {{Kepler-80}} and an {{Eighth Planet}} around {{Kepler-90}}},
  shorttitle = {Identifying {{Exoplanets}} with {{Deep Learning}}},
  author = {Shallue, Christopher J. and Vanderburg, Andrew},
  year = {2018},
  month = jan,
  journal = {AJ},
  volume = {155},
  number = {2},
  pages = {94},
  issn = {1538-3881},
  doi = {10.3847/1538-3881/aa9e09},
  urldate = {2025-09-02},
  langid = {english}
}

@article{shporer2017eb,
  title = {Three {{Statistically Validated K2 Transiting Warm Jupiter Exoplanets Confirmed}} as {{Low-mass Stars}}},
  author = {Shporer, Avi and Zhou, George and Vanderburg, Andrew and Fulton, Benjamin J. and Isaacson, Howard and Bieryla, Allyson and Torres, Guillermo and Morton, Timothy D. and Bento, Joao and Berlind, Perry and Calkins, Michael L. and Esquerdo, Gilbert A. and Howard, Andrew W. and Latham, David W.},
  year = {2017},
  month = sep,
  journal = {ApJL},
  volume = {847},
  number = {2},
  pages = {L18},
  issn = {2041-8205},
  doi = {10.3847/2041-8213/aa8bff},
  urldate = {2025-08-22},
  langid = {english}
}

@article{smith2012pdc,
  title = {Kepler {{Presearch Data Conditioning II}} - {{A Bayesian Approach}} to {{Systematic Error Correction}}},
  author = {Smith, Jeffrey C. and Stumpe, Martin C. and Cleve, Jeffrey E. Van and Jenkins, Jon M. and Barclay, Thomas S. and Fanelli, Michael N. and Girouard, Forrest R. and Kolodziejczak, Jeffery J. and McCauliff, Sean D. and Morris, Robert L. and Twicken, Joseph D.},
  year = {2012},
  month = sep,
  journal = {PASP},
  volume = {124},
  number = {919},
  pages = {1000},
  issn = {1538-3873},
  doi = {10.1086/667697},
  urldate = {2025-08-20},
  langid = {english}
}

@article{speagle2020dynesty,
  title = {Dynesty: A Dynamic Nested Sampling Package for Estimating {{Bayesian}} Posteriors and Evidences},
  shorttitle = {Dynesty},
  author = {Speagle, Joshua S},
  year = {2020},
  month = apr,
  journal = {\mnras},
  volume = {493},
  number = {3},
  pages = {3132--3158},
  issn = {0035-8711},
  doi = {10.1093/mnras/staa278},
  urldate = {2025-08-20}
}

@article{stassun2019tic,
  title = {The {{Revised TESS Input Catalog}} and {{Candidate Target List}}},
  author = {Stassun, Keivan G. and Oelkers, Ryan J. and Paegert, Martin and Torres, Guillermo and Pepper, Joshua and De Lee, Nathan and Collins, Kevin and Latham, David W. and Muirhead, Philip S. and Chittidi, Jay and {Rojas-Ayala}, B{\'a}rbara and Fleming, Scott W. and Rose, Mark E. and Tenenbaum, Peter and Ting, Eric B. and Kane, Stephen R. and Barclay, Thomas and Bean, Jacob L. and Brassuer, C. E. and Charbonneau, David and Ge, Jian and Lissauer, Jack J. and Mann, Andrew W. and McLean, Brian and Mullally, Susan and Narita, Norio and Plavchan, Peter and Ricker, George R. and Sasselov, Dimitar and Seager, S. and Sharma, Sanjib and Shiao, Bernie and Sozzetti, Alessandro and Stello, Dennis and Vanderspek, Roland and Wallace, Geoff and Winn, Joshua N.},
  year = {2019},
  month = oct,
  journal = {AJ},
  volume = {158},
  pages = {138},
  issn = {0004-6256},
  doi = {10.3847/1538-3881/ab3467},
  urldate = {2023-09-11}
}

@article{stumpe2012pdc,
  title = {Kepler {{Presearch Data Conditioning I}}---{{Architecture}} and {{Algorithms}} for {{Error Correction}} in {{Kepler Light Curves}}},
  author = {Stumpe, Martin C. and Smith, Jeffrey C. and Cleve, Jeffrey E. Van and Twicken, Joseph D. and Barclay, Thomas S. and Fanelli, Michael N. and Girouard, Forrest R. and Jenkins, Jon M. and Kolodziejczak, Jeffery J. and McCauliff, Sean D. and Morris, Robert L.},
  year = {2012},
  month = aug,
  journal = {PASP},
  volume = {124},
  number = {919},
  pages = {985},
  issn = {1538-3873},
  doi = {10.1086/667698},
  urldate = {2025-08-20},
  langid = {english}
}

@article{stumpe2014msmap,
  title = {Multiscale {{Systematic Error Correction}} via {{Wavelet-Based Bandsplitting}} in {{Kepler Data}}},
  author = {Stumpe, Martin C. and Smith, Jeffrey C. and Catanzarite, Joseph H. and Cleve, Jeffrey E. Van and Jenkins, Jon M. and Twicken, Joseph D. and Girouard, Forrest R.},
  year = {2014},
  month = jan,
  journal = {PASP},
  volume = {126},
  number = {935},
  pages = {100},
  issn = {1538-3873},
  doi = {10.1086/674989},
  urldate = {2025-08-20},
  langid = {english}
}

@article{tardugnopoleo2024,
  title = {{{NotPlaNET}}: {{Removing False Positives}} from {{Planet Hunters TESS}} with {{Machine Learning}}},
  shorttitle = {{{NotPlaNET}}},
  author = {Tardugno Poleo, Valentina and Eisner, Nora and Hogg, David W.},
  year = {2024},
  month = aug,
  journal = {AJ},
  volume = {168},
  number = {3},
  pages = {100},
  issn = {1538-3881},
  doi = {10.3847/1538-3881/ad5f29},
  urldate = {2025-09-02},
  langid = {english}
}

@article{tey2023astronettriage,
  title = {Identifying {{Exoplanets}} with {{Deep Learning}}. {{V}}. {{Improved Light-curve Classification}} for {{TESS Full-frame Image Observations}}},
  author = {Tey, Evan and Moldovan, Dan and Kunimoto, Michelle and Huang, Chelsea X. and Shporer, Avi and Daylan, Tansu and Muthukrishna, Daniel and Vanderburg, Andrew and Dattilo, Anne and Ricker, George R. and Seager, S.},
  year = {2023},
  month = feb,
  journal = {AJ},
  volume = {165},
  number = {3},
  pages = {95},
  issn = {1538-3881},
  doi = {10.3847/1538-3881/acad85},
  urldate = {2025-09-02},
  langid = {english}
}

@article{thompson2018kepler25,
  title = {Planetary {{Candidates Observed}} by {{Kepler}}. {{VIII}}. {{A Fully Automated Catalog}} with {{Measured Completeness}} and {{Reliability Based}} on {{Data Release}} 25},
  author = {Thompson, Susan E. and Coughlin, Jeffrey L. and Hoffman, Kelsey and Mullally, Fergal and Christiansen, Jessie L. and Burke, Christopher J. and Bryson, Steve and Batalha, Natalie and Haas, Michael R. and Catanzarite, Joseph and Rowe, Jason F. and Barentsen, Geert and Caldwell, Douglas A. and Clarke, Bruce D. and Jenkins, Jon M. and Li, Jie and Latham, David W. and Lissauer, Jack J. and Mathur, Savita and Morris, Robert L. and Seader, Shawn E. and Smith, Jeffrey C. and Klaus, Todd C. and Twicken, Joseph D. and Van Cleve, Jeffrey E. and Wohler, Bill and Akeson, Rachel and Ciardi, David R. and Cochran, William D. and Henze, Christopher E. and Howell, Steve B. and Huber, Daniel and Pr{\v s}a, Andrej and Ram{\'i}rez, Solange V. and Morton, Timothy D. and Barclay, Thomas and Campbell, Jennifer R. and Chaplin, William J. and Charbonneau, David and {Christensen-Dalsgaard}, J{\o}rgen and Dotson, Jessie L. and Doyle, Laurance and Dunham, Edward W. and Dupree, Andrea K. and Ford, Eric B. and Geary, John C. and Girouard, Forrest R. and Isaacson, Howard and Kjeldsen, Hans and Quintana, Elisa V. and Ragozzine, Darin and Shabram, Megan and Shporer, Avi and Aguirre, Victor Silva and Steffen, Jason H. and Still, Martin and Tenenbaum, Peter and Welsh, William F. and Wolfgang, Angie and Zamudio, Khadeejah A and Koch, David G. and Borucki, William J.},
  year = {2018},
  month = apr,
  journal = {ApJS},
  volume = {235},
  number = {2},
  pages = {38},
  issn = {0067-0049},
  doi = {10.3847/1538-4365/aab4f9},
  urldate = {2025-09-02},
  langid = {english}
}

@article{torres2011blender,
  title = {{{MODELING KEPLER TRANSIT LIGHT CURVES AS FALSE POSITIVES}}: {{REJECTION OF BLEND SCENARIOS FOR KEPLER-9}}, {{AND VALIDATION OF KEPLER-9}} d, {{A SUPER-EARTH-SIZE PLANET IN A MULTIPLE SYSTEM}}},
  shorttitle = {{{MODELING KEPLER TRANSIT LIGHT CURVES AS FALSE POSITIVES}}},
  author = {Torres, Guillermo and Fressin, Fran{\c c}ois and Batalha, Natalie M. and Borucki, William J. and Brown, Timothy M. and Bryson, Stephen T. and Buchhave, Lars A. and Charbonneau, David and Ciardi, David R. and Dunham, Edward W. and Fabrycky, Daniel C. and Ford, Eric B. and Gautier III, Thomas N. and Gilliland, Ronald L. and Holman, Matthew J. and Howell, Steve B. and Isaacson, Howard and Jenkins, Jon M. and Koch, David G. and Latham, David W. and Lissauer, Jack J. and Marcy, Geoffrey W. and Monet, David G. and Prsa, Andrej and Quinn, Samuel N. and Ragozzine, Darin and Rowe, Jason F. and Sasselov, Dimitar D. and Steffen, Jason H. and Welsh, William F.},
  year = {2010},
  month = dec,
  journal = {ApJ},
  volume = {727},
  number = {1},
  pages = {24},
  issn = {0004-637X},
  doi = {10.1088/0004-637X/727/1/24},
  urldate = {2025-09-02},
  langid = {english}
}

@article{twicken2018keplervalidation,
  title = {Kepler {{Data Validation I}}---{{Architecture}}, {{Diagnostic Tests}}, and {{Data Products}} for {{Vetting Transiting Planet Candidates}}},
  author = {Twicken, Joseph D. and Catanzarite, Joseph H. and Clarke, Bruce D. and Girouard, Forrest and Jenkins, Jon M. and Klaus, Todd C. and Li, Jie and McCauliff, Sean D. and Seader, Shawn E. and Tenenbaum, Peter and Wohler, Bill and Bryson, Stephen T. and Burke, Christopher J. and Caldwell, Douglas A. and Haas, Michael R. and Henze, Christopher E. and Sanderfer, Dwight T.},
  year = {2018},
  month = apr,
  journal = {PASP},
  volume = {130},
  number = {988},
  pages = {064502},
  issn = {1538-3873},
  doi = {10.1088/1538-3873/aab694},
  urldate = {2023-09-11},
  langid = {english}
}

@article{valizadegan2022exominer,
  title = {{{ExoMiner}}: {{A Highly Accurate}} and {{Explainable Deep Learning Classifier That Validates}} 301 {{New Exoplanets}}},
  shorttitle = {{{ExoMiner}}},
  author = {Valizadegan, Hamed and Martinho, Miguel J. S. and Wilkens, Laurent S. and Jenkins, Jon M. and Smith, Jeffrey C. and Caldwell, Douglas A. and Twicken, Joseph D. and Gerum, Pedro C. L. and Walia, Nikash and Hausknecht, Kaylie and Lubin, Noa Y. and Bryson, Stephen T. and Oza, Nikunj C.},
  year = {2022},
  month = feb,
  journal = {ApJ},
  volume = {926},
  number = {2},
  pages = {120},
  issn = {0004-637X},
  doi = {10.3847/1538-4357/ac4399},
  urldate = {2025-09-02},
  langid = {english}
}

@article{valizadegan2023exominer,
  title = {Multiplicity {{Boost}} of {{Transit Signal Classifiers}}: {{Validation}} of 69 {{New Exoplanets}} Using the {{Multiplicity Boost}} of {{ExoMiner}}},
  shorttitle = {Multiplicity {{Boost}} of {{Transit Signal Classifiers}}},
  author = {Valizadegan, Hamed and Martinho, Miguel J. S. and Jenkins, Jon M. and Caldwell, Douglas A. and Twicken, Joseph D. and Bryson, Stephen T.},
  year = {2023},
  month = jun,
  journal = {AJ},
  volume = {166},
  number = {1},
  pages = {28},
  issn = {1538-3881},
  doi = {10.3847/1538-3881/acd344},
  urldate = {2025-09-02},
  langid = {english}
}

@ARTICLE{valizadegan2025exominer,
       author = {{Valizadegan}, Hamed and {Martinho}, Miguel J.~S. and {Jenkins}, Jon M. and {Twicken}, Joseph D. and {Caldwell}, Douglas A. and {Maynard}, Patrick and {Wei}, Hongbo and {Zhong}, William and {Yates}, Charles and {Donald}, Sam and {Collins}, Karen A. and {Latham}, David and {Barkaoui}, Khalid and {Calkins}, Michael L. and {Carden}, Kylee and {Chazov}, Nikita and {Esquerdo}, Gilbert A. and {Guillot}, Tristan and {Krushinsky}, Vadim and {Nowak}, Grzegorz and {Rackham}, Benjamin V. and {Triaud}, Amaury and {Schwarz}, Richard P. and {Stephens}, Denise and {Stockdale}, Chris and {Watkins}, Cristilyn N. and {Wilkin}, Francis P.},
        title = "{ExoMiner++: Enhanced Transit Classification and a New Vetting Catalog for 2-minute TESS Data}",
      journal = {\aj},
         year = 2025,
        month = nov,
       volume = {170},
       number = {5},
          eid = {287},
        pages = {287},
          doi = {10.3847/1538-3881/ae03a4},
archivePrefix = {arXiv},
       eprint = {2502.09790},
 primaryClass = {astro-ph.EP},
       adsurl = {https://ui.adsabs.harvard.edu/abs/2025AJ....170..287V}
}

@article{vaneylen2018radiusvalley,
  title = {An Asteroseismic View of the Radius Valley: Stripped Cores, Not Born Rocky},
  shorttitle = {An Asteroseismic View of the Radius Valley},
  author = {Van~Eylen, V and Agentoft, Camilla and Lundkvist, M S and Kjeldsen, H and Owen, J E and Fulton, B J and Petigura, E and Snellen, I},
  year = {2018},
  month = oct,
  journal = {\mnras},
  volume = {479},
  number = {4},
  pages = {4786--4795},
  issn = {0035-8711},
  doi = {10.1093/mnras/sty1783},
  urldate = {2025-09-09}
}

@article{virtanen2020scipy,
  title = {{{SciPy}} 1.0: Fundamental Algorithms for Scientific Computing in {{Python}}},
  shorttitle = {{{SciPy}} 1.0},
  author = {Virtanen, Pauli and Gommers, Ralf and Oliphant, Travis E. and Haberland, Matt and Reddy, Tyler and Cournapeau, David and Burovski, Evgeni and Peterson, Pearu and Weckesser, Warren and Bright, Jonathan and {van der Walt}, St{\'e}fan J. and Brett, Matthew and Wilson, Joshua and Millman, K. Jarrod and Mayorov, Nikolay and Nelson, Andrew R. J. and Jones, Eric and Kern, Robert and Larson, Eric and Carey, C. J. and Polat, {\.I}lhan and Feng, Yu and Moore, Eric W. and VanderPlas, Jake and Laxalde, Denis and Perktold, Josef and Cimrman, Robert and Henriksen, Ian and Quintero, E. A. and Harris, Charles R. and Archibald, Anne M. and Ribeiro, Ant{\^o}nio H. and Pedregosa, Fabian and {van Mulbregt}, Paul and {SciPy 1. 0 Contributors}},
  year = {2020},
  month = feb,
  journal = {Nat. Methods},
  volume = {17},
  pages = {261--272},
  doi = {10.1038/s41592-019-0686-2},
  urldate = {2022-08-26}
}

@article{wang2024gpfc,
  title = {The {{GPU}} Phase Folding and Deep Learning Method for Detecting Exoplanet Transits},
  author = {Wang, Kaitlyn and Ge, Jian and Willis, Kevin and Wang, Kevin and Zhao, Yinan},
  year = {2024},
  month = mar,
  journal = {\mnras},
  volume = {528},
  number = {3},
  pages = {4053--4067},
  issn = {0035-8711},
  doi = {10.1093/mnras/stae245},
  urldate = {2025-09-02}
}

@article{wright2012hj,
  title = {{{THE FREQUENCY OF HOT JUPITERS ORBITING NEARBY SOLAR-TYPE STARS}}*},
  author = {Wright, J. T. and Marcy, G. W. and Howard, A. W. and Johnson, John Asher and Morton, T. D. and Fischer, D. A.},
  year = {2012},
  month = jun,
  journal = {ApJ},
  volume = {753},
  number = {2},
  pages = {160},
  issn = {0004-637X},
  doi = {10.1088/0004-637X/753/2/160},
  urldate = {2025-09-09},
  langid = {english}
}

@article{yu2019astronettriage,
  title = {Identifying {{Exoplanets}} with {{Deep Learning}}. {{III}}. {{Automated Triage}} and {{Vetting}} of {{TESS Candidates}}},
  author = {Yu, Liang and Vanderburg, Andrew and Huang, Chelsea and Shallue, Christopher J. and Crossfield, Ian J. M. and Gaudi, B. Scott and Daylan, Tansu and Dattilo, Anne and Armstrong, David J. and Ricker, George R. and Vanderspek, Roland K. and Latham, David W. and Seager, Sara and Dittmann, Jason and Doty, John P. and Glidden, Ana and Quinn, Samuel N.},
  year = {2019},
  month = jun,
  journal = {AJ},
  volume = {158},
  number = {1},
  pages = {25},
  issn = {1538-3881},
  doi = {10.3847/1538-3881/ab21d6},
  urldate = {2025-09-02},
  langid = {english}
}

@ARTICLE{Kunimoto2025,
       author = {{Kunimoto}, Michelle and {Bryson}, Steve and {Jaffee}, Drayson and {Rowe}, Jason F. and {Daylan}, Tansu and {Giacalone}, Steven and {Lissauer}, Jack J. and {Matesic}, Michael R.~B. and {Mullally}, Susan E. and {Eschen}, Yoshi Nike Emilia},
        title = "{LEO-Vetter: Fully Automated Flux- and Pixel-level Vetting of TESS Planet Candidates to Support Occurrence Rates}",
      journal = {\aj},
         year = 2025,
        month = nov,
       volume = {170},
       number = {5},
          eid = {280},
        pages = {280},
          doi = {10.3847/1538-3881/ae070a},
archivePrefix = {arXiv},
       eprint = {2509.10619},
 primaryClass = {astro-ph.EP},
       adsurl = {https://ui.adsabs.harvard.edu/abs/2025AJ....170..280K}
}




\appendix

\section{Summary plot example}

\begin{figure*}
\centering
\includegraphics[width=0.68\textwidth]{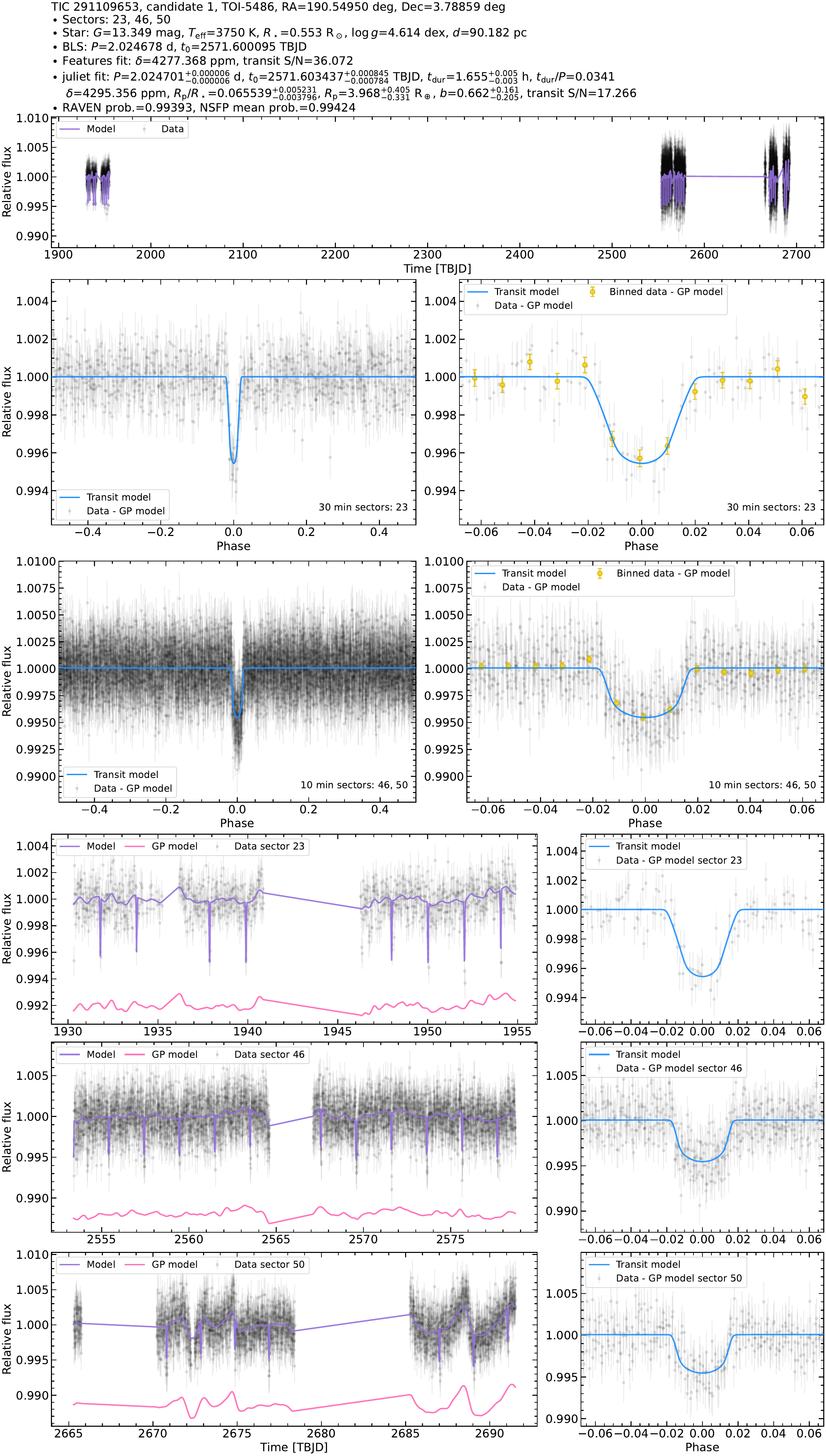}
\caption{Example of a summary plot of the validated candidate TOI-5486~b (TIC~291109653.01, recovered in BLS peak 1).
The first row shows the FFI light curves (PDCSAP flux) of all sectors used (black points) and the best-fitting \juliet model (purple line).
The second and third rows show the FFI data of the 30-minute (second row) and 10-minute (third row) sectors phase-folded to the candidate period (full phase on the left panel, zoom-in on the right panel), the same data binned to 30 minutes (yellow circles), and best-fitting transit model (blue line).
The rest of rows show, for each sector, the FFI data (black points), best-fitting \juliet model (purple line), and the GP part of the model (with an arbitrary offset added, pink line) on the left panels. The right panels show the GP-detrended, phase-folded sector data (black data points) and transit model (blue line).
}
\label{fig:summary}
\end{figure*}


\section{Additional tables}

\begin{table*}
\caption{Newly identified significant candidates detected as single transits by our pipeline in sectors 1-55. The Notes column details if further transit have been visually identified in sectors after 55 with the SPOC-FFI, main SPOC, and/or QLP reductions.}
\label{tab:mono}
{\setlength{\tabcolsep}{2pt}
\begin{tabular}{lcccccp{10cm}}
\hline
TIC & Candidate & BLS $t_0$ & NSFP & RAVEN & Sector & Notes \\
 &  & [TBJD]  &  mean prob. & prob. &  &  \\
\hline
103629155 & 3 & 1593.912017 & 0.976139 & 0.978489 & 10 & Single transit in sector 10 (FFI, QLP). Further observed in sector 64 (FFI, QLP, no transit seen), and to be observed in sectors 101 and 102. \\
258617228 & 1 & 2467.036849 & 0.957866 & 0.963600 & 42 & Single transit in sector 42 (FFI, QLP). Further observed in sectors 56 (FFI, QLP) and 83 (QLP). The QLP data of sector 56 shows a transit {of similar depth} $\sim368$~d after, but it falls on a data gap of the FFI data. Possible partial transit also seen in QLP sector 83, $\sim736$~d after the one in sector 56 (i.e. double the time gap between the first two transits), with the second half of the transit in a data gap.\\
356657070 & 1 & 1906.203867 & 0.910697 & 0.980421 & 22 & Single transit in sector 22 of the FFI. QLP data of the same sector shows a flux increase at the time of the apparent transit, which could indicate a false alarm. No further sectors.\\
434095530 & 1 & 2466.377152 & 0.909282 & 0.982688 & 42 & Single transit in sector 42 of the FFI. QLP data of the same sector might show a flux increase, which could indicate a false alarm, but the data generally shows a large scatter. Star further observed in sector 92 (QLP, no transit seen). \\
\hline
\end{tabular}
}
\end{table*}

\begin{table*}
\caption{Newly visually identified multi-candidate systems with a vetted or validated candidate. The table shows the two-planet \juliet fit parameters for the second candidate only if several transits are present in the FFI light curves.}
\label{tab:multi}
{\setlength{\tabcolsep}{2pt}
\begin{tabular}{lcccp{10.65cm}}
\hline
TIC & \multicolumn{3}{c}{Second candidate} & Notes \\
\cmidrule(lr){2-4}
    & \Porb [d] & $t_0$ [TBJD] & \Rp [\Rearth] &  \\
\hline
18942729 & $23.1108^{+0.0033}_{-0.0031}$ & $2505.8752^{+0.0045}_{-0.0044}$ & $3.64^{+0.35}_{-0.32}$ & Our BLS identified an ultra-short period super Earth in the third BLS peak ($P\simeq1.00$~d and $\Rp\simeq2.00\,\Rearth$, see also Sect. \ref{sec:usp}) which is in our validated sample. The only other significant peak in the BLS is peak 2, which has a period of $\sim13.7$~d and NSFP probability $\sim0.1$, likely due to the \tess sector length. Light curves of sectors 44, 45, and 46 show one extra transit each, significantly deeper than our BLS candidate. \\
77319217 & $32.365^{+0.011}_{-0.011}$ & $1426.8496^{+0.0075}_{-0.0077}$ & $4.05^{+0.34}_{-0.33}$ & Our BLS identified a $\simeq7.55$~d period signal with $\Rp\simeq3.11$~\Rearth in sectors 4, 5, 31, and 32, which is in our validated sample. We visually identified a larger outer candidate with $P\simeq32.36$~d and $\Rp\simeq4.05$~\Rearth in sectors 4 and 5 (one transit in each sector), while the planet does not transit in the data covered by sectors 31 and 32. This star will be observed again in sectors 98 and 107. \\
%
383482813 & $1.7480688^{+0.0000054}_{-0.0000056}$ & $1629.4858^{+0.0016}_{-0.0016}$ & $5.13^{+0.33}_{-0.32}$ & This star has a PC CTOI reported in \citet{montalto2023diamante} with $P\simeq2.01$~d and $\Rp\simeq15.06\,\Rearth$ (with a high impact parameter, $b\sim0.9$), which we recover in the first BLS peak and is in our vetted sample. The light curves of sectors 12 and 39 clearly show an inner, smaller candidate. This candidate is not recovered in the BLS because the other four selected peaks are all harmonics of the $\sim2$~d CTOI ($P/3$, $3P$, $P/4$, and $4P$), which is significantly deeper (and hence has a stronger BLS signal) than the inner candidate. The orbital parameters of the outer candidate derived from the original single-planet \juliet fit agree within $1\sigma$ with the new parameters derived here with the two-planet fit except for the impact parameter, now $b\sim1$, resulting in a larger radius of 24.58~\Rearth with large uncertainties (about $\pm5$~\Rearth). \\
\hline
\end{tabular}
\begin{tabular}{lp{16.5cm}}
TIC & Notes \\
\hline
131330900 & We identified one candidate in the first BLS peak with $P\simeq4.27$~d and $\Rp\simeq4.74\,\Rearth$, which is in our vetted sample. Sector 37 shows at least three transit-like features that could belong to further planets in the system, however their periods are unclear from the single sector. This star has only been observed in sector 37 and is scheduled to be observed in sector 102. \\
306990399 &  We recovered a known PC, TOI-5034.01, with $P\simeq8.70$~d and $\Rp\simeq3.99\,\Rearth$, in our vetted sample, from sectors 12, 13, and 39. Sector 39 shows three extra transit-like features that could be due to further planets in the system, however their periods are unclear. This star has been further observed in sectors 66 (in the FFI, main mission, and QLP) and 93 (main mission and QLP), but again these data do not show clear transits of possible extra candidates. This star will be observed again in sectors 100, 101, 102, 103, and 114. \\
404736082 & This star has a vetted candidate from sectors 42 and 53 with $P\simeq8.92$~d and $\Rp\simeq3.87\,\Rearth$. Sector 43 shows a single transit from an extra candidate. Further data from sectors 70 and 71 show two more transits, however it is unclear from the three events if they correspond to the same candidate. \\
408506314 & This star has a PC, TOI-6868.01, with $P\simeq3.41$~d and $\Rp\simeq11.92\,\Rearth$, which we recover in our vetted sample from the first BLS peak in data from sectors 42 and 43. Sector 42 shows a clear single transit of an extra candidate. Further data from sector 70 only show transits of the known TOI. \\
364561528 & This star has a CTOI reported by \citet{salinas2025transformer} with $P\simeq12.76$~d and $\Rp\simeq9.67\,\Rearth$ that we recover in our vetted sample from sectors 15, 16, and 55. The last transit of sector 55 overlaps with a transit from another candidate. Other extra transits from further QLP sectors (41, 56, 75, 76, 82, 83) constrain the period of this second candidate to $\sim27$~d. \\
\hline
\end{tabular}
}
\end{table*}


\section{Phase-folded light curves of new validated candidates}\label{sec:lc_val_new}

\begin{figure*}
\centering
\includegraphics[width=\textwidth]{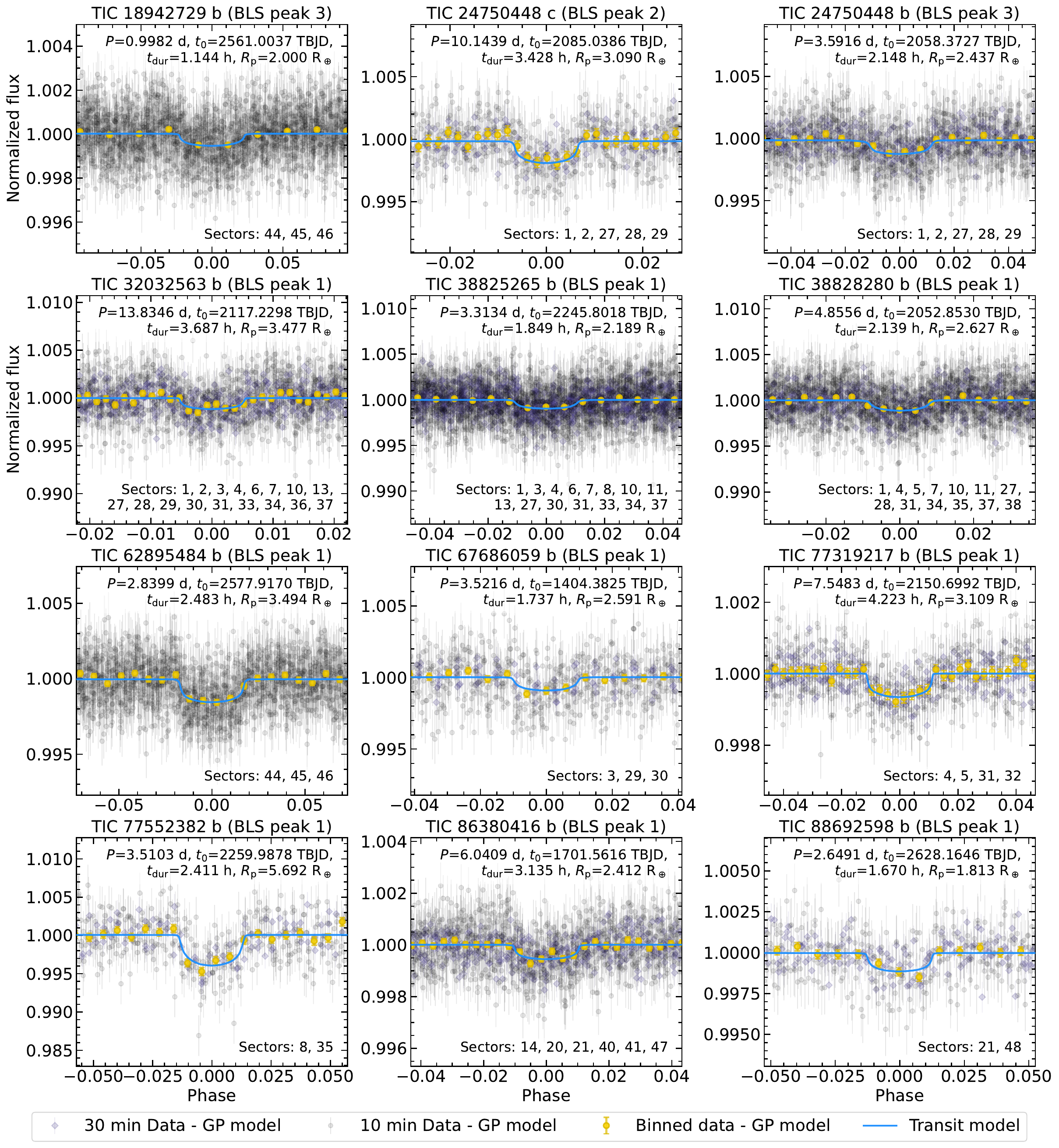}
\caption{Phase-folded light curves of all 31 new validated candidates first identified in this work. Grey circles and purple diamonds show the PDCSAP data from the 10- and 30-minute sectors, respectively, detrended by the GP model and phase-folded to the \Porb and $t_0$ from the \juliet fit. Yellow circles show the same data averaged to 30-minute bins. The transit model from the best-fitting \juliet parameters is shown as the solid blue line. Note that the model shown is that corresponding to the 10-minute cadence. The top inset text details the orbital period $P$, mid-transit time $t_0$, transit duration $t_\mathrm{dur}$, and planetary radius \Rp. The bottom inset text shows the sectors available for each target (up to sector 55).
}
\label{fig:lc_val_new0}
\end{figure*}

\begin{figure*}
\centering
\includegraphics[width=\textwidth]{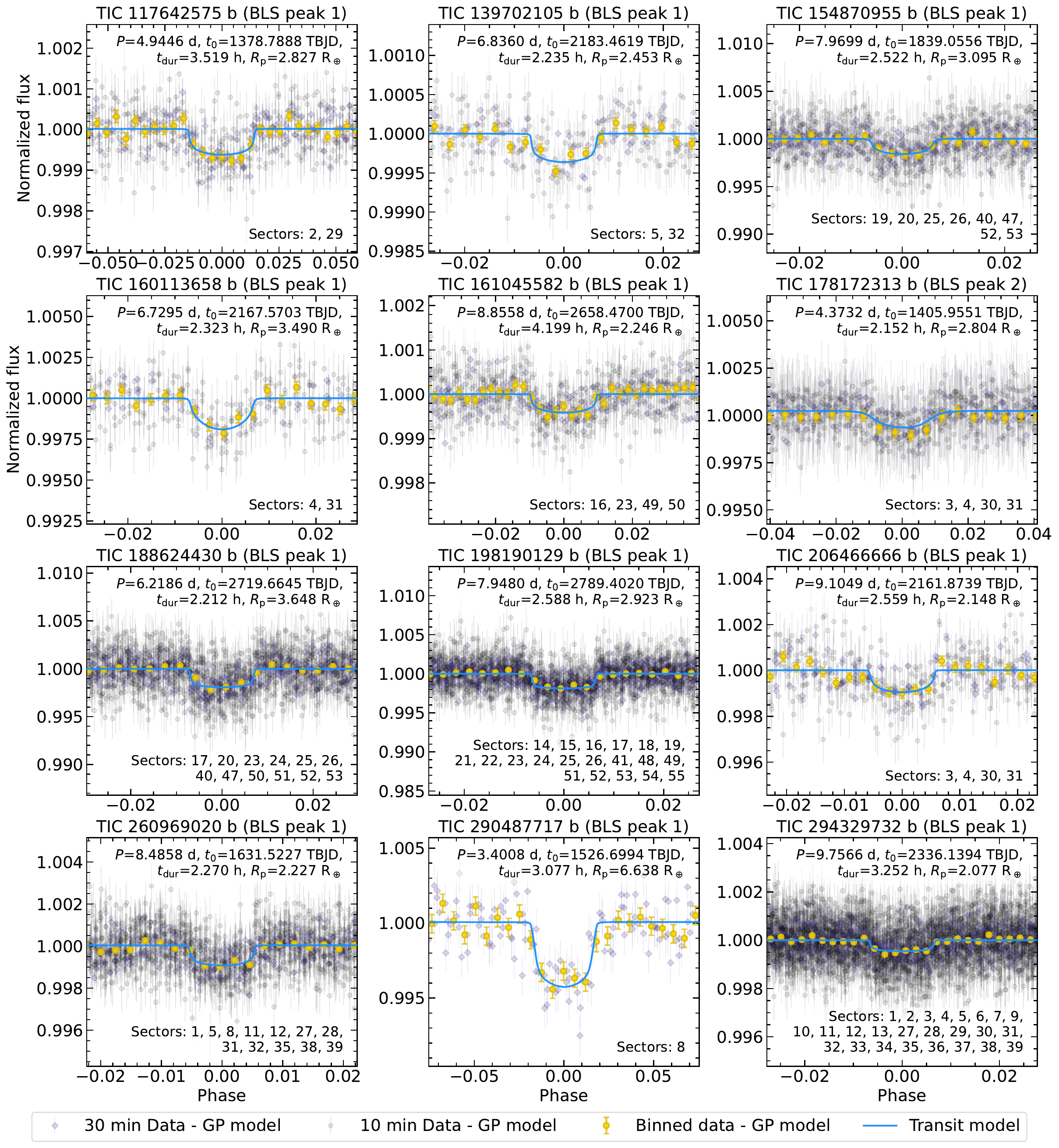}
\contcaption{}
\end{figure*}

\begin{figure*}
\centering
\includegraphics[width=\textwidth]{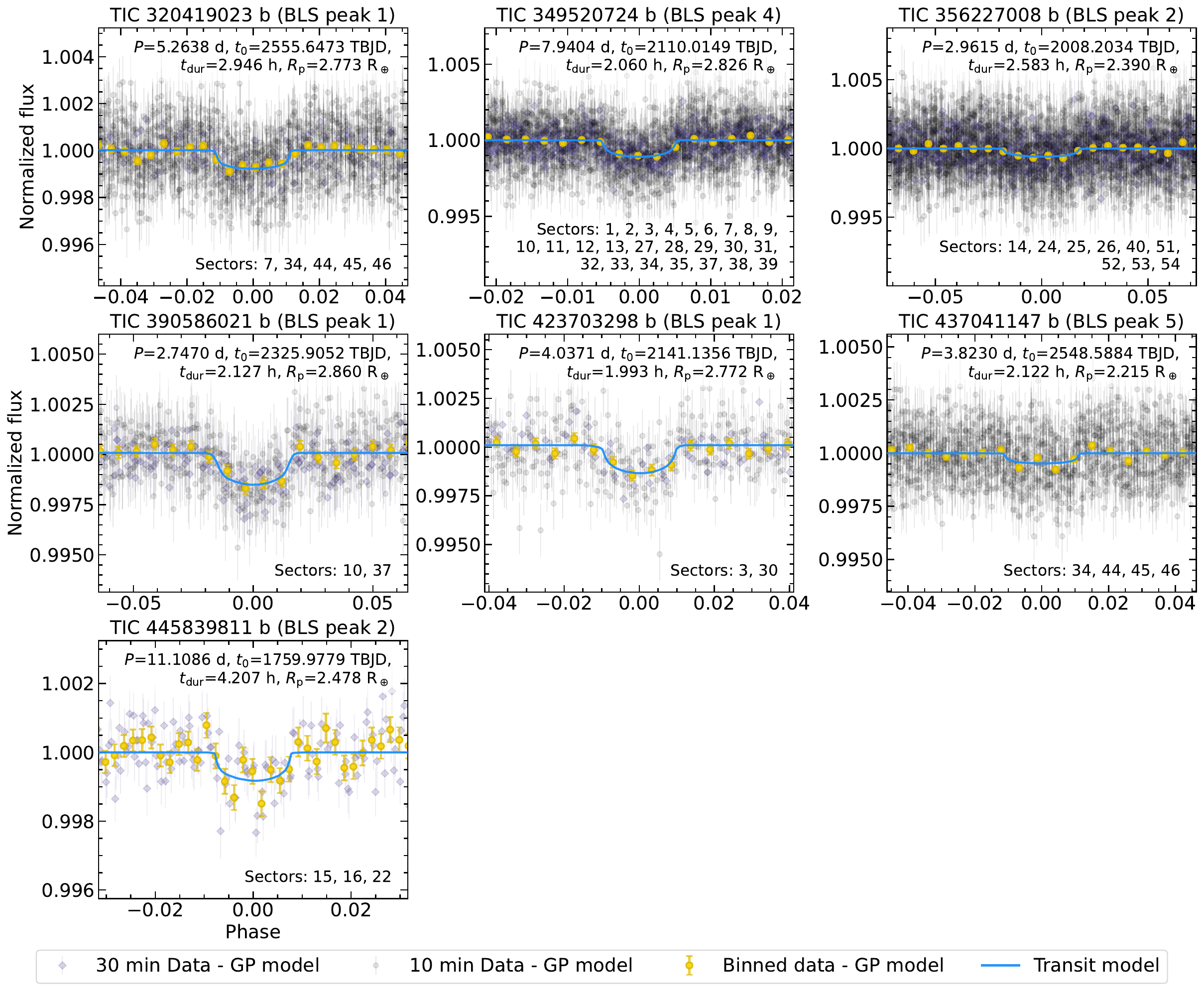}
\contcaption{}
\end{figure*}


\section{Vetted candidates in stars with known TOI/CTOI that partially match/do not match the reported parameters}\label{sec:cand_vet_toi_other_recovery}

\begin{itemize}
\item TIC 6404965 BLS peak 5: This star has a PC, TOI-3346.01, with $P\simeq4.87$~d, that we recover in the first BLS peak and is in our vetted sample. The candidate in the fifth BLS peak is also in the vetted sample and has a period of $\simeq6.49$~d, close to 4/3 of the TOI period, and matches only some of the transits corresponding to the TOI, making this candidate a complex harmonic of the true period.
\item TIC~20579360 BLS peak 1: This star has a PC, TOI-2532.01, with a reported $P\simeq16.72$~d just outside of the range of periods we searched. We recover its second harmonic ($P/2$), 8.36~d, in the first peak of our BLS search. A visual inspection of the light curves of sectors 10 and 37 (those used by our pipeline) shows two transits each separated by $\simeq$16.7~d. The missing transits needed in between those observed for the period to be $\simeq8.4$~d fall on the intra-sector gaps, which explains the recovered harmonic. Further data from sector 64 (not used by our pipeline) rules out the $\sim8$~d period.
\item TIC 23434737 BLS peak 1: This star has a PC, TOI-1203.01, with $P\simeq25.52$~d, outside of our search range. We recover a candidate with $P\simeq15.74$~d in the first BLS peak. A visual inspection of the light curves shows a single transit in sector 9, another single transit in sector 10, $\simeq25.5$~d after the first one, and another transit in sector 36, also corresponding to the period of $\simeq25.5$~d. Our BLS only recovered the transits in sectors 9 and 36 (and misses the one in sector 10), which also match a period of $\simeq15.74$~d, making this candidate a complex harmonic of the TOI period. The fourth BLS peak corresponds to the second harmonic of the TOI period, $\simeq12.77$~d, but this candidate is not in our vetted sample.
\item TIC 26999884 BLS peak 4: This star has a PC, TOI-7175.01, with $P\simeq2.99$~d, that we recover in the first BLS peak but is not in our vetted sample. The candidate in the fourth BLS peak is in the vetted sample and has a period of $\simeq1.00$~d, very close to 1/3 of the TOI period, and matches some of the transits corresponding to the TOI, making this candidate a likely harmonic of the true period.
\item TIC 37749396 BLS peak 2: This star has a CP, TOI-260, with $P\simeq13.48$~d, that we recover in the first BLS peak and is in our vetted sample. The candidate in the second BLS peak is also in the vetted sample, with a period of $\simeq11.10$, matching two transits of the known TOI and missing another transit. This makes this candidate a complex harmonic of the true signal.
\item TIC 39077574 BLS peak 1: This star has a FP, TOI-3825.01, with $P\simeq7.14$~d. We recover half the TOI period, $\simeq3.57$~d, in the first BLS peak in our vetted sample. The ExoFOP notes mention an eccentric EB with significant secondary ($\sim8000$~pppm, close to the primary depth). Therefore, our BLS is recovering both the primary and secondary, explaining the $P/2$ recovered.
\item TIC 72668830 BLS peak 2: This star has a PC, TOI-4296.01, with $P\simeq19.97$~d. We find a candidate with $P\simeq12.49$~d in the second BLS peak from sectors 3 and 30, close to 3/5 of the TOI period, and with a transit overlapping with that of the TOI. The light curve shows extra transits not matching either of the candidates.
\item TIC 73038411 BLS peak 2: This star has a PC, TOI-978.01, with $13.17$~d that we recover in the first BLS peak but is not in our vetted sample. The second BLS peak is in our vetted sample and has a period of $\simeq14.18$. A visual inspection of the light curve shows that only some of the transit are recovered, making this candidate a complex harmonic of the true signal.
\item TIC 73232401 BLS peak 2: This star has a CTOI, TIC~73232401.01, identified in \citet{salinas2025transformer}, with $P\simeq11.02$~d, which we recover in the first peak of the BLS, and it is also in our vetted sample. Our vetted sample also has a second candidate with $P\simeq11.35$~d in the second BLS peak, very close to the first peak. The depths of the two candidates are also very similar, $\sim1600$~ppm and $\sim1650$~ppm. A visual inspection of the light curves show that these two candidates correspond to the same transits, and our \texttt{juliet} fits for both converge on the same period of 11.01~d, consistent with the reported CTOI. Hence, this candidate is an artifact of the BLS.
\item TIC 88460991 BLS peak 2: This star has a PC, TOI-6392.01, with $P\simeq6.96$~d. We recover half of that period ($\simeq3.48$~d) in the second BLS peak, which is in our vetted sample. The first BLS peak recovers two times the TOI period. The true TOI period is not in any BLS peak, because it was probably removed during the BLS peak selection process.
\item TIC 95337971 BLS peak 2, 3, and 4: This star has a PC, TOI-5139.01, with $P\simeq5.95$~d that we recover in the first BLS peak and is in our vetted sample. We recover a third, a fifth, and a fourth of that period in the second, third, and fourth BLS peaks, respectively, all of which are in our vetted sample.
\item TIC 137020190 BLS peak 1: This star has a CTOI, 137020190.01, identified in \citet{salinas2025transformer}, with $P\simeq$23.06~d, outside of our search range. We recover the second harmonic in the first BLS peak, $P\simeq11.53$~d. Sector 22 shows two transits separated by $\sim23$~d and sector 49 shows a single transit (only sectors included in our search). The transits needed for the $\sim12$~d to be real fall on data gaps. Further data from sector 76 (not included in our pipeline) shows the same situation as sector 49. The QLP reduction suffers from the same data gap problem. It is unclear which one is the true period.
\item TIC 138126035 BLS peak 1: This star has a PC, TOI-2321.01 with very long period, $P\simeq716.35$~d, first identified as a single transit by Planet Hunters \tess \citet{eisner2021pht} and is now a duo-transit. We recover a candidate with $P\simeq12.57$~d in the first BLS peak. The recovered transits (at the end of sector 23, and another transit after a data gap in sector 50) match those in the archive and are separated separated by the reported period, which is a multiple of our recovered period of $P\simeq12.6$~d. 
\item TIC 142589416 BLS peak 2 (TIC~142589416.02): This star has a CTOI, TIC~142589416.01, identified in \citet{montalto2023diamante}, with $P\simeq2.10$~d, that we recover in the first BLS peak and is in our validated sample. In our vetted sample is BLS peak 2, with period $\simeq4.80$~d. This candidate does not seem to be an alias of the CTOI and the transits do not coincide, making this a possible new candidate.
\item TIC 147923561 BLS peak 2: This star has a PC, TOI-5715.02 with $P\simeq6.28$~d. We recover double this period, 12.56~d, in the second BLS peak. From a visual inspection of the light curves it is unclear which one is the true period since the transit is relatively shallow.
\item TIC 155005217 BLS peak 1: This star has a FP, TOI-1369.01, with a period of $P\simeq7.60$~d. Here, using data from sector 16, we recover this period in the first BLS peak, but the match with the mid-transit time fails. This is because the reported and our mid-transit times are separated in time, and the precision of the period and/or mid-transit time are not high enough to result in a match when propagated in time. Sector 16 shows tow clear transits, with the secondary eclipse falling in the intra-sector gap. This target has further data in sectors 56 and 57, outside of our search, which show clear secondary eclipses.
\item TIC 188589164 BLS peaks 2 and 5: This star has a known transiting planet, GJ~3929~b or TOI-2013.01, with $P\simeq2.62$~d, that we recover in the first BLS peak and is in our vetted sample. We also recover periods of $\simeq0.87$ and $\simeq6.54$~d in the second and fifth BLS peaks, corresponding roughly to 1/3 and 5/2 times the true TOI period. The transits of these two candidates coincide with those of the true TOI period, making them complex harmonics of that signal.
\item TIC 197807778 BLS peak 3: This star has a PC, TOI-4402.01, with $P\simeq3.70$~d that we recover in the first BLS peak and is in our vetted sample. We recover three times this period ($\simeq$11.10~d) in the third BLS peak, also in our vetted sample, which seems to be missing transits.
\item TIC 198153540 BLS peak 2: This star has two PCs, TOI-1798.01 and TOI-1798.02, with periods of $\simeq8.02$ and $\simeq0.44$~, respectively. We recover TOI-1798.01 in our first BLS peak in our validated sample. The candidate in the second BLS peak (in the vetted sample) is close to 2/3 the period of TOI-1798.01 ($\simeq$5.35~d) and some of its transits coincide, making this a complex harmonic of the true signal. Additionally, we also flagged this candidate has having low S/N.
\item TIC 204698337 BLS peak 1: This star has a PC, TOI-2535.01, with $P\simeq16.45$~d, just outside out BLS search range. We recover in the first BLS peak half that period, $\simeq8.24$~d, which is in our vetted sample.
\item TIC 209459275 BLS peak 5: This star has a CP, TOI-559~b, with $P\simeq6.98$~d that we recover in the first BLS peak in our vetted sample. The fifth BLS peak has a period of $\simeq13.19$~d and matches some of the transits corresponding to the true TOI period, making this candidate a complex harmonic of the true signal.
\item TIC 237101326 BLS peak 2: This star has a PC, TOI-4051.01, with $P\simeq1.54$~d. We recover the second harmonic in the second BLS peak, $P\simeq0.77$~d. We also recover double the reported period in the first BLS peak, $P\simeq3.07$~d, but this candidate is not in our vetted sample. The notes on ExoFOP mention that the period could be two times the reported one. From a visual inspection of the light curves, it is unclear which one is the true period since the transit is relatively shallow.
\item TIC 241062925 BLS peak 5: This star has PC, TOI-6345.01, with $P\simeq6.24$~d that we recover in the first BLS peak in our vetted sample. The fifth BLS peak has a period of $\simeq5.23$~d and matches some of the transits corresponding to the true TOI period, making this candidate a complex harmonic of the true signal.
\item TIC 245076932 BLS peak 1: This star has a CTOI, 245076932.01, identified in \citet{salinas2025transformer}, with $P\simeq$21.61~d, outside of our search range. We recover the second harmonic in the first BLS peak, $P\simeq10.81$~d. Both sector 11 and 38 (the only ones included in our search) show two transits separated by $\sim21.6$~d, with the extra transit required for the $\sim11$~d period to be true falling in the intra-sector gap. Further QLP data of sectors 37 and 64 rule out the $\sim11$~d period.
\item TIC 250864248 BLS peak 1: This star has a CTOI, 250864248.01, identified in \citet{montalto2023diamante}, with $P\simeq$16.45~d, outside of our search range. We recover the second harmonic in the first BLS peak, $P\simeq8.23$~d. Sector 18 (the only one used by our pipeline) shows two transits separated by $\sim16$~d, with the transit required for the period to be $\sim8$~d falling in the intra-sector gap. Further QLP data of sectors 58 and 85 rule out the $\sim8$~d period.
\item TIC 270471727 BLS peak 1: This star has a CTOI, 270471727.01, identified in \citet{melton2024dtarps2}, with $P\simeq$10.46~d. We recover the same period in the first peak of the BLS, but the mid-transit times do not match. Our mid-transit time is 2239.461110 TBJD, while the one in ExoFOP is 2458365.287 BJD. Using our recovered period of 10.456344~d, the two mid-transit times are $\sim1.6$~d apart (while our maximum required separation to count as recovered is 0.5~d). The two mid-transit time measurements are at very different times, separated by more than 6100 days, or more than 580 cycles. Therefore, it is possible that the period and/or mid-transit times precisions are not high enough to result in a match when propagated a large number of cycles. This candidate is also in our validated sample.
\item TIC 285094173 BLS peak 1: This star has a PC, TOI-1817.01, with $P\simeq13.40$~d. We recover half this period ($\simeq6.70$~d) in the first BLS peak. Notes on ExoFOP mention that the true period could be half of the reported one due data gap in sector 49. A visual inspection of the light curves of sectors 22 and 49 used here show transits matching the BLS period of $\simeq6.70$~d, which if true make this candidate actually correctly recovered.
\item TIC 294394558 BLS peak 2 (TOI-6484.02): This star has a PC, TOI-6484.01, with $P\simeq18.25$~d, outside of our search range. We find a candidate in the second BLS peak with $P\simeq10.90$~d, which is close to $5/3$ times the TOI period and is in our vetted sample.
We also recover the second harmonic of the TOI period in the first BLS peak, but this candidate is not in our vetted sample.
The transits of the two candidates do not overlap and their depths are different, making this a likely new candidate, see Sect. \ref{sec:new_toi}.
\item TIC 305048116 BLS peak 1: This star has a PC, TOI-362.01, with $P\simeq14.23$~d. We recover half this period ($\simeq7.12$~d) in the first BLS peak. A visual inspection of the light curves of sector 2 and 29 show transits corresponding to the $\simeq7.12$~d period. Two different user-uploaded planet parameters from the TFOPWG reported on ExoFOP show a period of $\simeq7.12$~d, making this candidate a correct recovery.
\item TIC 328012209 BLS peak 1: This star has a PC, TOI-2309.01, with $P\simeq20.27$~d, outside of our search range. We recover half this period ($\simeq10.14$~d) in the first BLS peak. A visual inspection of the light curve shows that one of the transit that constrains the short period is actually due to a trend.
\item TIC 330687113 BLS peak 5: This star has a KP, K2-334~b or TOI-5089.01, with $P\simeq5.11$~d, that we recover in the first BLS peak but it is not in our vetted sample. BLS peak 5 is in our vetted sample and corresponds to a fifth of the KP period ($\simeq1.02$~d), with some of the transits overlapping with those corresponding to the true period.
\item TIC 332564140 BLS peak 1, 3 and 4: This star has a PC, TOI-5400.01, with $P\simeq19.81$~d, outside of our search range. In our vetted sample, we recover half of the true period in BLS peak 1. BLS peaks 3 and 4, also in our vetted sample, recover only some of the true transits in a complex harmonic of the true period.
\item TIC 335540507 BLS peak 5: This star has a PC, TOI-5526.01, with $P\simeq21.73$~d, outside of our search range. We recover its second harmonic, $\simeq10.86$~d in the first BLS peak, but this candidate is not in our vetted sample. Sector 10 shows two transits separated by about 10.8 d. Sector 36 and 46 each show two transits separated by 21.7 d, with the extra transits needed for a period of $\sim10.9$~d falling on the intra-sector gap. All these data indicate that the true period is $\simeq$10.86~d rather than the reported $P\simeq21.73$~d. BLS peak 5, with a period of $\simeq11.44$~d, is in our vetted sample, but the transits seem to correspond to trends and the \juliet fits is unreliable.
\item TIC 337129672 BLS peak 1: This star has a FP (an eclipsing binary), TOI-4635.01, with $P\simeq49.11$~d, ourtside of our search range. We find a candidate with $P\simeq12.27$~d in the first BLS peak, which corresponds to the fourth harmonic ($P/4$) of the TOI period. The fourth harmonic of $\simeq49$~d is the first one within our search range. A visual inspection of the light curves of sector 42 and 43 show one transit each, separated by 49~d. The gaps in and between the light curves allow for missing transits at a period of $\simeq12.27$~d.
\item TIC 339207847 BLS peak 1: This star has a PC, TOI-3727.01, with $P\simeq$13.40~d. We recover its second harmonic, $\simeq$6.70~d, in the first BLS peak. A visual inspection of the light curve of sector 19 (the only sector used by our pipeline) shows two transits separated by $\simeq$13.4~d. The extra transit needed for the period to be $\simeq$6.7~d falls on the intra-sector gap. Further data from sector 59 rule out the $\simeq$6.7~d period.
\item TIC 346673534 BLS peak 1: This star has a PC, TOI-4084.01, with $P\simeq15.63$~d. We recover its second harmonic, $\simeq$7.82~d, in the first BLS peak. A visual inspection of the light curves of sectors 53 and 54 shows two transits each. The two transits of each light curve are separated by $\simeq15.6$~d, but the last transit of sector 53 and the first transit of sector 54 are separated by $\simeq7.8$~d. The extra transits needed in between those observed for the period of be of $\simeq7.8$~d fall on data gaps. Further QLP data from sector 40 also shows transits separated by $\simeq7.8$~d. All these data indicate that the true period is $\simeq$7.82~d rather than the reported $P\simeq15.63$~d.
\item TIC 376353509 BLS peak 4: This star has a PC, TOI-1648.01, with $P\simeq7.33$~d, that we recover in the second BLS peak and is in our vetted sample. In BLS peak 4 we recover a third of the TOI period.
\item TIC 381714186 BLS peak 4: This star has a PC, TOI-1839.01 with $P\simeq1.42$~d, and a CTOI with $P\simeq4.02$~d, that we recover in the first and second BLS peaks, respectively, and are in our validated sample. In our vetted sample we have BLS peak 4, with period $\simeq5.37$, close to 4.3 the CTOI period and with overlapping transits, making this a complex harmonic of the CTOI signal.
\item TIC 388909695 BLS peak 2: This star has a KP, WASP-134~b or TOI-5812.01 with $P\simeq0.15$~d, that we recover in the first BLS peak and is in our vetted sample (this star also has a PC with a single transit, TOI-5812.02). Also in our vetted sample we have BLS peak 2 with period $\simeq6.77$~d, close to 2/3 of the true TOI period with matching transits and that hence corresponds to some complex harmonic.
\item TIC 435868942 BLS peak 1 and 2: This star has a CTOI, TIC~435868942.01, with $P\simeq2.69$~d identified in \citet{melton2024dtarps2}. We have a match in period in our first BLS peak but the match in mid-transit time $t_0$ fails. It seems that the reported $t_0$ is offset in phase by 0.5 compared to our candidate, which results in the failed match, making this a recovered candidate. BLS peak 2 has a period if $\simeq8.07$~d and is also in our vetted sample. This candidate corresponds to three times the CTOI period and is a clear harmonic.
\item TIC 438260486 BLS peak 5: This star has a PC, TOI-5084.01, with $P\simeq5.83$~d that we recover in the second BLS peak but is not in our vetted sample. BLS peak 5, with period $\simeq13.10$~d, is in our vetted sample and seems to be due to trends and the \juliet fits is unreliable.
\item TIC 443582629 BLS peak 3: This star has a PC, TOI-5519.01, with $P\simeq5.85$~d that we recover in the first BLS peak and is in our vetted sample. BLS peak 3 is also in our vetted sample and corresponds to a third of the TOI period, with some of the transits overlapping with those corresponding to the true period.
\item TIC 445837596 BLS peak 5: This star has a PC, TOI-3896.01, with $P\simeq4.38$~d that we recover in the first BLS peak and is in our vetted sample. BLS peak 5 is also in our vetted sample and corresponds to a quarter of the TOI period, with some of the transits overlapping with those corresponding to the true period.
\item TIC 458424950 BLS peak 1: This star has a PC, TOI-3893.01, with $P\simeq$9.06~d. Our pipeline finds a candidate at the same period in the first BLS peak, but the match in mid-transit time failed. A visual inspection of the phase-folded light curves at the transit time reported in ExoFOP and the one found by our pipeline shows that the transit time reported in ExoFOP is offset by 0.5 in phase (i.e., the transit happens at phase 1, rather than phase 0), making this a recovered candidate.
\item TIC 464802921 BLS peak 2: This star has a CTOI, TIC~464802921.01, with $P\simeq1.66$~d identified in \citet{montalto2023diamante} that we recover in our first BLS peak but it is not in our vetted sample. The second BLS peak, with period $\simeq0.55$~d, is in our vetted sample. This candidate is close to a third of the CTOI period, has matching transits an is likely a harmonic.
\item TIC 469775147 BLS peak 1: This star has a CTOI, TIC~469775147.01, with $P\simeq2.61$~d identified in \citet{melton2024dtarps2}. We have a match in period in our first BLS peak but the match in mid-transit time $t_0$ fails. The reported $t_0$ is relatively far apart in time from our value, and if the periods and/or mid-transit times are not precise enough, they can fail to result in a match when propagated. The reported $t_0$ could also be offset in phase by 0.5 compared to our candidate, which could also be the cause of the failed match. Either option make this a recovered candidate.
\end{itemize}


\bsp	
\label{lastpage}
\end{document}